\documentclass[a4paper,11pt]{article}
\pdfoutput=1 
\usepackage{graphicx}  % needed for figures
\usepackage{dcolumn}   % needed for some tables
\usepackage{bm,relsize}        % for math
\usepackage{amssymb, amsmath}
\usepackage{textcomp}
\usepackage{wasysym}
\usepackage{slashed}
\usepackage{caption, subcaption}
\usepackage{multirow}

\usepackage{lipsum, color}
\usepackage[usenames,dvipsnames,svgnames]{xcolor}
\usepackage{braket}
\usepackage{jheppub} 
\usepackage{booktabs}
\usepackage{pdflscape}
\usepackage[utf8]{inputenc}

\newcommand{\C}{{\cal C}}
\newcommand{\Q}{{\cal Q}}
\newcommand{\CO}{\mathcal{O}}

\def\beq{\begin{equation}}
\def\eeq{\end{equation}}
 \def\be{\begin{equation}} \def\ee{\end{equation}}
\def\bea{\begin{eqnarray}} \def\eea{\end{eqnarray}}

\newcommand{\Tr}{{\rm Tr}}

\newcommand{\mutoenretvone}{{\tt Mu2e{\_}NRET{\_}v1}}
\newcommand{\mutoenretvtwo}{{\tt Mu2e{\_}NRET{\_}v2}}

\usepackage{caption}
\usepackage{subcaption}
\usepackage{multirow}
\usepackage[normalem]{ulem}
\allowdisplaybreaks

\begin{document}

\title{Effective theory tower for $\mu \to e$ conversion}

\author[a,b]{Wick Haxton,}
\author[a,b]{Kenneth McElvain,}
\author[b,c]{Tony Menzo,}
\author[a,d]{Evan Rule}
\author[c]{and Jure Zupan}
\affiliation[a]{Department of Physics, University of California, Berkeley, CA 94720, USA}
\affiliation[b]{Lawrence Berkeley National Laboratory, Berkeley, CA 94720, USA}
\affiliation[c]{Department of Physics, University of Cincinnati, Cincinnati, Ohio 45221, USA}
\affiliation[d]{Theoretical Division, Los Alamos National Laboratory, Los Alamos, NM 87545, USA}

\emailAdd{haxton@berkeley.edu}
\emailAdd{kenmcelvain@berkeley.edu}
\emailAdd{menzoad@mail.uc.edu}
\emailAdd{erule@berkeley.edu}
\emailAdd{zupanje@ucmail.uc.edu}

\date{\today}

\preprint{LA-UR-24-24937, N3AS-24-023}

\abstract{We present theoretical predictions for $\mu \rightarrow e$ conversion rates using a tower of effective field theories connecting the UV to nuclear physics scales. The interactions in nuclei are described using a recently developed nonrelativistic effective theory (NRET) that organizes contributions according to bound nucleon and muon velocities, $\vec{v}_N$ and $\vec{v}_\mu$, with $|\vec{v}_N| > |\vec{v}_\mu|$. To facilitate the top-down matching, we enlarge the set of Lorentz covariant nucleon-level interactions mapped onto the NRET operators to include those mediated by tensor interactions, in addition to the scalar and vector interactions already considered previously, and then match NRET nonperturbatively onto the Weak Effective Theory (WET). At the scale $\mu \approx 2$ GeV WET is formulated in terms of $u$, $d$, $s$ quarks, gluons and photons as the light degrees of freedom, along with the flavor-violating leptonic current. We retain contributions from WET operators up to dimension 7, which requires the full set of 26 NRET operators. The results are encoded in the open-source {\tt Python}- and {\tt Mathematica}-based software suite {\tt MuonBridge}, which we make available to the theoretical and experimental communities interested in $\mu \rightarrow e$ conversion. }

\maketitle

\section{Introduction}
The observation of neutrino oscillations \cite{Super-Kamiokande:1998kpq,Gonzalez-Garcia:2002bkq,Balantekin:2013tqa} shows that lepton flavor is not conserved in nature. This raises the question of whether similar flavor violation might occur among charged leptons. In the Standard Model (SM), neutrino loop amplitudes for charged lepton flavor violation (CLFV) are suppressed by the ratio of neutrino and $W$ masses, $\Delta m^2_\nu/m_W^2 \approx 10^{-24}$, yielding rates too small to be observed experimentally. 
 Consequently, any observation of CLFV would be evidence of new physics. There are many plausible modifications of the SM --- including supersymmetry, leptoquarks, heavy neutrinos, or a more complicated Higgs sector --- that could induce observable levels of CLFV.  Current upper bounds on CLFV branching ratios thus provide stringent constraints on such UV SM extensions \cite{Calibbi:2017uvl}.

Among the most sensitive CLFV measurements are those performed using stopped muons. The current best limits on the CLFV branching ratios $B(\mu^+\rightarrow e^+\gamma)<4.2\times 10^{-13}$ and $B(\mu^+\rightarrow e^+e^-e^+)<1.2\times 10^{-12}$ were obtained by MEG \cite{MEG:2016leq} and SINDRUM \cite{SINDRUM:1987nra}, respectively. These bounds, given at
90\% confidence level (C.L.), should be significantly improved by the ongoing experiments MEG II \cite{MEGII:2018kmf} and Mu3e \cite{Mu3e:2020gyw}. Upcoming experiments can also search for exotic CLFV decays of muons involving light new physics states \cite{Calibbi:2020jvd,Hostert:2023gpk,Jho:2022snj,Knapen:2023iwg,Knapen:2023zgi,Hill:2023dym,Echenard:2014lma,Perrevoort:2018ttp}. 

The main focus of the present work is the neutrinoless conversion of a muon bound to an atomic nucleus, $\mu^-+(A,Z)\rightarrow e^- + (A,Z)$.  Experiments under construction at Fermilab and J-PARC are expected to improve limits on the $\mu \rightarrow e$ conversion rate by approximately four orders of magnitude. Here we provide a systematic treatment of the theory of $\mu \to e$ conversion based on effective field theory, with the goal of creating a general analysis framework for relating experimental bounds to high-scale sources of CLFV.

The quantity extracted from experiment is the branching ratio for $\mu \rightarrow e$ conversion
\begin{equation}
\label{eq:B:mu2e}
B(\mu^-\rightarrow e^-)=\frac{\Gamma\left[\mu^-+(A,Z)\rightarrow e^-+(A,Z)\right]}{\Gamma\left[\mu^-+(A,Z)\rightarrow \nu_\mu +(A,Z-1)\right]},
\end{equation}
with respect to ordinary muon capture in a nucleus with $A$ nucleons and $Z$ protons.  SM muon capture rates are generally available for targets of interest, due to the long history of measurements \cite{PhysRevC.35.2212}. The most stringent current bounds on $\mu \rightarrow e$ branching ratios are $B(\mu^-+$Ti$\rightarrow e^-+$Ti$)<6.1 \times 10^{-13}$ \cite{Wintz:1998rp} and $B(\mu^-+$Au$\rightarrow e^-+$Au$)<7 \times 10^{-13}$ \cite{SINDRUMII:2006dvw}. Both of these 90\% C.L. limits were established by the SINDRUM II collaboration. By the end of this decade, the new Fermilab (Mu2e \cite{Mu2e:2014fns,Bernstein:2019fyh}) and J-PARC (COMET \cite{COMET:2018auw,COMET:2018wbw}) experiments are expected to reach branching ratio sensitivities of about 10$^{-17}$, probing new physics scales in excess of $10^4$ TeV. Both experiments will measure electrons produced from the conversion of a muon bound in the $1s$ atomic orbital of $^{27}$Al. 

The properties of nuclear targets --- the nucleon number $A$ and proton number $Z$, the ground-state spin and isospin, and their responses to operators that involve orbital angular momentum $\vec{\ell}$, spin $\vec{s}$, or spin-orbit correlations $\vec{\ell} \cdot \vec{s}$ --- affect the relationship between $\mu \rightarrow e$ conversion bounds and underlying sources of CLFV. Comparable branching ratio limits obtained from different nuclear targets will differ in their sensitivity to a given source of CLFV.  This is an attractive feature of $\mu \rightarrow e$ conversion that distinguishes this process from $\mu \rightarrow e \gamma$ or $\mu \rightarrow 3e$: by using several nuclear targets, one can place multiple constraints on the CLFV mechanism.  Were $\mu \rightarrow e$ conversion to be discovered in a given target, one could conduct additional studies with suitably selected targets to further characterize the CLFV source.

On the other hand, the interpretation of experimental limits on $\mu\to e$ conversion is, compared to $\mu\rightarrow e\gamma$ and $\mu\rightarrow 3e$, significantly more complicated due to the use of nuclear targets. Although the literature on $\mu\rightarrow e$ conversion is extensive\footnote{Table I of Ref. \cite{Haxton:2022piv} provides a relatively thorough overview of previous elastic $\mu \to e $ conversion studies \cite{Weinberg:1959zz,Marciano:1977cj,Shanker:1979ap,Kosmas:1990tc,Chiang:1993xz,Kosmas:1994ti,Czarnecki:1998iz,Siiskonen:2000nz,Kosmas:2001ia,Kosmas:2001ij,Kitano:2002mt,Cirigliano:2009bz,Crivellin:2017rmk,Bartolotta:2017mff,Cirigliano:2017azj,Davidson:2017nrp,Civitarese:2019cds,Rule:2021oxe,Heeck:2022wer,Cirigliano:2022ekw,Hoferichter:2022mna}.}, most existing work has focused on the special case of coherent conversion, where the leading response is governed by a scalar, isoscalar operator that sums coherently over all nucleons in the target. With this operator choice, the nuclear physics simplifies dramatically, allowing one to compute the $\mu\rightarrow e$ conversion rate using experimentally determined proton and neutron densities (see, e.g., Ref. \cite{Kitano:2002mt}).

While studies of the coherent operator have provided useful constraints on sources of CLFV~\cite{Cirigliano:2009bz,Cirigliano:2022ekw,Ardu:2024bua}, symmetry arguments show that six nuclear response functions arise in a general description of elastic $\mu \rightarrow e$ conversion \cite{Haxton:2022piv}. As we do not know a priori which CLFV source is realized in nature, a proper interpretation of experiment requires that such a general description be used.
The situation is reminiscent of another elastic process, the scattering of a heavy dark matter (DM) particle off a nucleus, the process exploited in DM direct-detection searches. Over the last decade, an attractive formalism was developed for DM direct detection based on a tower of effective theories that link the low-energy nuclear scale, where experiments are performed, to the UV scale where the new DM interactions reside. The starting point is the nonrelativistic effective theory (NRET)  \cite{Fan:2010gt,Fitzpatrick:2012ix}, consisting of all possible heavy DM-nucleon interactions that can be constructed from the available hermitian operators. This guarantees that all symmetry-allowed nuclear responses will be generated. The NRET is ultimately connected to UV theories of DM through a series of matchings involving a tower of effective field theories \cite{Bishara:2018vix,Baumgart:2022vwr}, encoded in a combination of \texttt{DirectDM} \cite{Bishara:2017nnn} and \texttt{DMFormFactor} \cite{Anand:2013yka} computer codes, that can then be used to make leading-order predictions for direct-detection scattering rates for (almost) any heavy DM theory.     

\begin{figure}[t]
    \centering
    \includegraphics[width=0.65\textwidth]{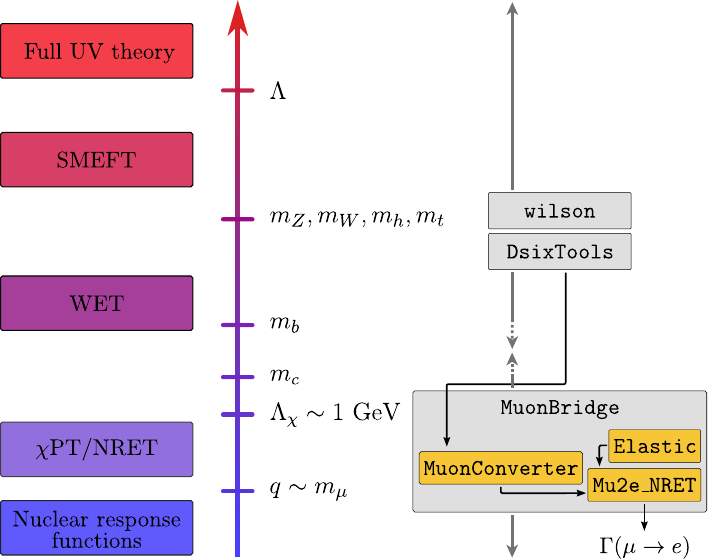}
    \caption{The effective field theories --- and their respective thresholds --- relevant for describing $\mu \to e$ conversion, as well as the corresponding computer codes (in grey boxes). The {\tt MuonBridge} repository,  introduced as part of this manuscript, consists of the interconnected packages {\tt MuonConverter}, {\tt Mu2e\_NRET}, and {\tt Elastic}  (orange boxes).
    \label{fig:diagram}}
\end{figure}

In this manuscript, we develop a similar bottom-up approach for $\mu \rightarrow e$ conversion, with the corresponding tower of EFTs shown in Fig.~\ref{fig:diagram}. Starting from a UV theory and integrating out heavy mediators, one recovers the SM augmented by higher dimension operators --- the so-called Standard Model Effective Field Theory (SMEFT). Only a subset of SMEFT operators are relevant for the problem at hand --- those that induce $\mu \rightarrow e$ transitions. At the electroweak scale the $Z$ and $W$ gauge bosons, the Higgs, and the top quark are integrated out, giving rise to Weak Effective Theory (WET) describing CLFV interactions with  currents constructed from quark, gluon or electromagnetic gauge fields, containing either five (above $b$ quark mass), four (above roughly the $c$ quark mass) or three ($u$, $d$, $s$) active quark flavors.  As our default choice we take the three-flavor WET at $\mu = 2$ GeV, a scale choice most commonly used in lattice QCD evaluations of the hadronic matrix elements.  The three-flavor WET can be matched nonperturbatively onto covariant nucleon-level interactions, describing physics below $\Lambda_\chi \approx m_\rho \approx 1$ GeV, where the dynamics due to the omitted degrees of freedom (such as the vector meson resonances) is included through the momentum exchange dependence of the single-nucleon form factors.  
Because $\mu \rightarrow e$ is characterized by the three-momentum transfer of size $q \approx m_\mu$, a
more efficient NRET can then be obtained by nonrelativistic reduction. The NRET  basis is compatible with the standard nonrelativistic many-body methods, like the shell model, that are typically employed in nuclear response function evaluations. 

The prerequisite nuclear-level NRET was developed recently in Refs.~\cite{Rule:2021oxe,Haxton:2022piv}, together with the accompanying public code \mutoenretvone, available in {\tt Mathematica} and {\tt Python}. In this work, we introduce an updated code, \mutoenretvtwo{}, with two major changes. First, we extend the code to include the leading muon-velocity-suppressed contributions, which were calculated in Ref.~\cite{Haxton:2022piv} but not included in \mutoenretvone. Second, to facilitate the matching between NRET and WET, we extend the set of covariant scalar- and vector-mediated interactions discussed in Ref.~\cite{Haxton:2022piv} to include tensor mediators. In addition, we provide a new open-source code {\tt MuonConverter} (also available in {\tt Python} and {\tt Mathematica}), that facilitates the connection between NRET and WET and provides an interface to SMEFT through optional linking to dedicated external codes \texttt{wilson} \cite{Aebischer:2018bkb} and \texttt{DsixTools} \cite{Celis:2017hod,Fuentes-Martin:2020zaz}, see Fig.~\ref{fig:diagram}. 
 
To allow for independent usage, {\tt MuonConverter}, {\tt Mu2e\_NRET}, and {\tt Elastic} --- a database of shell-model one-body density matrices that are required for the evaluation of NRET nuclear form factors --- are integrated together within a single parent repository, {\tt MuonBridge}, containing documentation, examples, and the appropriate instructions for assembling the independent components, depending on the specific goals of user.

The paper is organized as follows: In Sec.~\ref{sec:LEFT} we introduce a convenient WET operator basis up to dimension 7 for use in $\mu \to e$ conversion calculations. Sec.~\ref{sec:NRET} describes the NRET basis of 26 operators that arises in an expansion to linear order in $\vec{v}_N$ and $\vec{v}_\mu$, and the matching from WET to the NRET is performed in Sec.~\ref{sec:matching}. As an illustration, in Sec.~\ref{sec:NewPhysicsModels} we perform the EFT matching for two new physics models where $\mu\rightarrow e$ conversion is mediated by either heavy leptoquarks or light axion-like particles (ALPs). We also derive model-independent bounds on individual SMEFT operators. Sec.~\ref{sec:summary} contains our summary and outlook. 

Several appendices are included with further technical details:  App.~\ref{app:Public} describes the
{\tt MuonBridge} repository, App. \ref{app:more:WET:NRET} contains intermediate results for WET to NRET matching, including 
the mapping from WET onto the covariant Dirac basis
of 32 operators used as an intermediate step in the matching. 
App.~\ref{app:AFF} extends the results of Sec.~\ref{sec:matching} to the case of second-class currents. Details on the numerical evaluation of nucleon form factors are presented in App. \ref{app:nucleon:ff:values}. Finally, in App. \ref{app:wet-translation} we give the translation between our WET basis and the dimension-6 WET three-flavor basis of Ref.~\cite{Jenkins:2017jig}, which is used by the {\tt MuonConverter} to interface with other existing SMEFT/WET software. 

\section{Weak effective theory basis}
\label{sec:LEFT}
The tower of EFTs in Fig.~\ref{fig:diagram} relates UV-scale physics to the nuclear scale where $\mu\to e$ conversion experiments are performed. In this manuscript, we focus on the last three steps in the ladder of EFT matchings: the Weak Effective Theory (WET) at $\mu=2\,$GeV, the matching to NRET, and the prediction of conversion rates for nuclear targets of experimental interest using state-of-the-art shell-model methods to evaluate nuclear response functions. The formalism for relating the UV physics at an arbitrary high scale $\Lambda$ to the WET at $\mu=2\,$GeV is, on the other hand, well developed \cite{Jenkins:2013zja,Jenkins:2013wua,Alonso:2013hga,Jenkins:2017jig,Brivio:2017vri,Dekens:2019ept,Bern:2020ikv} and available in the form of public codes \cite{Aebischer:2018bkb,Fuentes-Martin:2020zaz,DiNoi:2022ejg}.

The part of the WET Lagrangian at $\mu=2$ GeV relevant for describing the $\mu\rightarrow e$ conversion process is given by\footnote{Our basis at $\mu=2\,$GeV follows from \cite{Liao:2020zyx}, though we choose instead the operator basis with currents of definite parity.  We keep only the subset of operators that give rise to $\mu\to e$ transitions and assume this is the only flavor-violating effect. }
\beq
\label{eq:WET}
{\cal L}_{\rm eff}^{\rm WET}= \sum_{a,d} \hat {\cal C}_a^{(d)} {\cal Q}_{a}^{(d)},
\eeq 
where the ${\cal Q}_a^{(d)}$ are CLFV operators of mass dimension $d$ (defined below) and the $\hat{{\cal C}}_a^{(d)}$ are dimensionful Wilson coefficients. By introducing an energy scale $\Lambda_{\text{CLFV}}$ associated with the CLFV physics, we can express these dimensionful Wilson coefficients in terms of dimensionless $\mathcal{O}(1)$ Wilson coefficients ${\cal C}^{(d)}_a$ as
\begin{equation}
\hat{\cal C}^{(d)}_a=\frac{{\cal C}^{(d)}_a}{\Lambda^{d-4}_{\text{CLFV}}}.
\end{equation}

In what follows, we retain operators up to and including dimension 7. The full set of dimension-5 operators consists of the magnetic and electric dipoles,
\begin{equation}
\label{eq:dim5:nf5:Q1Q2:light}
{\cal Q}_{1}^{(5)} = \frac{e}{8 \pi^2} (\bar e \sigma^{\alpha\beta}\mu)
 F_{\alpha\beta} \,, \qquad {\cal Q}_2^{(5)} = \frac{e }{8 \pi^2} (\bar
e \sigma^{\alpha\beta} i\gamma_5 \mu) F_{\alpha\beta} \,,
\end{equation}
where $F_{\alpha\beta}$ is the electromagnetic field strength tensor. The dimension-6 operators
are
\begin{align}
{\cal Q}_{1,q}^{(6)} & = (\bar e \gamma_\alpha \mu) (\bar q \gamma^\alpha q)\,,
 &{\cal Q}_{2,q}^{(6)} &= (\bar e\gamma_\alpha\gamma_5 \mu)(\bar q \gamma^\alpha q)\,, \label{eq:dim6:Q1Q2:light}
  \\ 
{\cal Q}_{3,q}^{(6)} & = (\bar e \gamma_\alpha \mu)(\bar q \gamma^\alpha \gamma_5 q)\,,
  & {\cal Q}_{4,q}^{(6)}& = (\bar
e \gamma_\alpha\gamma_5 \mu)(\bar q \gamma^\alpha \gamma_5 q)\,.\label{eq:dim6:Q3Q4:light}
\\
{\cal Q}_{5,q}^{(6)} & =  (\bar e \mu)( \bar q q)\,, 
&{\cal
  Q}_{6,q}^{(6)} &=  (\bar e i \gamma_5 \mu)( \bar q q)\,,\label{eq:dim6:Q5Q6:light}
  \\
{\cal Q}_{7,q}^{(6)} & = (\bar e \mu) (\bar q i \gamma_5 q)\,, 
&{\cal Q}_{8,q}^{(6)} & = (\bar e i \gamma_5 \mu)(\bar q i \gamma_5
q)\,, \label{eq:dim6:Q7Q8:light}  
 \\
{\cal Q}_{9,q}^{(6)} & =  (\bar e \sigma^{\alpha\beta} \mu) (\bar q \sigma_{\alpha\beta} q)\,, 
&{\cal Q}_{10,q}^{(6)} & = (\bar e  i \sigma^{\alpha\beta} \gamma_5 \mu)(\bar q \sigma_{\alpha\beta}
q)\,. \label{eq:dim6:Q9Q10:light} 
\end{align}
The quark label $q=u,d,s$ denotes one of the three light quark flavors. We assume that the CLFV interaction responsible for $\mu\rightarrow e$ conversion is flavor-conserving in the hadronic sector. 

The dimension-7 basis includes 8 operators that couple to gauge bosons
\begin{align}
{\cal Q}_1^{(7)} & = \frac{\alpha_s}{12\pi} (\bar e \mu)
 G^{a\alpha\beta}G_{\alpha\beta}^a\,, 
 & {\cal Q}_2^{(7)} &= \frac{\alpha_s}{12\pi} (\bar e i\gamma_5 \mu) G^{a\alpha\beta}G_{\alpha\beta}^a\,,\label{eq:dim7:Q1Q2:light}
 \\
{\cal Q}_3^{(7)} & = \frac{\alpha_s}{8\pi} (\bar e \mu) G^{a\alpha\beta}\widetilde
 G_{\alpha\beta}^a\,, 
& {\cal Q}_4^{(7)}& = \frac{\alpha_s}{8\pi}
(\bar e i \gamma_5 \mu) G^{a\alpha\beta}\widetilde G_{\alpha\beta}^a \,, \label{eq:dim7:Q3Q4:light}
\\
\label{eq:Q7:11}
\mathcal{Q}^{(7)}_{5} &= \frac{\alpha}{12\pi}(\bar e\mu)\,
F^{\alpha\beta} F_{\alpha\beta}\,, &\quad \mathcal{Q}^{(7)}_{6} &=
\frac{\alpha}{12\pi}(\bar{e}i\gamma_5\mu)\, F^{\alpha\beta}
F_{\alpha\beta}\,, \\
\label{eq:Q7:13}
\mathcal{Q}^{(7)}_{7} &=  \frac{\alpha}{8\pi}(\bar{e}\mu)\, F^{\alpha\beta} \tilde{F}_{\alpha\beta}\,, &\quad
\mathcal{Q}^{(7)}_{8} &=  \frac{\alpha}{8\pi}(\bar{e}i\gamma_5\mu)\, F^{\alpha\beta} \tilde{F}_{\alpha\beta}\,,
\end{align}
where $G^a_{\alpha\beta}$ is the gluon field strength tensor and $\tilde{G}_{\alpha\beta}=\frac{1}{2}\epsilon_{\alpha\beta\mu\nu}G^{\mu\nu}$ is its dual.\footnote{We use the convention $\epsilon_{0123}=+1$.} The electromagnetic dual $\tilde{F}_{\alpha\beta}$ is similarly defined. 
In order to complete the basis of dimension-7 CLFV operators, we introduce the following four-fermion operators with derivatives acting inside the lepton currents
\begin{align}
\label{eq:Q7:9}
\mathcal{Q}^{(7)}_{9,q} &=  (\bar{e}  \stackrel{\leftrightarrow}{i \partial}_\alpha \mu)\, (\bar{q} \gamma^\alpha q)\,, &\quad
\mathcal{Q}^{(7)}_{10,q} &=  (\bar{e}  i\gamma_5 \stackrel{\leftrightarrow}{i \partial}_\alpha \mu)\, (\bar{q} \gamma^\alpha q)\,, 
\\
\mathcal{Q}^{(7)}_{11,q} &=  (\bar{e}  \stackrel{\leftrightarrow}{i \partial}_\alpha \mu)\, (\bar{q} \gamma^\alpha \gamma_5 q)\,, &\quad
\mathcal{Q}^{(7)}_{12,q} &=  (\bar{e}  i\gamma_5 \stackrel{\leftrightarrow}{i \partial}_\alpha \mu)\, (\bar{q} \gamma^\alpha \gamma_5 q)\,, 
\\
\mathcal{Q}^{(7)}_{13,q} &=   \partial^\alpha (\bar{e} \gamma^{\beta}  \mu)\, (\bar{q} \sigma_{\alpha\beta}  q)\,, &\quad
\mathcal{Q}^{(7)}_{14,q} &=  \partial^\alpha (\bar{e} \gamma^{\beta} \gamma_5 \mu)\, (\bar{q} \sigma_{\alpha\beta}  q)\,, 
\\
\mathcal{Q}^{(7)}_{15,q} &= \partial^\alpha (\bar{e} \gamma^{\beta} \mu)\, (\bar{q}  i\sigma_{\alpha\beta} \gamma_5 q)\,, &\quad
\mathcal{Q}^{(7)}_{16,q} &=  \partial^\alpha (\bar{e} \gamma^{\beta} \gamma_5 \mu)\, (\bar{q}  i\sigma_{\alpha\beta} \gamma_5 q)\,,  
\end{align}
where $\bar{e} \stackrel{\leftrightarrow}{i \partial}_\nu
\mu=\bar{e} {i \partial}_\nu \mu-\bar{e}
\stackrel{\leftarrow}{i\partial}_\nu \mu$. 

Instead of the operators $\mathcal{Q}^{(7)}_{13,q}, \ldots, \mathcal{Q}^{(7)}_{16,q}$, one could use equations of motion to write a more symmetric operator basis, with the derivatives acting on the quark currents 
\begin{align}
\mathcal{Q}^{(7)}_{13,q}&=-(\bar{e} \gamma^{\alpha}  \mu)\, (\bar{q} \stackrel{\leftrightarrow}{i D}_\alpha  q) +  2 m_q \mathcal{Q}^{(6)}_{1,q} \,, 
\\
\mathcal{Q}^{(7)}_{14,q}&=-(\bar{e}  \gamma^{\alpha} \gamma_5 \mu)\, (\bar{q} \stackrel{\leftrightarrow}{iD}_\alpha q) + 2 m_q \mathcal{Q}^{(6)}_{2,q}  \,, 
\\
\mathcal{Q}^{(7)}_{15,q}&=-(\bar{e} \gamma^{\alpha} \mu)\, (\bar{q}  i \gamma_5 \stackrel{\leftrightarrow}{iD}_\alpha q)   \,, 
\\
\mathcal{Q}^{(7)}_{16,q}&=-(\bar{e}  \gamma^{\alpha} \gamma_5 \mu)\, (\bar{q}   i \gamma_5 \stackrel{\leftrightarrow}{iD}_\alpha q) \,.
\end{align}
This reformulation demonstrates that our dimension-7 basis is equivalent to the basis in Ref.~\cite{Liao:2020zyx} (when restricted to those operators that can mediate $\mu\rightarrow e$ conversion). Our basis is chosen to make the evaluation of hadronic matrix elements straightforward. 

\section{Nonrelativistic effective theory}
\label{sec:NRET}
The $\mu\rightarrow e$ conversion process results in momentum exchanges typical of the nuclear scale, where the natural degrees of freedom for describing the strong interaction are nucleons. To make contact with experiments, the CLFV light-quark operator basis of Sec.~\ref{sec:LEFT} must be matched to the basis of CLFV single-nucleon operators. Just as the WET basis is organized by a power-counting in mass dimension, the NRET basis must also be organized through an expansion in small dimensionless parameters. Before introducing the single-nucleon CLFV basis, we briefly review how the physics of $\mu\rightarrow e$ conversion motivates the particular form of the nuclear-scale effective theory \cite{Rule:2021oxe,Haxton:2022piv} used in this work.

\subsection{Kinematics}
The $\mu^-$ that gets captured in the nuclear Coulomb field quickly de-excites to the $1s$ orbital. The muon's binding energy $E_\mu^\mathrm{bind}$ (defined to be positive) and the muon's wave function $\psi_\mu$ can be determined by numerically solving the Dirac equation for a potential sourced by the experimentally known nuclear charge distribution.  To the precision we work, screening and other corrections arising from the Coulomb potential of the surrounding electron cloud can be ignored.

We are interested in $\mu\to e$ transitions where the nucleus remains in the ground state.\footnote{Experimentally, this is the preferred option as the resulting conversion electrons will be at the endpoint of the SM background from $\mu\to e \nu\bar\nu$ decays. In $\mu\to e$ conversions where the nucleus transitions to an excited state, the less energetic outgoing electron must compete with a larger SM background but may provide additional information about the underlying CLFV mechanism \cite{Haxton:2024ecp}.} 
The three-momentum $\vec q$ of the outgoing electron is then given by
\begin{equation}
\label{eq.vecq2}
\vec{q}^{\;2}=\frac{M_T}{m_{\mu}+M_T}\left[\left(m_\mu-E_\mu^\mathrm{bind}\right)^2-m_e^2\right],
\end{equation}
where we keep the first correction in $m_\mu/M_T$, with $M_T$, $m_\mu$, and $m_e$ being the masses of the target nucleus, muon, and electron, respectively. Once the energy of the electron is known, the outgoing electron wave function can be obtained from numerical solutions of the Dirac equation
in the Coulomb field generated by the extended nuclear charge. As the momentum transfer from the leptons to the nucleus is sufficient to require retention of several electron
partial waves, this has the potential to significantly complicate calculations.  Fortunately, the Coulomb-distorted waves can be very well approximated by much simpler plane waves, 
evaluated for a shifted effective momentum $q_\mathrm{eff}$ determined from the average value of the Coulomb potential near the nucleus \cite{Haxton:2022piv}. Numerically, the difference between $q$ and $q_{\rm eff}$ is $\approx 5\%$ in $^{27}$Al.

To motivate the form of the nuclear-scale effective theory, we note that:
\begin{enumerate}
\item The outgoing electron is highly relativistic, with $E_e \approx m_\mu$: the correction due to the muon binding energy $E_\mu^\mathrm{bind}$ is small ($E_\mu^\mathrm{bind}\approx 0.463$ MeV in $^{27}$Al). The electron velocity --- defined in the NRET as the Galilean-invariant velocity with respect to the center of mass of the final-state nucleus --- thus has magnitude 1 (in units of $c$)
and direction $\hat{q}$, the latter of which is, in principle, an observable.  This leaves only the nonrelativistic
velocities of the bound-state nucleons and muon to be treated as operators, a task ideally suited for NRET methods.

\item Relativistic corrections for the bound muon are roughly proportional to $Z\alpha /2$. For light nuclei such as $^{27}$Al, the muon is highly nonrelativistic, so that its Dirac wave function is dominated by its Schr\"odinger-like upper component.  The muon's velocity operator $\vec{v}_{\mu}$ thus enters the NRET as a correction generated by the suppressed lower component of the Dirac solution.  In the NRET, $\vec{v}_{\mu}$ is defined as the Galilean-invariant velocity associated with the motion of the bound muon with respect to the center of mass of the initial-state nucleus. The effects of $\vec{v}_{\mu}$ are limited to small numerical changes in nuclear form factors: the $\vec{v}_{\mu}$ operator plays no role in the selection rules that determine the nuclear response functions.

\item Nucleons bound in a nucleus are only mildly relativistic, with typical velocities $v_\mathrm{avg} \approx 0.1 $. We can therefore perform a nonrelativistic expansion of the nuclear charges and currents.  In the NRET, the nucleon velocity operator $\vec{v}_N$ stands
for the set of $A-1$ independent nucleon Jacobi velocities, e.g., the Galilean-invariant
velocities $\vec{v_N} \equiv \{  (\vec{v}_2-\vec{v}_1)/\sqrt{2}, ~(2\vec{v}_3-(\vec{v}_1+\vec{v}_2))/\sqrt{6},~\ldots \}$, where $\vec{v}_i$ is velocity operator for the $i$-th nucleon and $A$ is the nucleon number.  See \cite{Fitzpatrick:2012ix,Haxton:2022piv} for details.
\end{enumerate}

The available Hermitian operators that enter into the construction of the NRET CLFV operators are: $i \hat{q}$ where $\hat{q}$ is the velocity of the outgoing ultra-relativistic electron, the nucleon velocity operator $\vec{v}_N$, and the respective lepton and nucleon spin operators, $\vec{\sigma}_L$ and $\vec{\sigma}_N$. Extending this set of building-block operators to include the muon velocity operator $\vec{v}_\mu$ generates the relativistic corrections associated with the lower component of the muon's Dirac wave function.
\subsection{NRET basis}
Working to the first order in the nucleon velocity $\vec{v}_N$ and neglecting the muon velocity $\vec{v}_{\mu}$, there are 16 independent CLFV single-nucleon operators \cite{Rule:2021oxe,Haxton:2022piv}, 
\begin{subequations}
\label{eq:ops}
\begin{align}
\CO_1 &= 1_L ~1_N, & \CO^\prime_2 &= 1_L ~i \hat{q} \cdot \vec{v}_N,  \\
 \CO_3 &= 1_L~  i \hat{q} \cdot  \left[ \vec{v}_N \times \vec{\sigma}_N \right], & 
 \CO_4 &= \vec{\sigma}_L \cdot \vec{\sigma}_N,  \\
\CO_5 &=  \vec{\sigma}_L \cdot \left( i \hat{q} \times \vec{v}_N \right), &
 \CO_6&=  i \hat{q} \cdot \vec{\sigma}_L ~ i \hat {q} \cdot \vec{\sigma}_N,   \\
\CO_7 &= 1_L~ \vec{v}_N \cdot \vec{\sigma}_N, &
\CO_8 &= \vec{\sigma}_L\cdot \vec{v}_N,  \\ 
 \CO_9 &= \vec{\sigma}_L \cdot \left( i \hat{q} \times \vec{\sigma}_N \right), &
\CO_{10} &= 1_L~ i \hat{q} \cdot \vec{\sigma}_N, \\
\CO_{11} &= i \hat{q} \cdot \vec{\sigma}_L ~ 1_N,  &
\CO_{12} &= \vec{\sigma}_L \cdot \left[ \vec{v}_N \times \vec{\sigma}_N \right], \\ 
\CO^\prime_{13} &= \vec{\sigma}_L \cdot  \left( i \hat{q} \times \left[ \vec{v}_N \times \vec{\sigma}_N \right] \right),  &
\CO_{14} &= i  \hat{q} \cdot \vec{\sigma}_L ~ \vec{v}_N \cdot \vec{\sigma}_N,   \\
\CO_{15} &= i \hat{q} \cdot \vec{\sigma}_L~ i \hat{q} \cdot \left[ \vec{v}_N \times \vec{\sigma}_N \right], &
\CO^\prime_{16} &= i \hat{q} \cdot \vec{\sigma}_L ~i \hat{q} \cdot  \vec{v}_N. 
\end{align}
\end{subequations}
This operator basis matches closely the one previously derived for dark matter
direct detection \cite{Fitzpatrick:2012ix}; we distinguish with a prime the operators for which there are significant differences.
The NRET operators $\mathcal{O}_i$ are understood to act between Pauli spinors $\xi_s$ for muon, electron, and nucleons. The leptonic current operators $1_L$ or $\vec \sigma_L$ couple
the $1s$ muon wave function to the various distorted partial waves comprising the outgoing electron's wave function.

The corresponding effective interaction can be expressed as 
\begin{equation}
    \mathcal{L}_\mathrm{eff}^{\rm NRET}= \sum_{N=n,p} \sum_{i=1}^{16} c_i^N \mathcal{O}_i^N +\cdots,
    \label{eq:L_NRET}
\end{equation}
where ellipses denote omitted corrections due to operators generated
by $\vec{v}_\mu$, two-body currents that arise in nuclear systems, etc. We refer to the numerical coefficients $c_i^N$ as the low-energy constants (LECs) of the NRET. They are the analogs of the Wilson coefficients $\hat{\mathcal{C}}^{(d)}_a$ in the WET Lagrangian, Eq.~\eqref{eq:WET}, and carry dimensions of $1/(\mathrm{mass})^{2}$. Strictly speaking, the LECs are functions of $\vec q^{\;2}$; however, for a given target the momentum transfer in $\mu \to e$ conversion is a fixed quantity determined by kinematics, as in Eq. \eqref{eq.vecq2}. As a result, the NRET LECs are genuine constants capable of encoding the exact momentum dependence that arises, for example, from the exchange of light mediators with $m^2\lesssim \vec{q}^{\;2}$. Thus, the NRET formalism is ideally suited to describe not just heavy mediators but also light new physics such as the CLFV axion-like-particle scenario considered below in Sec. \ref{sec:PNGB}. 

In Eq.~\eqref{eq:L_NRET} we have also introduced the index $N$ to allow for the CLFV physics to couple differently to protons vs. neutrons. Equivalently, we can work in terms of isoscalar and isovector operators
\begin{equation}
    \mathcal{L}_\mathrm{eff}^{\rm NRET}= \sum_{\tau=0,1} \sum_{i=1}^{16} c_i^{\tau} \mathcal{O}_i t^{\tau}+\cdots,
    \label{eq:L_NRETv2}
\end{equation}
where $c_i^0=(c_i^p+c_i^n)/2$, $c_i^1=(c_i^p-c_i^n)/2$, and $t^0=1$, $t^1=\tau_3$ are the isospin matrices. 

Working to first order in the nucleon velocity $v_N$ and neglecting the muon velocity $v_\mu$, the $\mu\to e$ conversion rate is given by~\cite{Rule:2021oxe,Haxton:2022piv}
\beq
\begin{split}\label{eq:Gamma:mutoe}
\Gamma(\mu\to e)=\frac{1}{2 \pi}\frac{q_{\rm eff}^2}{1+q/M_T}&\big|\phi_{1s}^{Z_\mathrm{eff}}(\vec{0})\big|^2\\
\times\sum_{\tau,\tau'}\Bigg\{&\Big[R_{MM}^{\tau\tau'} W_{MM}^{\tau\tau'}(q_{\rm eff})+R_{\Sigma''\Sigma''}^{\tau\tau'} W_{\Sigma''\Sigma''}^{\tau\tau'}(q_{\rm eff})+R_{\Sigma'\Sigma'}^{\tau\tau'} W_{\Sigma'\Sigma'}^{\tau\tau'}(q_{\rm eff})\Big] \\
+ \frac{q_\mathrm{eff}^2}{m_N^2}&\Big[R_{\Phi''\Phi''}^{\tau\tau'}W_{\Phi''\Phi''}^{\tau\tau'}(q_\mathrm{eff})+R_{\tilde{\Phi}'\tilde{\Phi}'}^{\tau\tau'}W_{\tilde{\Phi}'\tilde{\Phi}'}^{\tau\tau'}(q_\mathrm{eff})+R_{\Delta\Delta}^{\tau\tau'}W_{\Delta\Delta}^{\tau\tau'}(q_\mathrm{eff})\Big] \\
- \frac{2q_\mathrm{eff}}{m_N}&\Big[ R_{\Phi''M}^{\tau\tau'}W_{\Phi''M}^{\tau\tau'}(q_\mathrm{eff}) + R_{\Delta\Sigma'}^{\tau\tau'}W_{\Delta\Sigma'}^{\tau\tau'}(q_\mathrm{eff}) \Big]
\Bigg\},
\end{split}
\eeq
where 
\begin{align}
R_{MM}^{\tau\tau'} &=c_1^\tau c_1^{\tau'*}+c_{11}^\tau c_{11}^{\tau'*},
\\
R_{\Sigma''\Sigma''}^{\tau\tau'} &=\big(c_4^\tau-c_6^\tau\big) \big(c_4^{\tau'}-c_6^{\tau'}\big)^*+c_{10}^\tau c_{10}^{\tau'*},\\
R_{\Sigma'\Sigma'}^{\tau\tau'} &=c_4^\tau c_4^{\tau'*}+c_9^\tau c_9^{\tau'*},\\
R_{\Phi''\Phi''}^{\tau\tau'}&=c_3^{\tau}c_3^{\tau' *}+(c_{12}^\tau - c_{15}^\tau)(c_{12}^{\tau' *}- c_{15}^{\tau' *}),\\
R_{\tilde{\Phi}'\tilde{\Phi}'}^{\tau\tau'}&=c_{12}^{\tau}c_{12}^{\tau' *}+c_{13}^\tau c_{13}^{\tau' *},\\
R_{\Delta\Delta}^{\tau\tau'}&=c_{5}^{\tau}c_{5}^{\tau' *}+c_{8}^\tau c_{8}^{\tau' *},\\
R_{\Phi'' M}^{\tau\tau'}&=\mathrm{Re}[c_3^\tau c_1^{\tau' *}-(c_{12}^\tau - c_{15}^\tau)c_{11}^{\tau' *}],\\
R_{\Delta\Sigma'}^{\tau\tau'}&=\mathrm{Re}[c_5^\tau c_4^{\tau' *}+c_8^\tau c_9^{\tau' *}],
\end{align}
and $\phi^{Z_\mathrm{eff}}_{1s}(\vec{0})$ is the $1s$ wave function of a muonic atom with effective charge $Z_{\rm eff}$, evaluated at the origin. The leptonic response functions $R_i^{\tau\tau'}$ are bilinears in the NRET LECs.  The specific combinations define what can (and thus what cannot) be determined about CLFV from elastic $\mu \rightarrow e$ conversion. Note that the nuclear response functions $W_i^{\tau\tau'}$ depend on the modified momentum of the outgoing electron wave $q_{\rm eff}$. 

In the long wavelength limit, $q_{\rm eff}\to 0$, the coherently enhanced response function $W_{MM}^{\tau\tau}$ counts the number of protons and neutrons in the nucleus,  while $W_{\Sigma'\Sigma'}^{\tau\tau'}$ and $W_{\Sigma''\Sigma''}^{\tau\tau'}$ measure the transverse and longitudinal nuclear spin responses, respectively. The velocity-dependent response functions $W_{\Phi''\Phi''}^{\tau\tau'}$, $W_{\tilde{\Phi}'\tilde{\Phi}'}^{\tau\tau'}$, and $W_{\Delta\Delta}^{\tau\tau'}$ appear in the rate formula multiplied by a factor of $q_\mathrm{eff}^2/m_N^2$, reflecting their origin as responses sensitive to the composite structure of the nucleus, generated by operators like the orbital angular momentum $\vec{\ell}$. As such, contributions from these responses vanish in the limit of a point-like nucleus, $q_\mathrm{eff}\rightarrow 0$.

One of these response functions, $W_{\Phi''\Phi''}^{\tau\tau'}$, becomes
coherent in nuclei like $^{27}$Al where one of two spin-orbit 
partner shells $j=\ell\pm 1/2$ is occupied \cite{Fitzpatrick:2012ix}. The response $\Phi''$, which is generically associated with tensor mediators, corresponds to the longitudinal projection of the nuclear spin-velocity current $\vec{v}_N\times\vec{\sigma}_N$ and can interfere with the charge multipole operator $M$. (Similarly, there can be interference between the transverse-magnetic response $\Delta$ and the transverse-electric response $\Sigma'$.) In $^{27}$Al, which is a nearly ideal target for maximizing the coherence of $\Phi''$, the interference term $W_{\Phi'' M}^{00}$ contributes $\approx 5\%$ of the total response (for equal NRET coefficients, e.g., $c_1^0=c_3^0$). Along with the usual coherent coupling to nuclear charge, these two distinct sources
of nuclear enhancements of operators lead to a hierarchy in the associated response functions for $^{27}$Al.  In the case of isoscalar couplings --- assuming that the relevant NRET coefficients are roughly equal in magnitude --- the hierarchy is
\beq
\label{eq:nuclear:response:scalings}
W_{MM}^{00}\sim {\mathcal O}(A^2)\gg \frac{q_{\rm eff}}{m_N}W_{M\Phi''}^{00}\gg \Big\{W_{\Sigma'\Sigma'}^{00},W_{\Sigma''\Sigma''}^{00},\frac{q_{\rm eff}^2}{m_N^2}W_{\Phi''\Phi''}^{00} \Big\}\gg\Big\{\frac{q_{\rm eff}^2}{m_N^2}W_{\Delta\Delta}^{00},\frac{q_{\rm eff}^2}{m_N^2}W_{\tilde{\Phi}'\tilde{\Phi}'}^{00}\Big\}.
\eeq

This assumes, as is the case for ${}^{27}\text{Al}$, that the nuclear ground state carries angular momentum $j \ge 1$, so that all nuclear response functions can contribute. The above hierarchy of nuclear response functions illustrates that the
nucleus can alter the na\"ive nucleon-level counting based on the small parameter $|\vec{v}_N|$; coherence can elevate operators to be of the allowed
strength.   

The six response functions (and the two interference terms) constitute the most general set of symmetry-allowed nuclear response functions. That is, the constraints of the nuclear ground state --- angular momentum, parity, and time-reversal --- restrict the operators that can contribute to the elastic $\mu\rightarrow e$ conversion process, so that the 16 leading single-nucleon operators embed into just six nuclear response functions and two interference terms. The general form is generated at ${\mathcal O}(v_N)$; the extension to
${\mathcal O}(v_\mu)$ adds small form factor corrections, but does not
change any of the essential features of the CLFV physics.

If one includes all first-order effects of velocity, whether associated with the nucleon or muon, the NRET operator basis in Eq. \eqref{eq:L_NRET} expands to \cite{Haxton:2022piv}
\begin{equation}
   \mathcal{L}_\mathrm{eff}^{\rm NRET}= \sum_{N=n,p} \sum_{~i=1}^{16} c_i^N \mathcal{O}_i^N + \sum_{N=n,p}\sum_{~i\in I }b_i^N\mathcal{O}^{f,N}_i,
    \label{eq:L_NRET:vmu}
\end{equation}
where the second sum is over ten new operators linear in $\vec{v}_\mu$, indexed by the
set $I=\{2,3,5,7,8,12,13,14,15,16\}$, which arise from the muon's lower component,
\begin{subequations}
\label{eq:ops2}
\begin{align}
{\CO}^{f \, \prime}_2 &= i \hat{q} \cdot \tfrac{\vec{v}_\mu}{2} ~1_N, &
{ \CO}^f_3 &=   i \hat{q} \cdot  \left[ \tfrac{\vec{v}_\mu}{2} \times \vec{\sigma}_L \right] ~1_N, \\
{\CO}^f_5 &=   \left( i \hat{q} \times \tfrac{\vec{v}_\mu}{2} \right)  \cdot \vec{\sigma}_N, &
{\CO}^f_7 &=  \tfrac{\vec{v}_\mu}{ 2} \cdot \vec{\sigma}_L~1_N,  \\
{\CO}^f_8 &= \tfrac{\vec{v}_\mu}{2} \cdot \vec{\sigma}_N, &
{\CO}^f_{12} &= \left[ \tfrac{\vec{v}_\mu}{2} \times \vec{\sigma}_L \right] \cdot \vec{\sigma}_N, \\ 
{\CO}^{f \, \prime}_{13} &=  \left( i \hat{q} \times \left[ \tfrac{\vec{v}_\mu}{2} \times \vec{\sigma}_L \right] \right)  \cdot \vec{\sigma}_N, &
{\CO}^f_{14} &= \tfrac{\vec{v}_\mu}{2} \cdot \vec{\sigma}_L  ~ i  \hat{q} \cdot \vec{\sigma}_N,   \\
{\CO}^f_{15} &= i \hat{q} \cdot \left[ \tfrac{\vec{v}_\mu}{2} \times \vec{\sigma}_L \right] ~ i \hat{q} \cdot \vec{\sigma}_N, &
{\CO}^{f \, \prime}_{16} &=  i \hat{q} \cdot  \tfrac{\vec{v}_\mu}{2} ~i \hat{q} \cdot \vec{\sigma}_N,
\end{align}
\end{subequations}
with the $b_i^N$ the associated LECs. Although there are only 10 additional operators, they are labeled in analogy with the 16 upper component operators: $\CO_i\leftrightarrow \CO_i^f$ under the exchange $\vec{v}_N\leftrightarrow \vec{v}_\mu/2$.
Here we employ $\vec{v}_\mu/2$ because this operator, when acting on the muon's upper Dirac component, generates the lower component. The complete expression for $\Gamma(\mu \to e)$ with all velocities handled through linear order is given in Eq. (B3) of Ref.~\cite{Haxton:2022piv}.  The new public computer codes discussed in App.~\ref{app:Public} are the first to properly include the effects of $\vec{v}_\mu$.

\section{Matching quarks and gluons to nucleons}
\label{sec:matching}
We turn next to the nonperturbative matching from WET to NRET, starting with single nucleon matrix elements. 

\subsection{Nucleon matrix elements}
For the nucleon matrix elements, we use a notation closely resembling that of Refs. \cite{Bishara:2017nnn,Bishara:2017pfq}\footnote{The difference is in the definitions of the tensor form factors, $\hat F_{T,i}^{q/N} \equiv F_{T,i}^{q/N} /m_q$, where $F_{T,i}^{q/N}$, with $i=0,1,2$, are the form factors in  \cite{Bishara:2017nnn,Bishara:2017pfq}. This choice reflects the normalization of the tensor current operators in WET basis, Section \ref{sec:LEFT}. For scalar and pseudoscalar form factors $F_S^{q/N}$, $F_P^{q/N}$ we keep the $m_q$ prefactors in the currents, since the form factors thus defined are more precisely known. Note that the definitions of $F_{\tilde \gamma}$ and $F_{\tilde G}$ form factors include minus signs in order to match \cite{Bishara:2017nnn,Bishara:2017pfq}, where the $\epsilon^{0123}=+1$ convention was used.}

\begin{align}
\label{vec:form:factor}  
\langle N'|\bar q \gamma^\mu q|N\rangle&=\bar u_N'\Big[F_1^{q/N}(q_{\rm rel.}^2)\gamma^\mu-\frac{i}{2m_N}F_2^{q/N}(q_{\rm rel.}^2) \sigma^{\mu\nu}q_\nu\Big]u_N\,,
\\
\begin{split}
\label{axial:form:factor}
\langle N'|\bar q \gamma^\mu \gamma_5 q|N\rangle&=\bar u_N'\Big[F_A^{q/N}(q_{\rm rel.}^2)\gamma^\mu\gamma_5-\frac{1}{2m_N}F_{P'}^{q/N}(q_{\rm rel.}^2) \gamma_5 q^\mu\Big]u_N\,,
\end{split}
\\
\label{scalar:form:factor}
\langle N'| m_q \bar q   q|N\rangle&= F_S^{q/N} (q_{\rm rel}^2)\, \bar u_N' u_N\,,
\\
\label{pseudoscalar:form:factor}
\langle N'|m_q \bar q  i \gamma_5 q|N\rangle&= F_P^{q/N} (q_{\rm rel}^2)\, \bar u_N' i \gamma_5 u_N\,,
\\
\label{CPeven:gluonic:form:factor}
\langle N'| \frac{\alpha_s}{12\pi} G^{a\mu\nu}G^a_{\mu\nu} |N\rangle&= F_G^{N} (q_{\rm rel.}^2)\, \bar u_N' u_N\,,
\\
\label{CPodd:gluonic:form:factor}
\langle N'| \frac{\alpha_s}{8\pi} G^{a\mu\nu}\tilde G^a_{\mu\nu}|N\rangle&= -F_{\tilde G}^{N} (q_{\rm rel.}^2)\, \bar u_N' i \gamma_5 u_N\,,
\\
\label{tensor:form:factor}
\begin{split}
\langle N'| \bar q \sigma^{\mu\nu} q |N\rangle&=  \bar u_N'\Big[\hat F_{T,0}^{q/N} (q_{\rm rel.}^2)\,  \sigma^{\mu\nu} -\frac{i}{2 m_N} \gamma^{[\mu}q^{\nu]} \hat F_{T,1}^{q/N} (q_{\rm rel.}^2) 
\\
&\qquad \qquad- \frac{i}{m_N^2} q^{[\mu}k_{12}^{\nu]} \hat F_{T,2}^{q/N} (q_{\rm rel.}^2) \Big] u_N\,,
\end{split}
\\
\label{CPeven:photon:form:factor}
\langle N'| \frac{\alpha}{12\pi} F^{\mu\nu}F_{\mu\nu} |N\rangle&= F_\gamma^{N} (q_{\rm rel.}^2)\, \bar u_N' u_N\,,
\\
\label{CPodd:photon:form:factor}
\langle N'| \frac{\alpha}{8\pi} F^{\mu\nu}\tilde F_{\mu\nu}|N\rangle&= -F_{\tilde \gamma}^{N} (q_{\rm rel.}^2)\, \bar u_N' i \gamma_5 u_N\,.
\end{align}
Here we shortened $\langle N'|=\langle N(k_2)| $, $| N\rangle= |
N(k_1)\rangle $, $\bar u_N'= \bar u_N(k_2)$, $u_N= u_N(k_1)$ and introduced
$q^\mu=k_1^\mu-k_2^\mu$, $k_{12}^\mu=k_1^\mu+k_2^\mu$, with
\beq
\label{eq:qrel}
q_{\rm rel.}^2\equiv q_\mu q^\mu\simeq -|\vec{q}\,|^2 \equiv - q^2 .
\eeq
 Compared to Ref.~\cite{Bishara:2017nnn}, the definition of the momentum exchange $q^\mu$ differs by a sign (but the definitions of form factors coincide). For covariant derivative we use $D_\mu\psi=(\partial_\mu+i e Q_\psi A_\mu )\psi$, where $Q_\psi$ is the electric charge. The antisymmetrized tensors are defined as $\gamma^{[\mu}q^{\nu]}=\gamma^{\mu}q^{\nu}-\gamma^{\nu}q^{\mu}$, and similarly for $q^{[\mu}k_{12}^{\nu]}$ and $\gamma^{[\mu}\slashed{q}\gamma^{\nu]}$. For CP-violating light new physics, additional form factors appear on  right-hand sides (RHS) of Eqs. (\ref{vec:form:factor}), (\ref{axial:form:factor}), and (\ref{tensor:form:factor}). These are expected to be small and are discussed in Appendix \ref{app:AFF}. The numerical values of the form factors for $\mu\rightarrow e$ conversion in $^{27}$Al are derived in Appendix \ref{app:nucleon:ff:values}. 

\subsection{Covariant nucleon-level interactions}
To facilitate matching between nuclear NRET and WET, we follow Ref. \cite{Haxton:2022piv} and introduce an intermediate step --- a set of covariant single-nucleon operators and corresponding LECs. We denote the resulting interaction  as 
\beq
\label{eq:d:Lagr}
{\cal L}_{\rm eff}^{\rm cov}=\sum_{N=n,p}\sum_{j=1}^{32} d_j^{N} {\cal L}_{{\rm int}}^{j,N}.
\eeq

The Lorentz covariant operators $\mathcal{L}_\mathrm{int}^{j,N}$ are listed in Appendix \ref{app:nret-decomp}, in the first columns of Tables \ref{tab:LWL} and \ref{tab:tensor_upper}. The operators in Table \ref{tab:LWL} appear in Ref.~\cite{Haxton:2022piv}, where all covariant interactions generated from scalar or vector mediators were enumerated. Here, we extend this set to include tensor-mediated interactions, listed in the first column of Table \ref{tab:tensor_upper}. 

The WET is matched onto ${\cal L}_{\rm eff}^{\rm cov}$, relating the LECs of the former to the CLFV coefficients $d_j^N$ of the nucleon-level
interaction that we employ below $\Lambda_\chi$.  In this
nonperturbative matching, we use the definitions of the nucleon matrix elements in Eqs.~\eqref{vec:form:factor}--\eqref{CPodd:photon:form:factor}, with the corresponding results for $d_j^N$ coefficients listed in Appendix \ref{app:WET:to:cov}, Eqs.~\eqref{eq:d1}--\eqref{eq:d32}. The $d_i^N$ coefficients --- expressed in the proton/neutron basis --- are readily converted into the isospin basis via
\begin{equation}
d_i^0=\frac{1}{2}\left(d_i^p+d_i^n\right),~~~~d_i^1=\frac{1}{2}\left(d_i^p-d_i^n\right).
\end{equation}

\subsection{The NRET}
\label{sec:NRET:matching}
Because the momentum scale in $\mu \rightarrow e$ conversion is set by $m_\mu \ll m_N$, we can perform a nonrelativistic reduction of the covariant operators, ${\cal L}_{{\rm int}}^{j,N}$, to obtain the NRET.  This reduces the number of nucleon-level
operators while generating an interaction compatible with standard nuclear calculations, which are
typically nonrelativistic. Since we consider only elastic $\mu \rightarrow e$ conversion, and since the 
nuclear recoil energy $\approx m_\mu^2/M_T$ is extremely small, the reduction can be done
with $q^0 \equiv 0$.

The relativistic reduction of operators $\mathcal{L}_\mathrm{int}^{1-20}$ --- the interactions arising from scalar and vector mediators --- to their
corresponding NRET forms was done in Ref.~\cite{Haxton:2022piv}. For 
convenience those results are summarized in Appendix \ref{app:nret-decomp},  with the resulting $\mathcal{O}_i^{N}$ and $\mathcal{O}_i^{f,N}$ displayed in 
Tables \ref{tab:LWL} and \ref{tab:LWL2}, respectively. The corresponding results from the nonrelativistic reduction of 12 new tensor-mediated interactions are given in Tables \ref{tab:tensor_upper} and \ref{tab:tensor_lower}, respectively. (We note that certain tensor
operators were also considered recently in \cite{Glick-Magid:2023uhk}.)
The final columns of these tables relate the relativistic $\{d_i^N\}$ to the appropriate $\{c_i^N,b_i^N\}$ NRET LEC combinations. 

Combining with the results for the $d_i^N$ in Appendix \ref{app:WET:to:cov}, we then obtain for the $c_i^N$
\begin{align}
\begin{split}\label{eq:c1}
c_1^{N}&=-\frac{\alpha}{\pi q}\hat {\cal C}_1^{(5)} \sum_q Q_q F_1^{q/N} +\sum_q \hat {\cal C}_{1,q}^{(6)} F_1^{q/N} +\sum_q \frac{1}{m_q} \hat {\cal C}_{5,q}^{(6)} F_S^{q/N}
\\
&-\frac{q}{m_N}\sum_q  \hat {\cal C}_{9,q}^{(6)} \big(\hat F_{T,0}^{q/N}-\hat F_{T,1}^{q/N}+ 4 \hat F_{T,2}^{q/N}\big)
\\
&+\hat {\cal C}_{1}^{(7)} F_G^{N}+\hat {\cal C}_{5}^{(7)} F_\gamma^{N}  + \big(q+m_+ \big)\sum_q \hat {\cal C}_{9,q}^{(7)} F_1^{q/N}
\\
& -\frac{q^2}{2m_N} \sum_q \hat {\cal C}_{13,q}^{(7)} \biggr[\hat F_{T,0}^{q/N}-\hat F_{T,1}^{q/N}+\biggr(4+\frac{q^2}{m_N^2}\biggr) \hat F_{T,2}^{q/N}\biggr],
\end{split}
\\
c_2^N&=i\Big[\sum_q \hat {\cal C}_{1,q}^{(6)} F_1^{q/N} +
m_+\sum_q \hat {\cal C}_{9,q}^{(7)} F_1^{q/N} +\frac{q^2}{2m_N} \sum_q \hat {\cal C}_{13,q}^{(7)} \Big(\hat F_{T,1}^{q/N}-4 \hat F_{T,2}^{q/N}\Big)\Big],
\\
c_3^N&=-2 \sum_q \hat {\cal C}_{9,q}^{(6)} \hat F_{T,0}^{q/N} -
 q \sum_q \hat {\cal C}_{13,q}^{(7)}  \Big( \hat F_{T,0}^{q/N}+\frac{q^2}{m_N^2} \hat F_{T,2}^{q/N}\Big),
 \\
 \begin{split}
c_4^{N}&=-\frac{\alpha}{2 \pi m_N}\hat {\cal C}_1^{(5)} \sum_q Q_q \big(F_1^{q/N} +F_2^{q/N}\big)-\frac{q}{2m_N}\sum_q \hat {\cal C}_{1,q}^{(6)}\big( F_1^{q/N} +F_2^{q/N}\big)
\\
&-\sum_q \hat {\cal C}_{4,q}^{(6)} F_A^{q/N} +2\sum_q \hat {\cal C}_{9,q}^{(6)} \hat F_{T,0}^{q/N}
-\frac{q}{2m_N} \big(m_+-q\big) \sum_q \hat {\cal C}_{9,q}^{(7)} \big(F_1^{q/N}+F_2^{q/N}\big)
\\
& +i (m_- -q) \sum_q \hat {\cal C}_{12,q}^{(7)} F_A^{q/N}- q \sum_q  \hat {\cal C}_{13,q}^{(7)} \biggr[\hat F_{T,0}^{q/N}+\frac{q^2}{4 m_N^2}\hat F_{T,1}^{q/N}\biggr],
\end{split}
\\
\begin{split}
c_5^N&=-\frac{\alpha}{\pi q}\sum_q \hat {\cal C}_{1}^{(5)} Q_q F_{1}^{q/N}- \sum_q \hat {\cal C}_{1,q}^{(6)} F_{1}^{q/N}
\\
& -(m_+-q) \sum_q \hat {\cal C}_{9,q}^{(7)} F_{1}^{q/N} -
 \frac{q^2}{2m_N} \sum_q \hat {\cal C}_{13,q}^{(7)}  \Big( \hat F_{T,1}^{q/N}-4 \hat F_{T,2}^{q/N}\Big),
 \end{split}
 \\
 \begin{split}
c_6^N&=-\frac{\alpha}{2 \pi m_N}\sum_q \hat {\cal C}_{1}^{(5)} Q_q \big(F_{1}^{q/N}+F_{2}^{q/N}\big)- \frac{q}{2m_N}\sum_q \hat {\cal C}_{1,q}^{(6)} \big(F_{1}^{q/N}+F_{2}^{q/N}\big)
\\
&- \frac{q m_+}{4m_N^2}\sum_q \hat {\cal C}_{4,q}^{(6)} F_{P'}^{q/N}- \frac{q}{2 m_N}\sum_q \frac{1}{m_q}\hat {\cal C}_{8,q}^{(6)} F_{P}^{q/N}
+\frac{q}{2m_N}\Big(\hat {\cal C}_{4}^{(7)} F_{\tilde G}^{N}+\hat {\cal C}_{8}^{(7)} F_{\tilde \gamma}^{N}\Big)
\\
& -\frac{(m_+-q)q}{2m_N} \sum_q \hat {\cal C}_{9,q}^{(7)} \big(F_{1}^{q/N}+F_{2}^{q/N}\big) - i q
  \sum_q \hat {\cal C}_{12,q}^{(7)}\Big(F_A^{q/N}- \frac{m_+m_-}{4m_N^2} F_{P'}^{q/N}\Big)
 \\
 &- q \sum_q  \hat {\cal C}_{13,q}^{(7)}  \Big( \hat F_{T,0}^{q/N}+\frac{q^2}{4m_N^2} \hat F_{T,1}^{q/N}\Big)+ i q \sum_q  \hat {\cal C}_{16,q}^{(7)} \hat F_{T,0}^{q/N},
  \end{split}
 \\
 c_7^N&=\sum_q \hat {\cal C}_{3,q}^{(6)} F_{A}^{q/N} +(m_++q)\sum_q \hat {\cal C}_{11,q}^{(7)} F_{A}^{q/N},
 \\
\begin{split}
c_8^N&=i\frac{\alpha}{\pi q}\sum_q \hat {\cal C}_{2}^{(5)} Q_q F_{1}^{q/N}- \sum_q \hat {\cal C}_{2,q}^{(6)} F_{1}^{q/N}
\\
& +i(m_--q) \sum_q \hat {\cal C}_{10,q}^{(7)} F_{1}^{q/N} -
 \frac{q^2}{2m_N} \sum_q \hat {\cal C}_{14,q}^{(7)}  \Big( \hat F_{T,1}^{q/N}-4 \hat F_{T,2}^{q/N}\Big),
 \end{split}
 \\
  \begin{split}
c_9^N&=i\frac{\alpha}{2 \pi m_N}\sum_q \hat {\cal C}_{2}^{(5)} Q_q \big(F_{1}^{q/N}+F_{2}^{q/N}\big)- \frac{q}{2m_N}\sum_q \hat {\cal C}_{2,q}^{(6)} \big(F_{1}^{q/N}+F_{2}^{q/N}\big)
\\
&- \sum_q \hat {\cal C}_{3,q}^{(6)} F_{A}^{q/N}- 2i\sum_q \hat {\cal C}_{10,q}^{(6)} \hat F_{T,0}^{q/N}
\\
& +i \frac{(m_--q)q}{2m_N} \sum_q \hat {\cal C}_{10,q}^{(7)} \big(F_{1}^{q/N}+F_{2}^{q/N}\big) - 
(m_+-q) \sum_q \hat {\cal C}_{11,q}^{(7)} F_{A}^{q/N}
 \\
 &- q \sum_q  \hat {\cal C}_{14,q}^{(7)}  \Big( \hat F_{T,0}^{q/N}+\frac{q^2}{4m_N^2} \hat F_{T,1}^{q/N}\Big),
  \end{split}
  \\
    \begin{split}
c_{10}^N&=i \sum_q \hat {\cal C}_{3,q}^{(6)} \Big(F_{A}^{q/N}-\frac{q \,m_-}{4 m_N^2}F_{P'}^{q/N}\Big)+ \frac{q}{2m_N}\sum_q\frac{1}{m_q}\hat {\cal C}_{7,q}^{(6)} F_P^{q/N}-2 \sum_q \hat {\cal C}_{10,q}^{(6)} \hat F_{T,0}^{q/N}
\\
& -\frac{q}{2m_N}\Big(\hat {\cal C}_3^{(7)} F_{\tilde G}^N+\hat {\cal C}_7^{(7)} F_{\tilde \gamma}^N\Big)+i
m_+ \sum_q \hat {\cal C}_{11,q}^{(7)} \Big(F_{A}^{q/N}-\frac{q m_-}{4 m_N^2} F_{P'}^{q/N}\Big)
 \\
 &- q \sum_q  \hat {\cal C}_{15,q}^{(7)}  \hat F_{T,0}^{q/N},
  \end{split}
 \\
  \begin{split}
c_{11}^N&=\frac{\alpha}{\pi q}\sum_q \hat {\cal C}_{2}^{(5)} Q_q F_{1}^{q/N}- i \sum_q \hat {\cal C}_{2,q}^{(6)} F_{1}^{q/N}
- \sum_q \frac{1}{m_q}\hat {\cal C}_{6,q}^{(6)} F_{S}^{q/N}
\\
&+ \frac{q}{m_N}\sum_q \hat {\cal C}_{10,q}^{(6)} \Big(\hat F_{T,0}^{q/N}-\hat F_{T,1}^{q/N}+4 \hat F_{T,2}^{q/N}\Big) -{\cal C}_2^{(7)}F_G^N-{\cal C}_6^{(7)} F_{\gamma}^N
\\
&- (m_-+q) \sum_q \hat {\cal C}_{10,q}^{(7)} F_{1}^{q/N} +i \frac{q^2}{2m_N} \sum_q  \hat {\cal C}_{14,q}^{(7)}  \Big[ \hat F_{T,0}^{q/N}-\hat F_{T,1}^{q/N}+\Big(4+\frac{q^2}{m_N^2} \Big)\hat F_{T,2}^{q/N}\Big],
  \end{split}  
\\
c_{12}^N&=-2 \sum_q \hat {\cal C}_{10,q}^{(6)} \hat F_{T,0}^{q/N} +
 q \sum_q \hat {\cal C}_{15,q}^{(7)}  \hat F_{T,0}^{q/N},  
\\
c_{13}^N&=-2 i \sum_q \hat {\cal C}_{9,q}^{(6)} \hat F_{T,0}^{q/N} +
 q \sum_q \hat {\cal C}_{16,q}^{(7)}  \hat F_{T,0}^{q/N},  
 \\
c_{14}^N&=- i \sum_q \hat {\cal C}_{4,q}^{(6)} F_{A}^{q/N} -
 (m_-+q) \sum_q \hat {\cal C}_{12,q}^{(7)}  F_{A}^{q/N},   
 \\
c_{15}^N&= i q \sum_q\hat {\cal C}_{14,q}^{(7)} \Big(\hat F_{T,0}^{q/N} +\frac{q^2}{m_N^2}\hat F_{T,2}^{q/N}\Big)+
 q \sum_q \hat {\cal C}_{15,q}^{(7)} \hat F_{T,0}^{q/N},  
 \\
 \label{eq:c16N}
c_{16}^N&=i\frac{\alpha}{\pi q} \sum_q \hat {\cal C}_{2}^{(5)} Q_q F_{1}^{q/N} -
i q \sum_q \hat {\cal C}_{10,q}^{(7)}  F_{1}^{q/N},   
\end{align}
where all the form factors are understood to be evaluated at $q_{\rm rel.}^2=-q_{\rm eff}^2$, and we defined
\beq
\label{eq:m+-}
m_\pm=m_\mu\pm m_e.
\eeq
Similarly, one can express the $b_i^N$ coefficients in terms of $\hat \C_i^{(d)}$. While the $b_i^N$ coefficients  are included in our numerical results, as well as in the {\tt MuonBridge} computer code, they generate only subleading corrections. As a result, they are not relevant to the present qualitative discussion, and we do not show their matching explicitly. 

Note that the NRET LECs $c_i^N$ are in general complex. First of all, the Wilson coefficients $\hat C_i^{(d)}$ can be complex, given that the lepton-flavor-violating WET operators in Eq. \eqref{eq:WET} are not hermitian (note though, that they are defined in such a way that they would have been hermitian, had the leptonic currents been flavor conserving).  However, even if $\hat C_i^{(d)}$ are taken to be real, the $c_i^N$ are still complex in general, due to the appearance of explicit factors of $i$ in Eqs. \eqref{eq:c1}--\eqref{eq:c16N}.  These can be traced back to the fact that $i \hat q$ used to construct the NRET operators actually represents two operators, the electron velocity and the three-momentum transfer, see the discussion in Appendix A of Ref. \cite{Haxton:2022piv}.

\begin{figure}[t]
    \centering
    \includegraphics[width=1.0\textwidth]{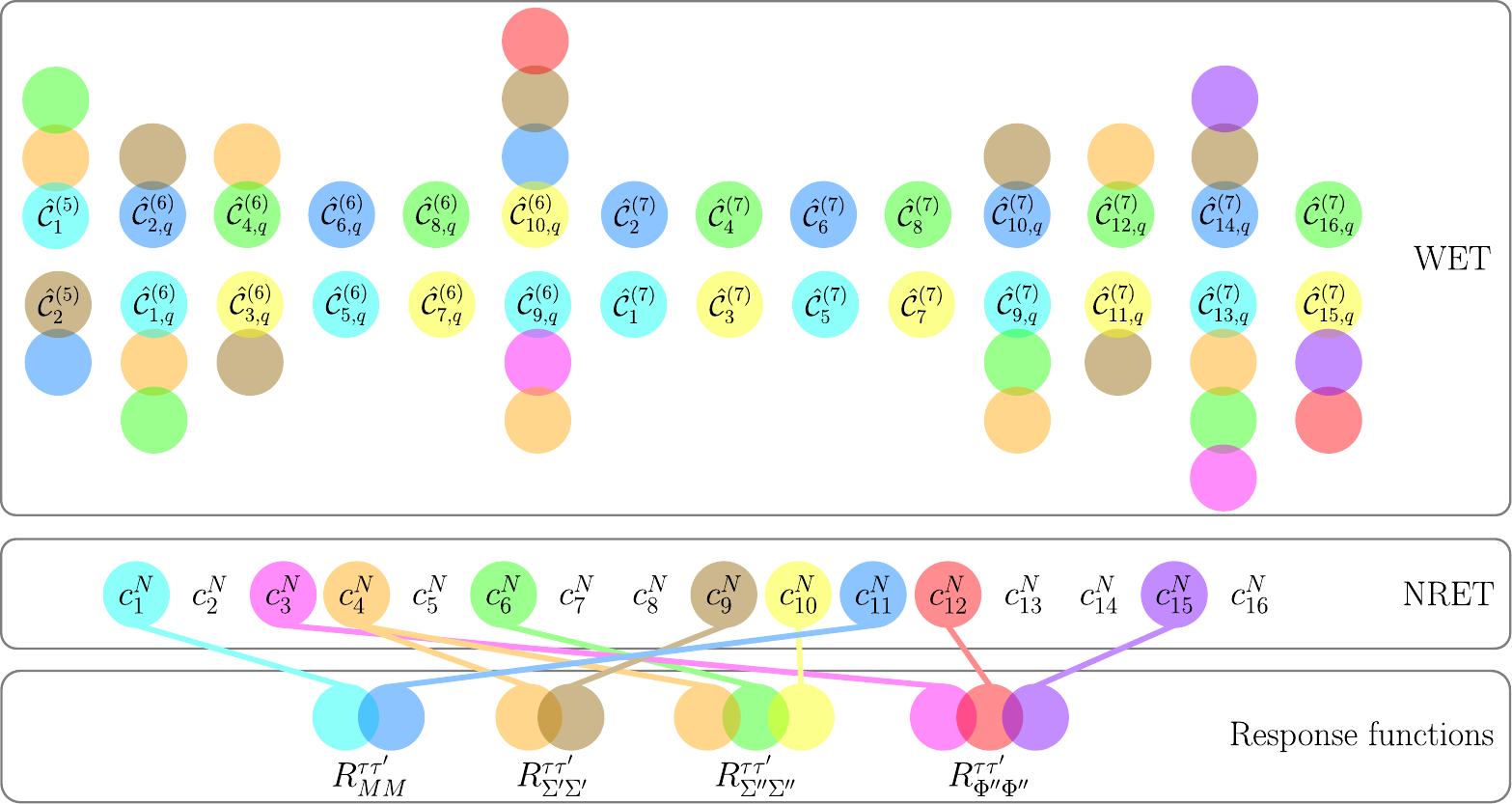}
    \caption{A graphical representation of how different WET operators \eqref{eq:WET} contribute to the $\mu\to e$ conversion rate, in which we keep the terms that remain in the $v_N, v_\mu\to 0$ limit, as well as the coherently enhanced but velocity-suppressed contribution from the $W_{\Phi''\Phi''}$ response function. The NRET coefficients $c_i^N$ receive contributions from WET Wilson coefficients $\hat \C_i^{(d)}$, as denoted with the corresponding colored circles (only the $c_i^N$ that contribute to the numerically leading response functions are considered). These then enter the leptonic response factors $R_{OO'}^{\tau\tau'}$ that multiply various  nuclear response functions, as indicated. For instance, $\hat\C_1^{(5)}$ contributes to $c_1^N, c_4^N$ and $c_6^N$, and these lead to $W_{MM}^{\tau\tau'}$,  $W_{\Sigma'\Sigma'}^{\tau\tau'}$,   $W_{\Sigma''\Sigma''}^{\tau\tau'}$ nuclear responses.   
    \label{fig:NR:WET:relations}
    }
\end{figure}

Using the results for $c_i^N$ in Eqs.~\eqref{eq:c1}--\eqref{eq:c16N}, together with the rate formula in Eq. \eqref{eq:Gamma:mutoe} and the known nuclear response functions, $W_a^{\tau\tau'}$, one can obtain predictions for the $\mu\to e$ conversion rate in terms of the WET Wilson coefficients $\hat \C_i^{(d)}$. Here, we focus on the leading contribution to $\Gamma(\mu\to e)$, i.e., those terms that are nonzero in the $v_N, v_\mu\to 0$ limit, supplemented by the coherently enhanced but velocity suppressed contribution from the $W_{\Phi''\Phi''}$ response function, which in general contributes at the same level as the spin-dependent nuclear response functions, cf., Eq.~\eqref{eq:nuclear:response:scalings}. The graphical representation of how the WET Wilson coefficients $\hat \C_i^{(d)}$ map onto the leading prediction for the $\mu \to e$ conversion rate is shown in Fig.~\ref{fig:NR:WET:relations}.

We make several phenomenologically relevant observations:
\begin{itemize}
\item Dimension-5 WET operators $\Q_{1,2}^{(5)}$, i.e., the transition magnetic and electric moments, induce the coherently enhanced spin-independent, $W_{MM}^{\tau\tau'}$, as well as the spin-dependent, $W_{\Sigma'\Sigma'}^{\tau\tau'}$, $W_{\Sigma''\Sigma''}^{\tau\tau'}$, nuclear responses. At dimension 6 also $W_{\Phi''\Phi''}^{\tau\tau'}$ is generated (although only from $\hat \C_{9,q}^{(6)}$ and $\hat \C_{10,q}^{(6)}$).
\item
Even if only a single Wilson coefficient $\hat C_i^{(d)}$ is nonzero, this always results in a nonzero contribution from at least one of the numerically leading nuclear response functions. 
\item All 16 NRET operators are generated by the dimension $d\leq 7$ WET basis. At dimension $d\leq 6$, all NRET operators except $\mathcal{O}_{15}^N$ are generated.
\item
Not all NRET operators  ${\mathcal O}_i^N$ in Eq. \eqref{eq:L_NRET:vmu} contribute to the above numerically leading nuclear response functions: $c_5^N$, $c_8^N$, and $c_{13}^N$ contribute only to numerically subleading nuclear responses, whereas $c_2^N$, $c_7^N$, $c_{14}^N$, and $c_{16}^N$ do not contribute at all to elastic $\mu\rightarrow e$ conversion due to the parity and time-reversal properties of the nuclear response functions that they generate \cite{Haxton:2022piv}. 
\item
Working at this numerically leading order, certain WET Wilson coefficients contribute to just a single NRET operator:\footnote{We ignored for simplicity the quark flavor in this counting. A more precise statement is that a certain linear combination of $\hat C^{(6)}_{5,u}, \hat C^{(6)}_{5,d}, \hat C^{(6)}_{5,s}, \hat C_{1}^{(7)}, \hat C_{5}^{(7)}$ enters $c_1^p$, while a different linear combination enters $c_1^n$, and similarly for the other $c_i^N$.}  the Wilson coefficients $\hat C^{(6)}_{5,q}, \hat C_{1}^{(7)}, \hat C_{5}^{(7)}$ only contribute to $c_1^N$;  $\hat C_{6,q}^{(6)}, \hat C_{2}^{(7)}, \hat C_{6}^{(7)}$ only to $c_{11}^N$;  $\hat C_{7,q}^{(6)}, 
\hat C_{3}^{(7)}, \hat C_{7}^{(7)}$ only to $c_{10}^N$;  and $\hat C_{8,q}^{(6)}, \hat C_{4}^{(7)}, \hat C_{8}^{(7)}$ only to $c_{6}^N$. The Wilson coefficients only contributing to a single $c_i^N$ cannot be distinguished from each other using measurements on different targets, since they always enter in the same linear combination, unless the much smaller $v_\mu$ suppressed relativistic corrections and the small corrections from the $q^2$ dependence of nucleon form factors can also be taken into account (i.e., the total theoretical error on the prediction for $\Gamma(\mu \to e)$ reaches the level where these corrections become relevant). 
\item
The $q_{\rm eff}$ dependence of CLFV coefficients $R_{MM}^{\tau\tau'}$,  $R_{\Sigma'\Sigma'}^{\tau\tau'}$,   $R_{\Sigma''\Sigma''}^{\tau\tau'}$ and $R_{\Phi''\Phi''}^{\tau\tau'}$ comes from the $q$ dependence of $c_i^N$ in Eqs. \eqref{eq:c1}-\eqref{eq:c16N} (for heavy new physics the WET Wilson coefficients $\C_i^{(d)}$ are $q$ independent). Since $q_{\rm eff}$ depends only mildly  on the chosen target nucleus, the $c_i^N$ entering the predictions for the $\mu\to e$ rate are effectively constant, i.e., independent of the nuclear target. In this limit there are only 12 combinations of $c_i^N$, i.e., the combinations multiplying $R_{MM}^{\tau\tau'}$,  $R_{\Sigma'\Sigma'}^{\tau\tau'}$,   $R_{\Sigma''\Sigma''}^{\tau\tau'}$ and $R_{\Phi''\Phi''}^{\tau\tau'}$ for $\tau^{(')}=0,1$ that can be measured. 
\end{itemize}

At a future time when $\mu\to e$ conversion has been discovered in several different targets, one could use the ``bottom-up" approach to efficiently encode the CLFV physics in the NRET LECs, which would then serve
as constraints on generic UV models. Given existing uncertainties in nuclear response calculations --- perhaps
in the range of tens of percent --- details like the variation of $q_{\rm eff}$ with target could be
neglected. (If our ability to compute nuclear responses were ever to reach the few percent level,
 one would need to exercise more care.)  In general, one would not be able to determine individual
 LECS, but only the combinations that appear in the
 expressions for the $R^{\tau \tau^\prime}_{O O^\prime}$. To do more, one would need additional 
 constraints on CLFV, beyond those available from
 elastic $\mu \rightarrow e$ conversion.
 
 The NRET employs single-nucleon currents and charges, but could in principle be extended to include 
 a similar set of Galilean-invariant two-body operators. For certain cases, like the Rayleigh operators, these two-body currents can be important \cite{Ovanesyan:2014fha}.  Due to the averaging properties of nuclei, 
 most of the effects of such contributions (assuming they are not treated explicitly) would be absorbed into the one-body LECs, making them effective, as
 discussed in Sec. IIIG of Ref. \cite{Haxton:2022piv}. While most of the missing physics would thus be properly incorporated into the fitted LECs, one would need to account for the
 associated renormalization and operator mixing, before relating empirically determined LECs to the predictions
 of an underlying model.

\section{Sample new physics models}\label{sec:NewPhysicsModels}
Next we turn to a few examples of new physics models that can lead to $\mu\to e$ conversions. In the ``top-down'' approach, our results can be used to obtain predictions for $\Gamma(\mu \to e)$ in a particular UV model. In this case, one must first match the UV model onto SMEFT by integrating out the mediators at scale $\mu\approx m_{\rm mediator}$, then RG evolve in SMEFT down to $\mu\approx m_W$, match onto WET, and finally RG evolve down to $\mu=2$ GeV. At this point one can then use our results, encoded in the {\tt MuonBridge} code suite, to obtain the prediction for the $\mu\to e$ transition rate. In Sec. \ref{sec:bounds:Wilson} we first illustrate the expected reach of the upcoming $\mu\to e$ experiments in terms of bounds on single SMEFT Wilson coefficients. In Secs. \ref{sec:leptoquarks} and \ref{sec:PNGB}, on the other hand, we use concrete new physics models. From the large array of possible UV examples, we choose two that best highlight the strengths of our systematic EFT based approach: $\mu\to e$ conversions induced by leptoquark exchanges, which lead to scalar, vector, and tensor currents (Sec. \ref{sec:leptoquarks}), and $\mu\to e$ conversions induced by light pseudoscalar/ALP exchanges, which can also be covered by our formalism, but now with $q^2$-dependent Wilson coefficients $\hat {\cal C}_i^{(d)}$ (Sec. \ref{sec:PNGB}).

\subsection{Bounds on UV Wilson coefficients}
\label{sec:bounds:Wilson}
We start by considering the case where the UV theory generates a single SMEFT operator above the electroweak scale. This is a standard approach in phenomenological EFT analysis used to estimate the scales that are being probed by experiments, without consideration for the ability or inability of plausible new physics models to realize these single-operator contributions. 

\begin{figure}[p!h!]
    \centering
    \includegraphics[width=1.0\textwidth]{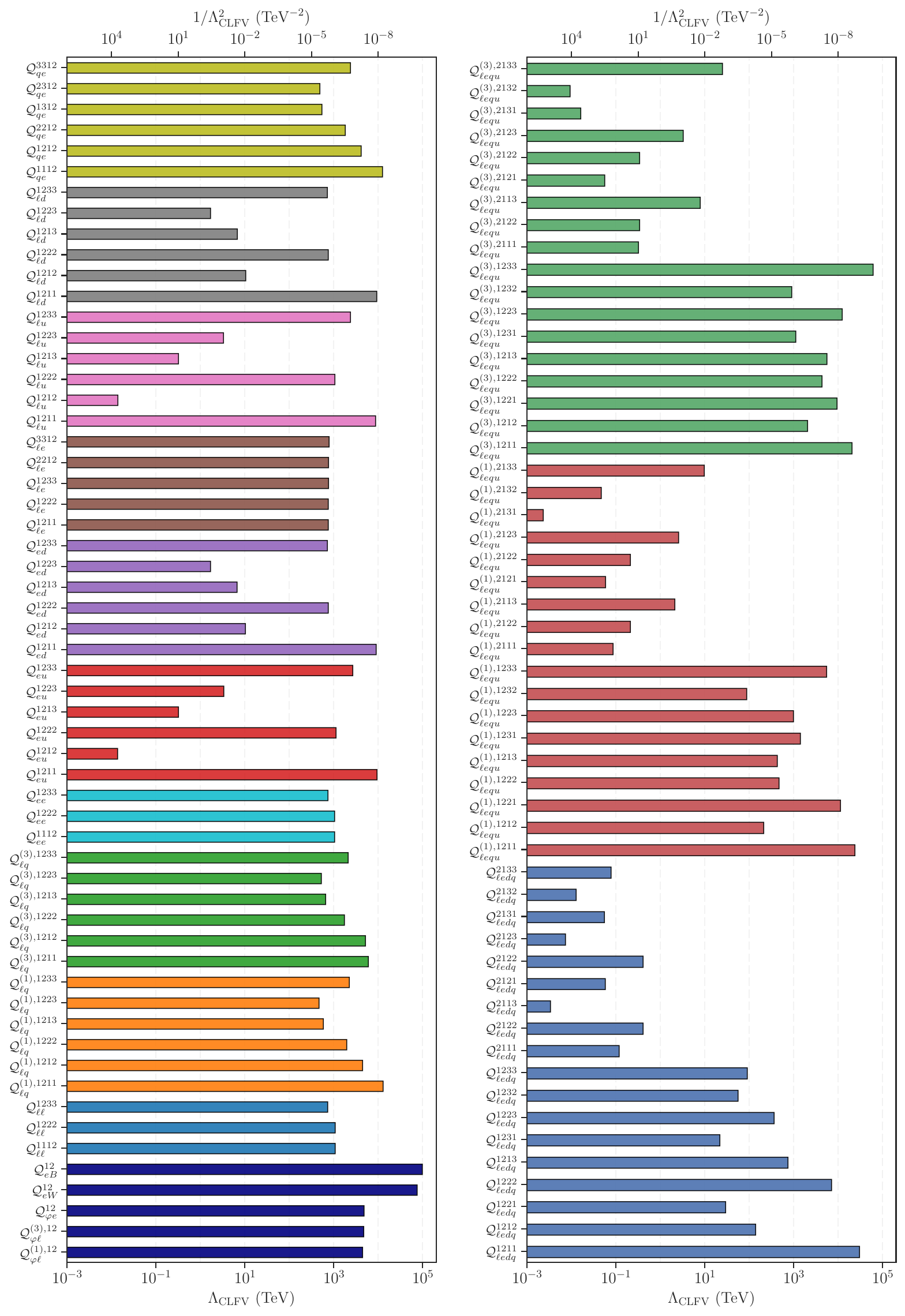}
    \caption{The energy scales probed by dimension-six SMEFT operators assuming the projected experimental limits of the future Mu2e and COMET experiments, $B(\mu^- + \mathrm{Al}\to e^- + \mathrm{Al}) < 10^{-17}$. The notation is as in \cite{Grzadkowski:2010es,Celis:2017hod}.
    }\label{fig:EFT:bounds}
\end{figure}

For concreteness, let us consider a UV theory that, after integrating out the heavy degrees of freedom at a scale $\mu=\Lambda$ ($\sim$ heavy mediator mass), generates just a single dimension-6 SMEFT operator in the Warsaw basis
\beq
{\cal L}_{\text{SMEFT,UV}}= \frac{{\cal C}_i}{\Lambda^2}{\cal Q}_i.
\eeq
To arrive at the prediction for the  $\mu \rightarrow e$ conversion rate, as described in Sec.~\ref{sec:NRET}, one first needs to RGE evolve from the scale $\mu= \Lambda$ to $\mu=2\,$GeV using the tower of EFTs, from SMEFT to WET with $n_f=5$ to WET with $n_f=4$ flavors followed by a tree-level matching to WET with $n_f=3$, as shown in Fig.~\ref{fig:diagram}. We perform the one-loop RG running using \texttt{wilson} \cite{Aebischer:2018bkb} and then use  \texttt{MuonConverter} to obtain the prediction for the $\mu\rightarrow e$ conversion rate.
Using the projected experimental sensitivity $B(\mu^- + \mathrm{Al}\to e^- + \mathrm{Al}) < 10^{-17}$, we show in Fig.~\ref{fig:EFT:bounds} the implied bounds on  $\Lambda_{\text{CLFV},i}=\Lambda/\sqrt{{\cal C}_i}$, if no signal is found. We perform the calculation for each dimension-6 CLFV SMEFT operator ${\cal Q}_i$ in the Warsaw basis that can induce $\mu\to e$ transitions. 

For each operator in  Fig.~\ref{fig:EFT:bounds}, the bound is obtained by identifying the UV scale  $\Lambda_{\text{CLFV},i}$ that, after one-loop RGE evolution to $\mu=2$\,GeV, produces a conversion rate saturating the projected single-event sensitivities of the upcoming Mu2e and COMET experiments.
The magnitude of the initial Wilson coefficient, assuming $\mathcal{C}_i =1$, is set by $\Lambda_{\text{CLFV},i}^{-2}$ and scanned over ten equally spaced points in the range $\Lambda_{\text{CLFV},i}^{-2} \in 10^{-17} - 10^{-4}$ GeV$^{-2}$ corresponding to new physics scales $\Lambda_{\text{CLFV},i} \sim 100 \text{ GeV} - 10^5$ TeV. The resulting conversion rates are then interpolated as a function of $\Lambda_{\text{CLFV},i}$ and a root-finding algorithm is utilized to obtain the scale at which the conversion rate crosses the projected experimental limits. The running of each operator is solved exactly via numerical integration of the one-loop RG equations from $\mu = \Lambda$ to $\mu = 2$ GeV using \texttt{wilson}. In agreement with the leading log resummation, the matchings between different thresholds are performed at tree level.  

\clearpage

Operator-mixing induced by the one-loop running allows for non-zero limits to be placed on the purely leptonic and off-diagonal semi-leptonic SMEFT operators. This treatment is consistent as long as the logarithmically enhanced RG running contributions to dimension-six operators are numerically leading and the finite terms from loop-level matching can be ignored. 

\subsection{Leptoquarks}
\label{sec:leptoquarks}
Next, we use our computational framework to analyze the $\mu \rightarrow e$ conversion rate in the context of an explicit UV model. Specifically, we assume that  the $\mu \rightarrow e$ conversion is generated by tree-level exchange of a leptoquark scalar $R_2$, which then leads to vector, scalar and tensor interactions, making it an ideal showcase for quantifying the relative magnitudes and correlations between different nuclear responses. The leptoquark $R_2$ is in the $(\mathbf{3}, \mathbf{2},7/6)$ representation of the SM gauge group, so that the interaction Lagrangian is given by  \cite{Dorsner:2016wpm}
\beq
\begin{split}\label{eq:LQ_R2}
{\cal L} & \supset -y_{2\,ij}^{RL}\, \bar u_R^{i} R^a_2 \epsilon^{ab} L_L^{j,b}+y_{2\,ij}^{LR}\,\bar e_R^{i} R_2^{a\,*} Q_L^{j,a}+{\rm h.c.},
\end{split}
\eeq
where the summation over flavor indices, $i,j=1,2,3$, and electroweak SU(2) indices $a,b=1,2$ is implicit, and we do not display the contraction of color indices. 

\begin{figure}
    \centering
    \includegraphics[scale=0.80]{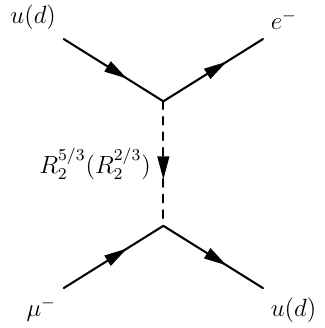}
    \caption{Diagram for tree-level contributions to $\mu\rightarrow e$ conversion mediated by scalar leptoquark $R_2$. Leptoquark superscripts indicate the electric charge of the exchanged particle.
    }
    \label{fig:LQ_diagrams}
\end{figure}

Integrating out the leptoquark at tree level, cf.~Fig.~\ref{fig:LQ_diagrams}, gives the following nonzero contributions to the SMEFT Wilson coefficients (we use the notation of Refs. \cite{Grzadkowski:2010es,Celis:2017hod}),
\begin{subequations}
\label{eq:LQ:all}
    \begin{align}
        C_{\ell u}^{12ii}&=-\frac{1}{2m_{\text{LQ}}^2}y_{2\,i2}^{RL}y_{2\,i1}^{RL*},
        \label{eq:C:ellu:12ii}
        \\
         C_{q e}^{ii12}&=-\frac{1}{2m_{\text{LQ}}^2}y_{2\,2i}^{LR*}y_{2\,1i}^{LR},
         \label{eq:C:qe:ii12}
         \\
        C^{(1),12ii}_{\ell equ}&=2C^{(3),12ii}_{\ell equ}=-\frac{1}{2m_{\text{LQ}}^2}y_{2\,2i}^{LR*}y_{2\,i1}^{RL*},
        \label{eq:C:(1):12ii}
        \\
        C^{(1),21ii}_{\ell equ}&=2C^{(3),21ii}_{\ell equ}=-\frac{1}{2m_{\text{LQ}}^2}y_{2\,i2}^{LR}y_{2\,1i}^{RL},
        \label{eq:C(1):21ii}
    \end{align}
\end{subequations} 
where $m_{\text{LQ}}$ is the leptoquark mass.
The Wilson coefficients in Eqs. \eqref{eq:C:ellu:12ii}, \eqref{eq:C:qe:ii12} multiply vector four-fermion operators $(\bar \ell_1 \gamma_\mu \ell_2)(\bar u_i \gamma^\mu u_i)$ and $(\bar q_i \gamma^\mu q_i)(\bar e_1 \gamma_\mu e_2)$, respectively, while the Wilson coefficients in Eqs. \eqref{eq:C:(1):12ii}, \eqref{eq:C(1):21ii} with the superscript $(1)$ $[(3)]$ multiply scalar [tensor] four-fermion operators of the form $(\bar \ell_1^a e_2) \epsilon_{ab}(\bar q_i^b u_i)$ [$(\bar \ell_1^a \sigma_{\mu\nu}e_2) \epsilon_{ab}(\bar q_i^b\sigma^{\mu\nu} u_i)$] and with $1\leftrightarrow 2$ lepton flavor indices exchanged.

The above effective interactions introduce both spin-independent as well as spin-dependent nuclear responses, when running down to low energies and matching onto NRET. 
Ignoring for the moment the effect of RG running, integrating out the leptoquark results in the SMEFT operators in Eq. \eqref{eq:LQ:all} and gives the following nonzero contributions to the WET Wilson coefficients:
\begin{align}
 \hat {\cal C}_{1,u}^{(6)} &= -\hat {\cal C}_{4,u}^{(6)} =-\frac{1}{8m_{\text{LQ}}^2} \lambda_{+},
 \\
 \hat {\cal C}_{2,u}^{(6)} &= -\hat {\cal C}_{3,u}^{(6)} =-\frac{1}{8m_{\text{LQ}}^2}\lambda_{-},
 \\
 \hat {\cal C}_{1,d_j}^{(6)} &= \hat {\cal C}_{2,d_j}^{(6)} =-\hat {\cal C}_{3,d_j}^{(6)} =-\hat {\cal C}_{4,d_j}^{(6)} =-\frac{1}{8m_{\text{LQ}}^2}\lambda_{j}',
 \\
 \hat {\cal C}_{5,u}^{(6)} &= -\hat {\cal C}_{8,u}^{(6)} =2 \hat {\cal C}_{9,u}^{(6)}= \frac{1}{8m_{\text{LQ}}^2} \lambda_{RL+},
 \\
 \hat {\cal C}_{6,u}^{(6)} &= \hat {\cal C}_{7,u}^{(6)}=2  \hat {\cal C}_{10,u}^{(6)} =\frac{i}{8m_{\text{LQ}}^2} \lambda_{RL-}.
\end{align}
 The up-quark couplings are 
\beq
\lambda_{\pm}=c_{1L} c_{2L}^*\pm c_{2R} c_{1R}^{*}, \qquad \lambda_{RL\pm}=c_{2R} c_{1L} \pm c_{2L}^*c_{1R}^{*},
\eeq
where (for $i=1,2$)
\beq
c_{iL}=( y_2^{LR} V^\dagger)_{i1},  \qquad c_{iR}=y_{2\,1i}^{RL},
\eeq
The coupling to the down quarks is (for $j=1,2$ down quark flavors)
\beq
\lambda_{j}'=y_{2\,1j}^{LR}y_{2\,2j}^{LR*}.
\eeq

The NRET LECs that lead to coherently enhanced nuclear responses (again, ignoring RGE effects) are  given by
\begin{align}
\begin{split}
c_1^N=&-\frac{1}{8m_{\text{LQ}}^2}\Big\{\lambda_{+} F_1^{u/N}- \lambda_{RL+}\Big[\frac{1}{m_u}F_S^{u/N}-\frac{q}{2m_N}\big(\hat{F}_{T,0}^{u/N}-\hat{F}_{T,1}^{u/N}+4 \hat{F}_{T,2}^{u/N}\Big)\Big]\Big\}
\\
&-\frac{1}{8m_{\text{LQ}}^2}\Big(\lambda_1' F_1^{d/N} +\lambda_2' F_1^{s/N} \Big),
\end{split}
\\
\begin{split}
c_{11}^N=&-i c_1^N\big(\lambda_+\to \lambda_-, \lambda_{RL+}\to \lambda_{RL-}\big),
\end{split}
\end{align}
while the LECs that contribute to spin-dependent nuclear responses $W_{\Sigma'\Sigma'}^{\tau\tau'}$, $W_{\Sigma''\Sigma''}^{\tau\tau'}$ are
\begin{align}
\begin{split}
c_4^N=&\frac{1}{8m_{\text{LQ}}^2}\Big\{\lambda_{+} \Big[-F_A^{u/N}+\frac{q}{2 m_N}\Big(F_1^{u/N}+F_2^{u/N}\Big)\Big]+ \lambda_{RL+} \hat{F}_{T,0}^{u/N} \Big\}
\\
&+\frac{1}{8m_{\text{LQ}}^2}\sum_{j} \lambda_j' \Big[-F_A^{d_j/N}+\frac{q}{2m_N}\Big(F_1^{d_j/N}+F_2^{d_j/N}\Big)\Big]
\end{split}
\\
\begin{split}
c_6^N=&\frac{1}{8m_{\text{LQ}}^2}\frac{q}{2m_N} \Big[\lambda_{+}\Big(F_1^{u/N}+F_2^{u/N}-\frac{m_+}{2m_N} F_{P'}^{u/N}\Big)+ \frac{\lambda_{RL+}}{m_u}F_P^{u/N}\Big]
\\
&+\frac{1}{8m_{\text{LQ}}^2} \frac{q}{m_N}\sum_j \lambda_j' \Big(F_1^{d_j/N}+F_2^{d_j/N}-\frac{m_+}{2 m_N}F_{P'}^{d_j/N}\Big)
\end{split}
\\
\begin{split}
c_9^N=&\frac{1}{8m_{\text{LQ}}^2}\Big\{\lambda_{-} \Big[\frac{q}{2 m_N}\Big(F_1^{u/N}+F_2^{u/N}\Big)-F_A^{u/N}\Big]+\lambda_{RL-} \hat{F}_{T,0}^{u/N} 
\Big\}
\\
&+\frac{1}{8m_{\text{LQ}}^2}\sum_j \lambda_j' \Big[\frac{q}{2m_N}\Big(F_1^{d_j/N}+F_2^{d_j/N}\Big)-F_A^{d_j/N}\Big],
\end{split}
\\
\begin{split}
c_{10}^N=&\frac{i}{8m_{\text{LQ}}^2}\Big[\lambda_{-} \Big(F_A^{u/N}-\frac{q m_-}{4 m_N^2} F_{P'}^{u/N}\Big)-\lambda_{RL-}\Big(\hat{F}_{T,0}^{u/N} -\frac{q}{2m_N}\frac{1}{m_u} F_P^{u/N}\Big)\Big]
\\
&+\frac{i}{8m_{\text{LQ}}^2}\sum_j\lambda_j'\Big(F_A^{d_j/N}-\frac{q m_-}{4m_N^2}F_{P'}^{d_j/N}\Big),
\end{split}
\end{align}
and those that generate the coherent but velocity-suppressed $W_{\Phi''\Phi''}^{\tau\tau'}$ nuclear response are
\begin{align}
c_3^N&=-\frac{1}{8m_{\text{LQ}}^2} \lambda_{RL+} \hat{F}_{T,0}^{u/N} ,
\\
c_{12}^N&=-\frac{i}{8m_{\text{LQ}}^2} \lambda_{RL-} \hat{F}_{T,0}^{u/N}.
\end{align}
In addition, the NRET LECs $c_5^N$, $c_8^N$, and $c_{13}^N$ are nonzero but contribute only to numerically subleading nuclear response functions. The NRET LECs $c_2^N$, $c_7^N$, and $c_{14}^N$ are also nonzero, but do not contribute to the elastic conversion process due to the parity and time-reversal symmetries of the nuclear ground state.
Numerically, 
\beq
\label{eq:c1(11):LQ}
    c_{1(11)}^p+c_{1(11)}^n=\frac{-1(+i)}{8m_{\text{LQ}}^2}\left\{2.91(\lambda_\pm +\lambda_1')-\left[6.8(2.2)-0.26(15)\right]\lambda_{RL{\pm}}+\left[2.2(2.1)\times 10^{-4}\lambda_2'\right]\right\}.
\eeq
In summary, i) the LQ model leads to many different NRET coefficients, including those leading to velocity suppressed but coherently enhanced $\Phi''$ nuclear response, and ii) it is possible that the leading, spin-independent, contribution could be accidentally small due to cancellations between different contributions, though that is not a generic situation (cf. Eq. \eqref{eq:c1(11):LQ}). These two qualitative features persist also once the RGE running is included, as we demonstrate below using a numerical analysis. 

The leptoquark model introduced in Eq. \eqref{eq:LQ_R2} depends on one dimensionful quantity, $m_{\text{LQ}}$, and 36 dimensionless parameters  (i.e., 18 complex coefficients $y_{2\, ij}^{RL,LR}$). In matching onto SMEFT in Eq.~\eqref{eq:LQ:all}, we already limited the discussion to $\mu \to e$ transitions and anticipated that only flavor-conserving quark currents are relevant in nuclear transitions.  

\begin{figure}[t!]
    \centering
    \includegraphics[width=1.0\textwidth]{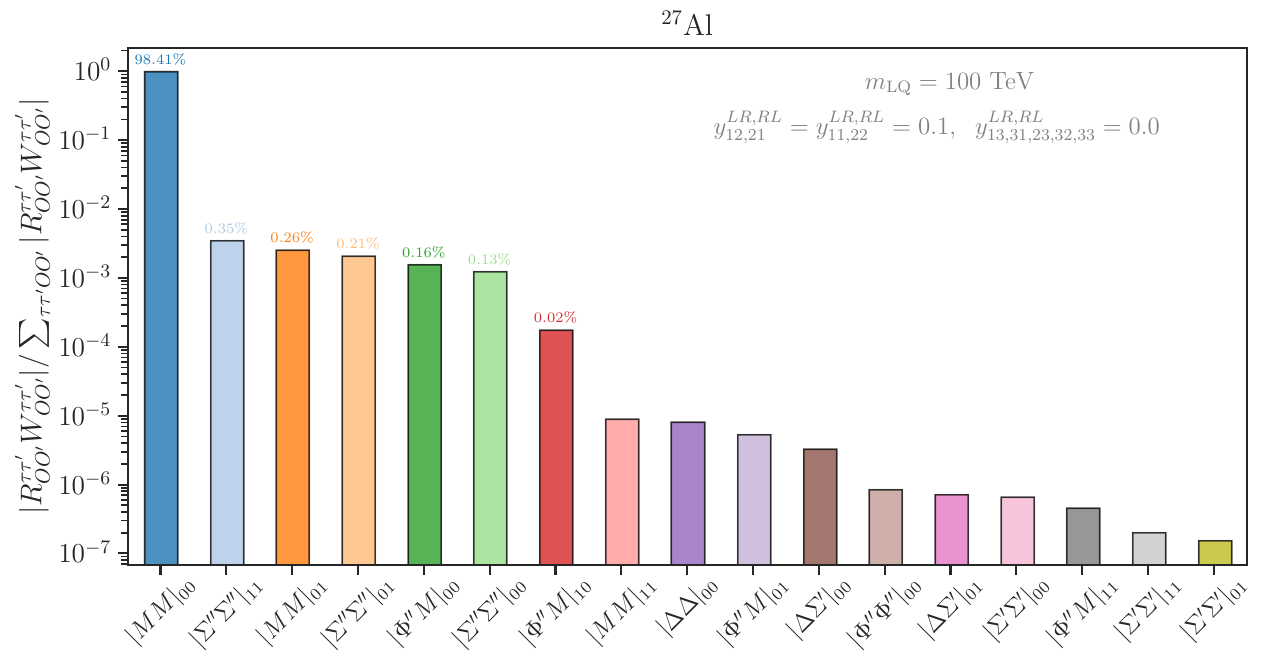}
    \caption{A decomposition of the $\mu \rightarrow e$ conversion rate in terms of contributions from different nuclear response components classified by the type of response ($OO'$) as well as the isoscalar/isovector component ($\tau \tau'$), for  $^{27}$Al target, at a single representative point in the parameter space of the $R_2$ leptoquark model, Eq.~\eqref{eq:LQ_R2}. 
    }
    \label{fig:LQ_rate_decomp}
\end{figure}

In our numerical analysis, we focus on conversion in $^{27}$Al and explore the 18-dimensional complex Yukawa parameter space. The SMEFT coefficients at the scale $\mu=m_{\rm LQ}$, Eq.~(\ref{eq:LQ:all}), are RG evolved to $\mu = 2$ GeV using \texttt{wilson}, after which we utilize our WET to NRET matching expressions to compute the $\mu \to e$ conversion rate as described in Secs.~\ref{sec:LEFT}--\ref{sec:matching}. 
The conversion rate is proportional to the product of leptonic $R^{\tau\tau'}_{OO'}$ and nuclear $W^{\tau\tau'}_{OO'}$ response functions summed over all nuclear responses ($OO'$) and isospin ($\tau, \tau' = 0,1$) components (see Eq.~(\ref{eq:Gamma:mutoe})). To better understand which nuclear responses are numerically relevant it is useful to analyze the sum in terms of its summands. We define the quantities
\begin{equation}
    |OO'|_{\tau \tau'} \equiv \frac{|R_{OO'}^{\tau \tau'}W_{OO'}^{\tau \tau'}|}{\sum_{\tau \tau' O O'} |R_{OO'}^{\tau \tau'}W_{OO'}^{\tau \tau'}|}, \quad (OO')_{\tau \tau'} \equiv \frac{R_{OO'}^{\tau \tau'}W_{OO'}^{\tau \tau'}}{\sum_{\tau \tau' O O'} R_{OO'}^{\tau \tau'}W_{OO'}^{\tau \tau'}},
\end{equation}
such that $|OO'|_{\tau \tau'}$ denotes the normalized, non-negative fractional contribution of the $R^{\tau \tau'}_{OO'} W_{OO'}^{\tau\tau'}$ component to the total rate and $(OO')_{\tau \tau'}$ denotes the normalized, signed fractional contribution. 
Figure \ref{fig:LQ_rate_decomp} shows the relative strengths of each response at a representative point in the leptoquark Yukawa parameter space, $m_{\rm LQ}=100\,$TeV, $y_{12,21}^{LR,RL}=y_{11,22}^{LR,RL}=0.1$, $y_{13,31,23,32,33}^{LR,RL}=0$.
We find that the isoscalar component of the scalar response $|MM|_{00}$ dominates the rate, as expected, but additionally that the coherent tensor-scalar interference response $|\Phi''M|_{00}$ is comparable in magnitude to the longitudinal nuclear spin response $|\Sigma'' \Sigma''|_{00}$ (but where the total summed longitudinal response dominates).
For $\mu\to e$ conversion on $^{63}$Cu nuclei, at the same point in parameter space, we find that the total tensor-scalar interference response instead dominates over longitudinal spin response --- the former making up $\sum_{\tau\tau'}|\Phi''M|_{\tau\tau'}\approx 0.23\%$ of the total rate compared to $\sum_{\tau\tau'}|\Sigma''\Sigma''|_{\tau\tau'}\approx 0.17\%$ from the latter. 

\begin{figure}[t!]
    \centering
    \includegraphics[width=0.95\textwidth]{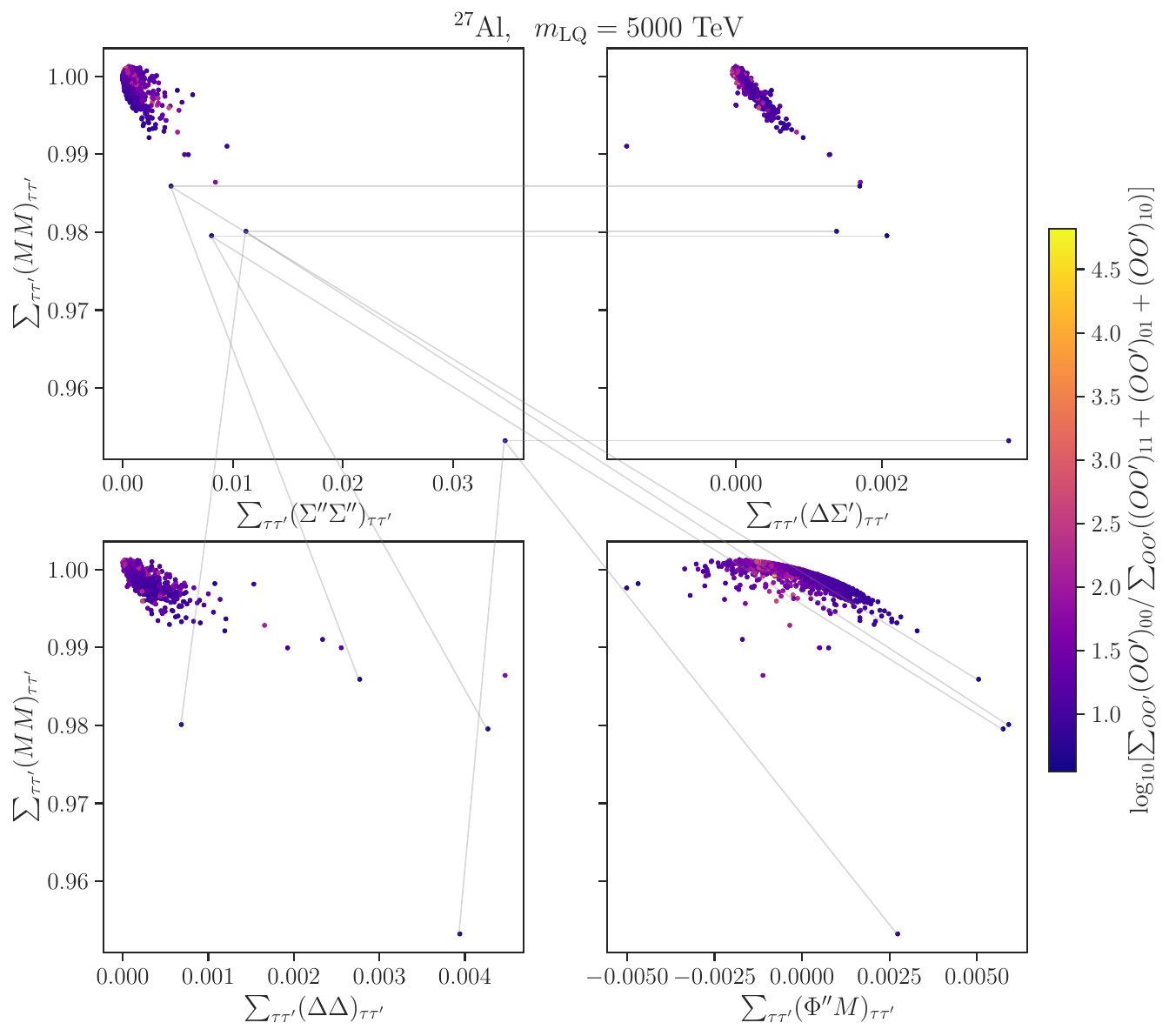}
    \caption{A comparison of the isoscalar nuclear response components over $2 \times 10^4$ parameterizations of the $R_2$ leptoquark model introduced in Eqs.~(\ref{eq:LQ_R2}), (\ref{eq:LQ:all}). Each of the 18 dimensionless complex parameters $y^{RL,LR}_{2\,ij}$ are randomly sampled over the unit complex circle and the leptoquark mass is held fixed at $5000$ TeV. The color denotes the logarithm of the ratio of the summed isoscalar components and the sum of isovector components for all contributing nuclear responses.  Whenever the scalar response contributions fall below $\approx 98\%$ of the total rate, the gray lines are used to illustrate how the components are distributed. The $\mu \rightarrow e$ conversion rates for the shown points lie in the range $10^{-17} < B(\mu^-+\mathrm{Al}\to e^- + \mathrm{Al}) < 10^{-11}$ (we do not denote which of the points are already excluded).
    }
    \label{fig:LQ_isoscalar_components}
\end{figure}

In Fig.~\ref{fig:LQ_isoscalar_components} we illustrate the correlations between the leading isospin-summed nuclear response $(MM)$ and the next-to-leading $(\Sigma'' \Sigma''), (\Delta \Sigma'), (\Delta \Delta)$, and $(\Phi''M)$ nuclear responses over $2 \times 10^4$ samples of the 18-dimensional $y_{2\,ij}^{RL,LR}$ parameter space. Additional nuclear response components, such as $(\Sigma' \Sigma'),(\Phi'' \Phi'')$, and $(\tilde{\Phi}'\tilde{\Phi}')$ are also generated but contribute $< 10^{-5}\%$ to the total rate and thus are not shown.
Each point represents a unique parameterization of the model where the 18 complex Yukawas have been sampled independently and uniformly over the unit complex circle such that
\begin{equation}
    \left|y_{2\,ij}^{RL,LR}\right|^2 = \text{Re}\left(y_{2\,ij}^{RL,LR}\right)^2 + \text{Im}\left(y_{2\,ij}^{RL,LR}\right)^2 \leq 1,
\end{equation}
while leptoquark mass is set to $m_{\rm LQ}=5\cdot 10^3$\,TeV. 
We find that the scalar $(MM)$ and longitudinal-spin $(\Sigma''\Sigma'')$ nuclear responses dominate the total rate and are highly correlated across parameter space. 
The remaining nuclear responses generally provide $<1\%$ contributions to the total rate. 
However, in some parts of parameter space, where contributions from the scalar response falls below $\approx 98\%$, these responses become enhanced with $\mathcal{O}(1\%)$--level contributions that may be measurable, as depicted by the gray lines between the figure sub-panels that connect like-parameterizations.

While outside the scope of the current work, a detailed analysis exploring the similarities and differences of nuclear responses between different target nuclei (see, for example, \cite{Heeck:2022wer}) as well as UV completions is easily accommodated by our computational framework.

\subsection{ALP exchanges} \label{sec:PNGB}
The above formalism for computing $\Gamma(\mu \to e)$ can, with trivial modifications, also be used in scenarios where the mediators are light. We illustrate this in the case of $\mu\to e $ conversion induced by the exchange of a light axion-like particle (ALP) with mass $m_a$. The only required change is that now the Wilson coefficients $\hat C_i^{(d)}$ in WET Lagrangian \eqref{eq:WET} become $q^2$ dependent. As long as one is only interested in $\mu \to e$ conversion, the fact that WET  is strictly speaking no longer the correct effective field theory, since there is an additional light degree of freedom --- the ALP, makes no practical difference since all the relevant effects of the ALP are absorbed in $\hat C_i^{(d)}(q^2)$. 

 The ALP interactions with the SM fields start at dimension 5 and are given by 
\beq
{\cal L}_{\rm ALP}\supset - \frac{\alpha_s}{8\pi}\frac{1}{f_a} a G_{\alpha\beta}^a\tilde G^{a\alpha\beta}+\frac{\partial_\alpha a}{2 f_a} \bar e\gamma^\alpha \big(C_\ell^V+C_\ell^A \gamma_5) \mu+ \sum_{q=u,d,s}  C_q^A \frac{\partial_\alpha a}{2 f_a} \bar q\gamma^\alpha  \gamma_5 q,
\eeq
where we only display the couplings relevant for $\mu \to e$ conversion.  The tree level ALP exchange induces $\mu\to e$ transition, inducing the following $\hat \C_i^{(d)}$ coefficients
\begin{align}
\label{eq:C7q:dim6:ALP}
\hat \C_{7,q}^{(6)}&=-i \frac{m_q m_-}{2 f_a^2}\frac{C_q^A C_\ell^V}{q^2+m_a^2},
\\
\hat \C_{8,q}^{(6)}&=\frac{m_q m_+}{2 f_a^2}\frac{C_q^A C_\ell^A}{q^2+m_a^2},
\\
\hat \C_{3}^{(7)}&=-i \frac{ m_-}{2 f_a^2}\frac{C_\ell^V}{q^2+m_a^2},
\\
\label{eq:C4:dim7:ALP}
\hat \C_{4}^{(7)}&=\frac{m_+}{2 f_a^2}\frac{ C_\ell^A}{q^2+m_a^2},
\end{align}
where we used $q_{\rm rel}^2\simeq -q^2$, cf. Eq.~\eqref{eq:qrel}. Since these are higher dimension operators, suppressed by two powers of the UV scale, $\propto 1/f_a^2$, other dimension 6 contributions from the complete new physics model could be relevant as well, a possibility that we ignore in this example (see, however, Ref. \cite{Fuyuto:2023ogq}).
 More importantly for our purposes is that the $\hat C_i^{(d)}$ are, as anticipated, now explicitly $q^2$ dependent. Since these coefficients are induced in the IR, no RGE running needs to be included in the numerical analysis. That is, the values of the $\hat C_i^{(d)}$ coefficients in \eqref{eq:C7q:dim6:ALP}-\eqref{eq:C4:dim7:ALP} are already given at $\mu=2$\,GeV. 

The resulting NRET coefficients are 
\begin{equation}
    \begin{split}
        c_6^N&=\frac{q}{2m_N}\frac{m_+}{2f_a^2}\frac{C_{\ell}^A}{q^2+m_a^2}\left[-\sum_q C_q^AF_P^{q/N}+F_{\tilde{G}}^N\right],\\
        c_{10}^N&=i\frac{q}{2m_N}\frac{m_-}{2f_a^2}\frac{C_{\ell}^V}{q^2+m_a^2}\left[-\sum_q C_q^AF_P^{q/N}+F_{\tilde{G}}^N\right],
    \end{split}
\end{equation}
which both contribute to the longitudinal spin response function $W_{\Sigma''\Sigma''}^{\tau\tau'}$.

\section{Conclusions}
\label{sec:summary}
Next-generation experiments, such as Mu2e at Fermilab \cite{Mu2e:2014fns,Bernstein:2019fyh} and COMET at J-PARC \cite{COMET:2018auw,COMET:2018wbw}, are expected to advance limits on $\mu \rightarrow e$ conversion rate by four orders of magnitude, to about $B(\mu^- + \mathrm{Al}\rightarrow e^- + \mathrm{Al})\lesssim 10^{-17}$.  What will the new limits --- or a nonzero signal --- tell us about the new physics responsible for CLFV?

The experiments are done at low energy, using nonrelativistic nuclear targets, yet probe new BSM physics associated with UV energy scales. Effective field theory is a powerful technique for bridging between low energies and the UV, thereby connecting experimental constraints to BSM models.  Recently, a nonrelativistic nucleon-level effective theory (NRET) was constructed and then embedded in a series of nuclei, allowing limits extracted from different nuclear targets to be meaningfully compared.  The NRET can be organized according to a hierarchy of dimensionless small parameters, $y=(qb/2)^2 > |\vec{v}_N| > |\vec{v}_\mu| > |\vec{v}_T|$.  The operator expansion through order $\vec{v}_N$ was shown to generate the most
general form of the nuclear $\mu \rightarrow e$ conversion rate, while the retention of $\vec{v}_\mu$ adds form factor corrections associated with the muon's lower component \cite{Rule:2021oxe,Haxton:2022piv}.  Open-source {\tt Mathematica} and {\tt Python} codes named \mutoenretvone{} for calculating nuclear $\mu \rightarrow e$ conversion rates were released with \cite{Haxton:2022piv}, using the NRET basis obtained by expanding through order $\vec{v}_N$.

In this paper, we connected this NRET to a CLFV weak effective theory (WET) in which the degrees of freedom are the light quarks ($u$, $d$, $s$), gluons, and photons. We also updated and extended the accompanying nucleon-level computer codes, to fully support this matching. These codes are collected in the repository {\tt MuonBridge}, consisting of {\tt MuonConverter} and \mutoenretvtwo{} computer codes (available in both {\tt Mathematica} and {\tt Python} versions), as well as of the {\tt Elastic} repository (see Appendix \ref{app:Public} for details). {\tt MuonConverter} matches from WET to the intermediate step of relativistic covariant nucleon interactions, while the new script \mutoenretvtwo{} extends \mutoenretvone{} to support this matching.  This
was done by adding to the script's existing 16 operators the 10 additional   $\vec{v}_\mu$ suppressed NRET operators already identified in \cite{Haxton:2022piv}.
Furthermore, selected tensor-mediated interactions were added to the scalar- and vector-mediated interactions already present in \mutoenretvone{}.  These are needed in the WET matching at dimension 7. Finally, the repository {\tt Elastic} is a database of shell-model one-body density matrices needed for nuclear form-factor calculations. In principle, this repository can be utilized for other problems such as the calculation of dark matter direct detection rates. 

Our choice for the WET basis, consisting of dimension-5, -6, and -7 operators, was motivated by the problem at hand; the operators are built out of quark, gluon and photon currents with definite parity, which simplifies the calculation of nucleon matrix elements. The basis is consistent with the complete WET basis of Ref.~\cite{Liao:2020zyx}, when the latter is restricted to the operators that can mediate $\mu\rightarrow e$ conversion.  {\tt MuonBridge} contains an example of translation between our WET basis and provides the interface to external SMEFT softwares such as \texttt{wilson} \cite{Aebischer:2018bkb}, and \texttt{DsixTools} \cite{Celis:2017hod,Fuentes-Martin:2020zaz}, that can be used to perform RG running, see Appendix \ref{app:Public} for further details. 

In conclusion, over the next decade the experimental community will be making a major effort to improve our understanding of CLFV. In anticipation of these experiments, it is important to develop theory tools that can treat the particle and nuclear physics of $\mu \rightarrow e$ conversion as completely and accurately as possible. As discussed in \cite{Haxton:2022piv}, most past work on $\mu \rightarrow e$ conversion has focused on one or two of the $16+10$ NRET operators, employed schematic nuclear response functions, and simplified the leptonic physics through partial-wave truncations and other steps that are not well justified.   The formalism developed in \cite{Rule:2021oxe,Haxton:2022piv} and encoded in \mutoenretvtwo{} addresses all of these issues, and now {\tt MuonConverter} connects this low-energy formalism to 
the light-quark and gluon WET as well as to higher energy EFTs.  The completeness of the WET and NRET operator bases ensures that they can faithfully encode the low-energy consequences of any UV CLFV theory.

\section*{Acknowledgments} 
The authors are grateful to Andre Walker-Loud for helpful discussions. WH, KM, and ER acknowledge support by the U.S. Department of Energy under grants DE-SC0004658 and DE-AC02-05CH11231 and by
the National Science Foundation under cooperative agreement 2020275. JZ and TM acknowledge support in part by the DOE grant DE-SC1019775, and the NSF grant OAC-2103889. TM acknowledges support in part by the U.S. Department of Energy, Office of Science, Office of Workforce Development for Teachers and Scientists, Office of Science Graduate Student Research (SCGSR) program. The SCGSR program is administered by the Oak Ridge Institute for Science and Education for the DOE under contract number DE‐SC0014664. ER is supported by the U.S. Department of Energy through the Los Alamos National Laboratory. Los Alamos National Laboratory is operated by Triad National Security, LLC, for the National Nuclear Security Administration of U.S. Department of Energy (Contract No. 89233218CNA000001). JZ acknowledges support of a Visiting Miller Professorship, UC Berkeley. JZ acknowledges support from the Simons Foundation Targeted Grant 920184 to the Fine Theoretical Physics Institute. JZ acknowledges support from the Boston University theoretical physics group during completion of this work. 

\appendix

\section{Public code}
\label{app:Public}
Below we outline the structure and purpose of the accompanying public code available in \texttt{Python} and \texttt{Mathematica} at
\begin{center}\label{url:MuonBridge}
 \href{https://github.com/Berkeley-Electroweak-Physics/MuonBridge}{https://github.com/Berkeley-Electroweak-Physics/MuonBridge}\,.
\end{center} 
 The full code repository, termed \texttt{MuonBridge}, is composed of three independent sub-repositories, namely \texttt{Elastic}, \texttt{Mu2e\_NRET}, and \texttt{MuonConverter} each with dependencies on the former i.e., 
 \texttt{MuonConverter} is dependent on \texttt{Mu2e\_NRET}, which is dependent on \texttt{Elastic}. 
 
 \texttt{Elastic} is a database containing ground-state-to-ground-state one-body density matrices computed using the nuclear shell-model code \texttt{BIGSTICK} \cite{Johnson:2013bna,Johnson:2018hrx} for a variety of relevant isotopes for muon-to-electron conversion. 

The repository \texttt{Mu2e\_NRET} contains two versions: \mutoenretvone{} and \mutoenretvtwo{}. \mutoenretvone, originally developed in \cite{Haxton:2022piv}, provides functionality for computing branching ratios and decay rates for nuclear muon to electron conversion. The current release, \mutoenretvtwo, extends the original code by including the effects that arise from tensor-mediated exchanges as well as form-factor corrections induced by the muon's velocity operator $\vec{v}_\mu$.  These additions were necessary to support the current ``top-down" WET reduction as discussed in the main text. 
A top-level {\tt Python} ({\tt Mathematica}) notebook, \texttt{Mu2e\_v2.ipynb}({\tt .nb}) provides an example of typical usage and input. 
The {\tt Mathematica} version offers both an interactive and manual (in the form of an association) input interface while the {\tt Python} version requires manual input in the form of a dictionary or YAML file containing the required parameters. The parameters for both languages include:
\begin{enumerate}
    \item A target isotope choice. For an updated list of supported target isotopes, please refer to the current repository.
    \item A shell-model interaction, used to select one-body density matrices. A detailed list of supported shell-model interactions for specific target nuclei can be found in Table XIII of \cite{Haxton:2022piv}.
    \item An optional harmonic oscillator length scale $b$ in units of fm.
    \item An optional response function option to generate an analytic nuclear response function $W(y)$ (available with {\tt Mathematica} only) and/or to generate plots of the response functions.   With this option, one also specifies the isospin with one of the options:  isoscalar, isovector, proton-only, or neutron-only couplings.
    \item The leptonic scale $m_L$, which should match the leptonic scale used by \texttt{MuonConverter} so that LEC interpretation matches.
    \item The relativistic LECs, i.e., the $d^N_i$ coefficients of the Lorentz-covariant EFT defined in Eq.~(\ref{eq:d:Lagr}). Only non-zero values need to be specified.
    \item An optional override of the default ordinary muon capture rate to be used in the branching ratio calculation. The default values are obtained through a weighted average of the measurements compiled in \cite{PhysRevC.35.2212}.
\end{enumerate}
Additional documentation and annotated examples of input files are included in the repository.

The main purpose of \texttt{MuonConverter} is to provide an interface between external EFT software such as \texttt{wilson} \cite{Aebischer:2018bkb}, \texttt{DsixTools} \cite{Celis:2017hod,Fuentes-Martin:2020zaz}, etc, and the \texttt{Mu2e\_NRET} software developed in \cite{Haxton:2022piv}. This interface extends the functionality of the original \texttt{Mu2e\_NRET} code and allows for full top-down (or bottom-up) phenomenological studies of muon-to-electron conversion in the field of a target nucleus. Explicitly, in conjunction with external EFT software, \texttt{MuonConverter} can be used to compute the influence of UV charged-lepton-flavor-violating operators on the predictions for branching and capture ratios reported by experimental collaborations.

Both versions of \texttt{MuonConverter} (Python and Mathematica) are comprised of four modular components:
\begin{enumerate}
    \item Numerical inputs --- all numerical inputs are stored within an associative array that can be modified by the user upon intialization of the \texttt{MuonConverter} class. The parameters and their default values can be found in \texttt{parameters.py}(\texttt{.wl}).
    \item Form factors --- the form factor expressions required for the WET to NRET matching, Eqs.~\eqref{vec:form:factor}-\eqref{CPodd:photon:form:factor}, and whose numerical values are derived in App.~\ref{app:AFF} can be found in \texttt{form\_factors.py}(\texttt{.wl}). For maximum flexibility, the default form factor values may be manually overwritten within \texttt{parameters.py}(\texttt{.wl}). 
    \item Matching --- to facilitate the WET to NRET matching, \texttt{MuonConverter} utilizes the matching expressions derived in Eqs.~(\ref{eq:d1})-(\ref{eq:d32}) for the relativistic $d_i$ coefficients (the $d_i$ coefficients are automatically translated to the nonrelativistic $c_i$, $b_i$ coefficients within \texttt{Mu2e\_NRET}). The matching expressions, as well as their translation to the isospin basis, can be found in \texttt{hadronization.py}(\texttt{.wl}).
    \item Interfacing --- given an array of WET coefficients (in units of GeV$^{-2}$)\footnote{The dimensionful input allows for the support of multiple CLFV scales $\Lambda, \Lambda', \Lambda'', \ldots$, if desired.}, the interface with \texttt{Mu2e\_NRET}, utilizing external and internal basis translations as well as the matching expressions implemented in \texttt{hadronization.py}(\texttt{.wl}), can be found in \texttt{MuonConverter.py}(\texttt{.wl}).
\end{enumerate}

 To facilitate interfacing with external EFT software we provide a representation of our WET basis (defined in Eqs.(\ref{eq:dim5:nf5:Q1Q2:light})-(\ref{eq:Q7:9})) up to dimension-six\footnote{At the time of writing, commonly used EFT software \cite{Aebischer:2018bkb,Celis:2017hod,Fuentes-Martin:2020zaz} performing RG evolution and matching above and below the electroweak scale only support operators up to dimension-six. Because our basis also includes matching expressions for dimension-seven operators, future versions of these codes supporting higher dimensional operators can be accommodated straightforwardly by \texttt{MuonConverter}.} in the naming conventions of the Wilson coefficient exchange format \texttt{WCxf} \cite{Aebischer:2017ugx}. In addition, we provide an explicit translation between our basis and the relevant subset of flavor-violating operators in the Jenkins, Manohar, Stoffer (JMS) three-flavor WET basis \cite{Jenkins:2017jig, Aebischer:2017ugx} which we will conventionally use as the `reference basis' when interfacing with external codes. For more details on this translation, see App.~\ref{app:wet-translation}.

As an example of typical usage, consider an arbitrary UV model defined above the electroweak scale that has been matched onto a SMEFT basis, run down to $\sim 2$ GeV, matched onto a three flavor WET basis, and translated to the JMS three-flavor WET basis. The output of the previously described procedure will be a data-structure\footnote{\texttt{MuonConverter} utilizes \texttt{Python} dictionaries and \texttt{Mathematica} associations as input.} consisting of Wilson coefficient names and values in the WET--3 JMS basis, e.g.,
\begin{equation}\label{eq:wet3jms}
    \texttt{\{`VeuLL\_1211':$10^{-13}$, $\cdots$\}}.
\end{equation}

Both the \texttt{Python} and \texttt{Mathematica} versions of \texttt{MuonConverter} take Wilson coefficient names and values as well as the momentum transfer $q^\mu \equiv (\Delta E, q_x, q_y, q_z)$ as input (in addition to the atomic and nuclear input required by \texttt{Mu2e\_NRET}, please see App.~C of \cite{Haxton:2022piv} or the example notebooks and documentation within the repository for more information and explicit examples). It is assumed by default that the input Wilson coefficients and values are given in the WET-3 JMS basis.\footnote{However, it is also possible to directly input coefficients from our WET basis, Eqs.~(\ref{eq:dim5:nf5:Q1Q2:light})-(\ref{eq:Q7:9}), by passing the following argument to the relevant functions: \texttt{basis = `HMMRZ'}.} Upon initialization, the input Wilson coefficient data is automatically translated to the basis defined in Eqs.~(\ref{eq:dim5:nf5:Q1Q2:light})--(\ref{eq:Q7:9}). The heart of \texttt{MuonConverter} is the matching expressions derived in Eqs.~(\ref{eq:d1})--(\ref{eq:d32}) which relate the Wilson coefficients from an EFT of relativistic quarks and gluons to an EFT of relativistic nucleons. Using these expressions, the translated dictionary can be straightforwardly `hadronized' and fed into \texttt{Mu2e\_NRET} where these coefficients are mapped to the NRET basis using the expressions derived in Tables \ref{tab:LWL}--\ref{tab:tensor_lower}. The final output is the conversion rate $\Gamma(\mu^- + A \rightarrow e^- + A)$ in units of $s^{-1}$ and corresponding branching ratio $B(\mu^- + A\rightarrow e^- + A)$. 

For example, consider the WET-3 JMS data structure given in Eq.~(\ref{eq:wet3jms}), with all coefficients except \texttt{VeuLL\_1211} set to zero, in an aluminum target with momentum transfer four vector $q^\mu = (0, 0, 0, 0.11081)$ GeV. Internally, the translation to the \texttt{MuonConverter} basis gives non-zero $\mathcal{C}_{1,u}^{(6)}, \mathcal{C}_{2,u}^{(6)}, \mathcal{C}_{3,u}^{(6)},$ and $\mathcal{C}_{4,u}^{(6)}$ coefficients. These WET coefficients then generate non-zero $d^{(N)}_2, d^{(N)}_4, d^{(N)}_5, d^{(N)}_6, d^{(N)}_7, d^{(N)}_{13}, d^{(N)}_{14}, d^{(N)}_{15}$ covariant coefficients upon hadronization. Finally, after feeding the covariant coefficients to \texttt{Mu2e\_NRET}, the conversion rate and capture ratio are given by\footnote{Note that, because the Wilson coefficients in \cite{Haxton:2022piv} are normalized using the weak scale, the conversion rate and the capture ratio returned by the \texttt{Mu2e\_NRET} code must be rescaled by $v^4 / \Lambda^4$ where $v = \left(\sqrt{2} G_F\right)^{-1/2} = 246.2$ GeV is the Higgs vacuum expectation value and $G_F$ is the Fermi constant. This rescaling is automatically incorporated into the rate computation functions within \texttt{MuonConverter}.}
\begin{equation}
    \Gamma(\mu^-+\text{Al} \rightarrow e^- +\text{Al}) = 4.19\times 10^{-10}~\text{s}^{-1}, \qquad B(\mu^-+\mathrm{Al} \rightarrow e^- + \mathrm{Al}) = 6.0 \times 10^{-16}.
\end{equation}

For details on the running and matching procedure down to $\sim 2$ GeV, we refer the reader to the documentation of the respective external EFT software of choice. For additional details, documentation, and examples showcasing the usage of \texttt{MuonConverter} with external EFT software see the public repository whose link is provided at the beginning of this section. 

\begin{landscape}
\begin{figure}[p!h!]
    \includegraphics[width=1.5\textwidth]{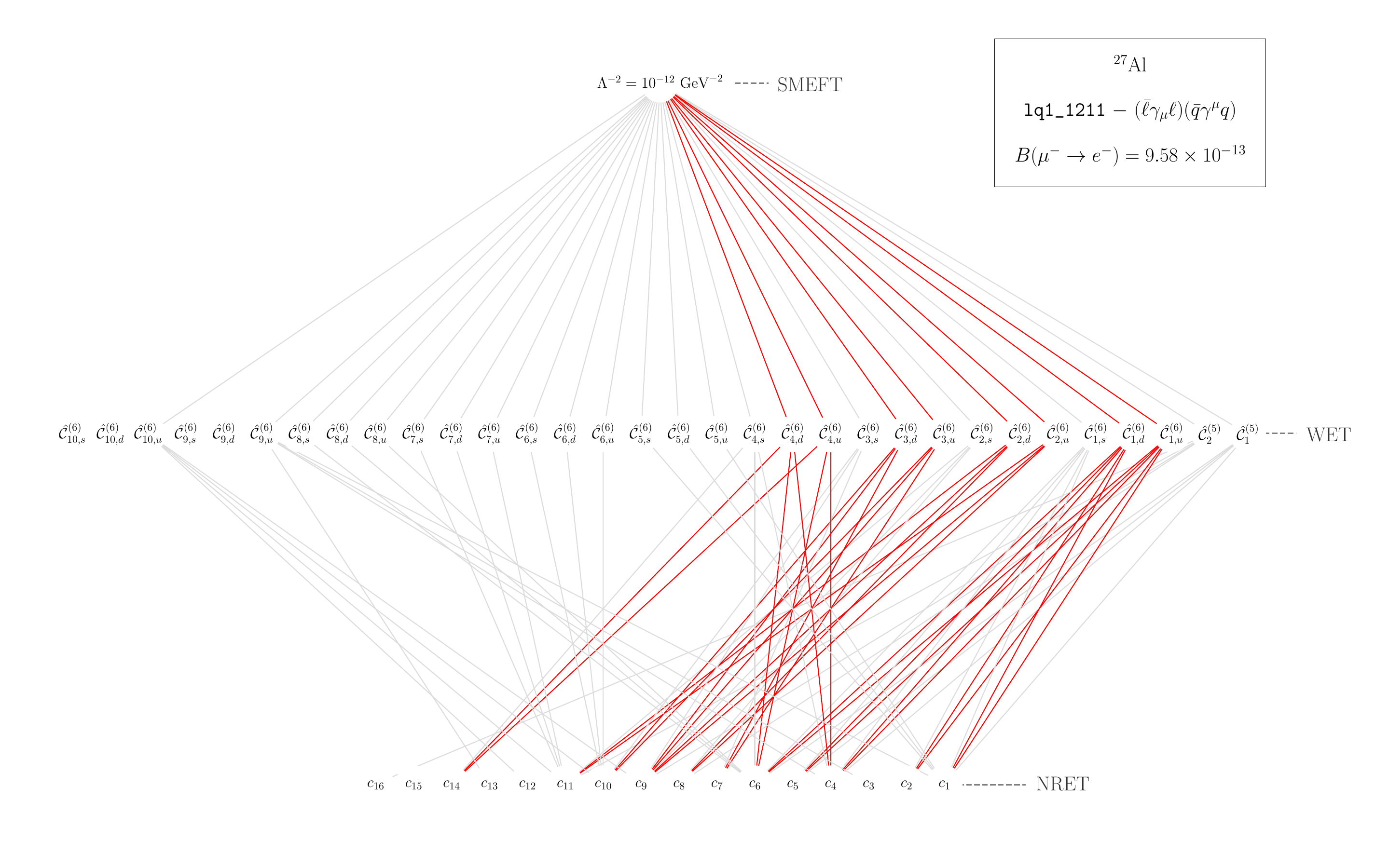}
    \caption{A simplified overview of the mapping and matching computations performed by \texttt{wilson} $+$ \texttt{MuonBridge} for a UV theory defined by a single CLFV SMEFT operator with a Wilson coefficient set to $1/\Lambda^2$ at scale $\mu_{\rm UV}=\Lambda$. The red lines denote the largest contributions to the WET coefficients and are chosen by selecting all coefficients with an absolute value within one order-of-magnitude of the largest coefficient. 
    }\label{fig:MC_map}
\end{figure}
\end{landscape}

A schematic outline of the computation performed by the \texttt{MuonBridge} software suite can be seen in Fig.~\ref{fig:MC_map}. The graph shows how a single CLFV SMEFT operator, defined at a scale $\Lambda = 10^3$ TeV, is mapped through our WET basis and down to the final NRET basis, where the conversion rate is computed. The one-loop RGE from $\mu = \Lambda$ to $\mu = 2$ GeV generates many WET operators, however, only a small portion are numerically relevant --- as depicted by the red edges of the graph.

\section{Intermediate results for WET to NRET matching}
\label{app:more:WET:NRET}
In this appendix we collect further details on the matching from WET to NRET effective theories, where, following Ref. \cite{Haxton:2022piv} we use as an intermediate step a set of Lorentz covariant CLFV operators that are the products of flavor-changing lepton currents and single-nucleon currents. In Sec. \ref{app:WET:to:cov} we first show the results for the nonperturbative matching from WET to these covariant interactions, followed in Sec. \ref{app:nret-decomp} by the nonrelativistic reduction to NRET. Some of the results were already derived in Ref. \cite{Haxton:2022piv} and are here merely reproduced for reader's convenience (Tables \ref{tab:LWL} and \ref{tab:LWL2}), while Tables \ref{tab:tensor_upper} and \ref{tab:tensor_lower} contain new results, required for tensor currents. 

\subsection{From WET to covariant nucleon interactions}
\label{app:WET:to:cov}

In this subsection we collect the results for the $d_j^{(N)}$ coefficients in the effective covariant interaction Lagrangian, Eq. \eqref{eq:d:Lagr}. The basis of covariant operators that we use is given in the first columns of Tables \ref{tab:LWL} and \ref{tab:tensor_upper}. While the basis is overcomplete, this does not cause any problems given that we only use the covariant interactions as the intermediate step in the matching from WET to NRET. The goal is to write down an effective Lagrangian, valid for interactions of external lepton currents with a single nucleon sector. This means that we replace the operators that contain quark currents with equivalent nucleon-level currents, using expressions for nucleon form factors in Eqs. \eqref{vec:form:factor}-\eqref{CPodd:photon:form:factor}. For instance, 
\beq
\bar q \gamma^\mu q\to \bar N\Big[F_1^{q/N}(q^2)\gamma^\mu-\frac{i}{2m_N}F_2^{q/N}(q^2) \sigma^{\mu\nu}q_\nu\Big] N\,,
\eeq
where $N$ are Dirac fields for nucleons, and the arrow means that the LHS and RHS give the same result when sandwiched between single nucleon states, $\langle N'|\ldots|N\rangle$, as is easily checked using Eq. \eqref{vec:form:factor}. Since this is an operator identity, $q_\nu$ should also be interpreted as a derivative, i.e., the operator of the form $\big(\bar e\ldots \mu\big)\big(\bar N\ldots q_\nu N\big)$ should be interpreted as $\big(\bar e\ldots \mu\big) i\partial_\nu \big(\bar N\ldots N\big)$, and similarly for lepton currents containing $q^\mu$ (for simplicity we do not show the Lorentz structures explicitly). The reason for this somewhat unconventional notation is that then the covariant interactions have exactly the same form as in Ref. \cite{Haxton:2022piv}.

As usual, we make use of equations of motion when performing the matching. Furthermore, in deriving the expressions for the $d_i^N$ coefficients the following identity proves to be useful 
\begin{align}
\label{eq:useful:quark:rel}
\gamma^{[\mu}\slashed{q}\gamma^{\nu]}&=2i\epsilon^{\mu\nu\lambda\rho}q_{\lambda}\gamma_{\rho}\gamma_5,
\end{align}
as well as the following expressions
\begin{align}
\begin{split}
m_q\bar{q}\sigma_{\alpha\beta}iq^{\beta}q&\to \left(F_{T,0}^{q/N}-\frac{q^2}{m_N^2}F_{T,2}^{q/N}\right)\bar{N}i\sigma_{\alpha\beta}q^{\beta}N
\\
&\quad+\frac{q^2}{2m_N}\left(F_{T,1}^{q/N}-4F_{T,2}^{q/N}\right)\bar{N}\gamma_\alpha N,
\end{split}
\\
\epsilon_{\alpha\beta\mu\nu}q^{\beta}m_q\bar{q}\sigma^{\mu\nu}q&\to 2iF_{T,0}^{q/N}\bar{N}\sigma_{\alpha\beta}\gamma_5q^{\beta}N-4F_{T,3}^{q/N}\left(2q_{\alpha}\bar{N}i\gamma_5 N+i\frac{q^2}{m_N}\bar{N}\gamma_{\alpha}\gamma_5N\right).
\end{align}
Making use of the relation \eqref{eq:useful:quark:rel} we can arrive at the following relations between different possible tensor operators
\begin{align}
\left(\bar{e}\gamma^{[\alpha}q^{\beta]}\mu\right)\left(\bar{N}\sigma_{\alpha\beta}\gamma_5N\right)&=\frac{1}{2}\left(\bar{e}\gamma^{[\alpha}\slashed{q}\gamma^{\beta]}\gamma_5\mu\right)\left(\bar{N}\sigma_{\alpha\beta}N\right),
\\
\left(\bar{e}\gamma^{[\alpha}q^{\beta]}\gamma_5\mu\right)\left(\bar{N}\sigma_{\alpha\beta}\gamma_5N\right)&=\frac{1}{2}\left(\bar{e}\gamma^{[\alpha}\slashed{q}\gamma^{\beta]}\mu\right)\left(\bar{N}\sigma_{\alpha\beta}N\right),
\\
\left(\bar{e}\gamma^{[\alpha}\slashed{q}\gamma^{\beta]}\mu\right)\left(\bar{N}\sigma_{\alpha\beta}\gamma_5 N\right)&=2\left(\bar{e}\gamma^{[\alpha}q^{\beta]}\gamma_5\mu\right)\left(\bar{N}\sigma_{\alpha\beta}N\right),
\\
\left(\bar{e}\gamma^{[\alpha}\slashed{q}\gamma^{\beta]}\gamma_5\mu\right)\left(\bar{N}\sigma_{\alpha\beta}\gamma_5 N\right)&=2\left(\bar{e}\gamma^{[\alpha}q^{\beta]}\mu\right)\left(\bar{N}\sigma_{\alpha\beta}N\right),
\label{eq:tensor_parity}
\end{align}
so that we are left with 4 independent operators instead of 8. 

While these 4 operator structures are not independent of the operators $\mathcal{L}^{1-20}_{\rm int}$, we do not attempt to reduce the basis in Tables \ref{tab:LWL} and \ref{tab:tensor_upper} further. For example, by a simple rearrangement, we may write
\begin{equation}
\begin{split}
\mathcal{L}^{29}_{\rm int}&=\frac{i}{2m_L}\left(\bar{e}\gamma^{[\alpha}q^{\beta]}\mu\right)\left(\bar{N}\sigma_{\alpha\beta} N\right)\\
&=\frac{1}{m_L}\left(\bar{e}\gamma^{\alpha}\mu\right)\left(\bar{N}i\sigma_{\alpha\beta}q^{\beta}N\right)\\
&=\frac{m_N}{m_L}\mathcal{L}^6_{\rm int},
\end{split}
\end{equation}
so that $\mathcal{L}^{29}_{\rm int}$ could have always be traded for $\mathcal{L}^6_{\rm int}$. In this regard, the basis furnished by $\mathcal{L}^{1-32}_\mathrm{int}$ is overcomplete. However, as already stated, since it is primarily used to translate between the WET basis and the NRET basis, the overcompleteness is of little practical consequence.
The above equality also highlights the fact that in Tables \ref{tab:LWL}--\ref{tab:tensor_lower}, the relativistic operators in which derivatives act on the leptonic fields contain a spurious mass scale $m_L$, whose only role is to make all the operators of the same mass dimension. After matching to WET the $m_L$ dependence drops out, as demonstrated explicitly in Section \ref{sec:NRET:matching}.

Finally, the $d_j^{(N)}$ coefficients in the covariant single-nucleon interaction Lagrangian in Eq. \eqref{eq:d:Lagr} are explicitly given by
\begin{align}
\begin{split}\label{eq:d1}
d_1^{N}&=\sum_q \frac{1}{m_q} \hat {\cal C}_{5,q}^{(6)} F_S^{q/N}+\hat {\cal C}_{1}^{(7)} F_G^{N}+\hat {\cal C}_{5}^{(7)} F_\gamma^{N},
\end{split}
\\
\begin{split}
d_2^{N}&=-i\frac{m_{-}}{2m_N}\sum_q \hat{\cal C}_{3,q}^{(6)}F_{P'}^{q/N}+\sum_q \frac{1}{m_q} \hat {\cal C}_{7,q}^{(6)} F_P^{q/N}-\hat {\cal C}_{3}^{(7)} F_{\tilde G}^{N}-\hat {\cal C}_{7}^{(7)} F_{\tilde \gamma}^{N}\\
&\quad-i\frac{m_+m_-}{2m_N} \sum_q \hat {\cal C}_{11,q}^{(7)} F_{P'}^{q/N},
\end{split}
\\
\begin{split}
d_3^{N}&=\sum_q \frac{1}{m_q} \hat {\cal C}_{6,q}^{(6)} F_S^{q/N}+\hat {\cal C}_{2}^{(7)} F_G^{N}+\hat {\cal C}_{6}^{(7)} F_\gamma^{N},
\end{split}
\\
\begin{split}
d_4^{N}&=\frac{m_+}{2m_N}\sum_q\hat{\cal C}_{4,q}^{(6)}F_{P'}^{q/N}+\sum_q \frac{1}{m_q} \hat {\cal C}_{8,q}^{(6)} F_P^{q/N}-\hat {\cal C}_{4}^{(7)} F_{\tilde G}^{N}-\hat {\cal C}_{8}^{(7)} F_{\tilde \gamma}^{N}
\\
&-i \frac{m_+m_-}{2m_N}\sum_q \hat {\cal C}_{12,q}^{(7)} F_{P'}^{q/N},
\end{split}
\\
\begin{split}
d_5^{N}&=\sum_q \hat {\cal C}_{1,q}^{(6)} F_1^{q/N} +
m_+\sum_q \hat {\cal C}_{9,q}^{(7)} F_1^{q/N} - \sum_q\frac{q_{\rm rel.}^2}{2m_N} \hat {\cal C}_{13,q}^{(7)} \Big(\hat F_{T,1}^{q/N}-4 \hat F_{T,2}^{q/N}\Big),
\end{split}
\\
d_6^{N}&= - \frac{1}{2} \sum_q \hat {\cal C}_{1,q}^{(6)} F_2^{q/N}-\frac{1}{2}m_+ \sum_q \hat {\cal C}_{9,q}^{(7)} F_2^{q/N},
\\
d_7^{N}&= \sum_q \hat {\cal C}_{3,q}^{(6)} F_A^{q/N}+m_+\sum_q \hat {\cal C}_{11,q}^{(7)} F_A^{q/N},
\\
d_8^{N}&=0,
\\
d_9^{N}&= -\frac{\alpha}{\pi}\hat {\cal C}_1^{(5)} \frac{m_L}{q_{\rm rel.}^2} \sum_q Q_q F_1^{q/N} - m_L \sum_q \hat {\cal C}_{9,q}^{(7)} F_1^{q/N},
\\
d_{10}^{N}&= \frac{\alpha}{2 \pi}\hat {\cal C}_1^{(5)}\frac{m_L}{q_{\rm rel.}^2} \sum_q Q_q F_2^{q/N}+\frac{m_L}{2} \sum_q \hat {\cal C}_{9,q}^{(7)} F_2^{q/N},
\\
d_{11}^N&= -m_L \sum_q \hat {\cal C}_{11,q}^{(7)} F_{A}^{q/N},
\\
d_{12}^N&=0,
\\
\begin{split}
d_{13}^{N}&= \sum_q \hat {\cal C}_{2,q}^{(6)} F_1^{q/N} -
im_- \sum_q \hat {\cal C}_{10,q}^{(7)} F_1^{q/N}-\sum_q \frac{q_{\rm rel.}^2}{2m_N} \hat {\cal C}_{14,q}^{(7)} \Big(\hat F_{T,1}^{q/N}-4 \hat F_{T,2}^{q/N}\Big),
\end{split}
\\
d_{14}^{N}&=-\frac{1}{2} \sum_q \hat {\cal C}_{2,q}^{(6)} F_2^{q/N}
+\frac{i}{2}m_- \sum_q \hat {\cal C}_{10,q}^{(7)} F_2^{q/N},
\\
\begin{split}
d_{15}^{N}&=\sum_q \hat {\cal C}_{4,q}^{(6)} F_A^{q/N} -i m_-\sum_q \hat {\cal C}_{12,q}^{(7)} F_A^{q/N},
\end{split}
\\
d_{16}^{N}&=0,
\\
d_{17}^{N}&=  \frac{\alpha}{\pi}\hat {\cal C}_2^{(5)}\frac{m_L}{q_{\rm rel.}^2}\sum_q Q_q F_1^{q/N}+m_L \sum_q \hat {\cal C}_{10,q}^{(7)} F_1^{q/N},
\\
d_{18}^{N}&= -\frac{\alpha}{2 \pi}\hat {\cal C}_2^{(5)}\frac{m_L}{q_{\rm rel.}^2}\sum_q Q_q F_2^{q/N}- \frac{1}{2}m_L \sum_q \hat {\cal C}_{10,q}^{(7)} F_2^{q/N},
\\
d_{19}^{N}&=  m_L \sum_q \hat {\cal C}_{12,q}^{(7)} F_A^{q/N},
\\
d_{20}^{N}&= 0,
\\
d_{21}^{N}&= \sum_q  \hat {\cal C}_{9,q}^{(6)} \hat F_{T,0}^{q/N},
\\
d_{22}^{N}&=- \sum_q  \hat {\cal C}_{9,q}^{(6)} \hat F_{T,1}^{q/N},
\\
d_{23}^{N}&=-\sum_q  \hat {\cal C}_{9,q}^{(6)} \hat F_{T,2}^{q/N},
\\
d_{24}^{N}&=0,
\\
d_{25}^{N}&=\sum_q \hat {\cal C}_{10,q}^{(6)} \hat F_{T,0}^{q/N},
\\
d_{26}^{N}&=-\sum_q  \hat {\cal C}_{10,q}^{(6)} \hat F_{T,1}^{q/N},
\\
d_{27}^{N}&=-\sum_q  \hat {\cal C}_{10,q}^{(6)} \hat F_{T,2}^{q/N},
\\
d_{28}^{N}&=0,
\\
d_{29}^{N}&=-m_L\sum_q \hat{\cal C}_{13,q}^{(7)}\Big(\hat F_{T,0}^{q/N}-\frac{q_{\rm rel.}^2}{m_N^2}\hat F_{T,2}^{q/N}\Big),
\\
d_{30}^{N}&=-m_L\sum_q \hat{\cal C}_{14,q}^{(7)}\Big(\hat F_{T,0}^{q/N}-\frac{q_{\rm rel.}^2}{m_N^2}\hat F_{T,2}^{q/N}\Big),
\\
d_{31}^{N}& = 
\frac{m_L}{4}\sum_q \C^{(7)}_{16,q} \hat F^{q/N}_{T,0},
\\ \label{eq:d32}
d_{32}^{N}&=
\frac{m_L}{4}\sum_q  \hat{\cal C}_{15,q}^{(7)}  \hat F_{T,0}^{q/N}.
\end{align}
Note that all the form factors depend on $q_{\rm rel}^2=-q^2$, Eq.~\eqref{eq:qrel}, see a more detailed discussion in Appendix \ref{app:nucleon:ff:values}.
As before, we shortened the notation above by introducing $m_\pm=m_\mu\pm m_e$, cf. Eq. \eqref{eq:m+-}.
Note that $m_-$ vanishes in the limit when electron and muon masses are the same, $m_e \to m_\mu$. 

Six of the single-nucleon coefficients, $d_8^N$, $d_{12}^N$, $d_{16}^N$, $d_{20}^N$, $d^N_{24}$, and $d^N_{28}$ are zero. They
are associated with the second-class currents --- nucleon currents 
that have opposite time-reversal parity from the quark currents that generate them. Although suppressed, they could be generated by CP-violating light new physics, see further discussion in App. \ref{app:AFF}.

\subsection{NRET  decomposition of covariant interactions}
\label{app:nret-decomp}

When performing the nonrelativistic reduction of the covariant interactions in Eq. \eqref{eq:d:Lagr}, we follow closely Ref. \cite{Haxton:2022piv}. In particular, we work to linear orders in $v_N$ and $v_\mu$, while the nonrelativistic reduction is performed on matrix elements $\langle e, N'|{\cal L}_{\rm eff}^{\rm cov} |\mu, N\rangle$. The results in Ref. \cite{Haxton:2022piv} were limited to the first 20 ${\cal L}_{{\rm int}}^{j,N}$, since these involve scalar and vector currents. Working to linear order in $v_N$, but to ${\mathcal O}(v_\mu^0)$,  the nonrelativistic reduction of ${\cal L}_{{\rm int}}^{j,N}$, $j=1,\ldots,20$, gives rise to the NRET operator combinations listed in Table \ref{tab:LWL} (with slight abuse of notation, $N$ now denotes the corresponding Dirac four-component spinors, see caption for details). We observe that these involve only a subset of the 16 operators constituting the complete ${\mathcal O}(v_N)$ NRET operator basis in  Eq.~(\ref{eq:ops}): the operators $\CO_3$, $\CO_{12}$, $\CO_{13}^\prime$, and $\CO_{15}$ do not appear. 
If the contributions of ${\mathcal O}(\vec{v}_\mu)$ are added, thereby including contributions from the muon's lower component, the additional NRET operators in Eq.~(\ref{eq:ops2}) are generated, with the results collected in Table \ref{tab:LWL2}.  Note that extending the nonrelativistic reduction to ${\mathcal O}(v_\mu)$ does not
generate the missing NRET operators.  The net effect of $\vec{v}_\mu$ is to modify [at ${\mathcal O}(\sim$ 5\%)] the nuclear response functions $W_{i}^{\tau\tau'}$, all of which were already present in the expansion to order ${\mathcal O}(\vec{v}_N)$: this is a consequence of the fact that the emitted electron is ultra-relativistic, which
guarantees that the contribution of $\vec{v}_\mu$ to the leptonic current will always be just a correction. The results of the above nonrelativistic reduction at ${\mathcal O}(v_N)$ were 
encoded in \mutoenretvone{} version of the public code that accompanied Ref. \cite{Haxton:2022piv}.

\begin{table*}[!]
 \caption{ \label{tab:LWL}Dirac forms of the CLFV amplitudes $\mathcal{L}_\mathrm{int}^j$ are related to linear combinations of the Pauli forms (the operators  $\CO_i$).
Bjorken and Drell spinor and
gamma matrix conventions are used.  Here $\chi_e=\left( \begin{array}{c} \xi_s \\ \vec{\sigma}_L \cdot \hat{q} ~\xi_s \end{array} \right)$, 
$\chi_\mu=\left( \begin{array}{c} \xi_s \\ 0  \end{array} \right)$, and $N=\left( \begin{array}{c} \xi_s \\ {\vec{\sigma}_N \cdot \vec{v}_N \over 2} \xi_s \end{array} \right)$.  The Dirac forms are expanded to first order in  $\vec{v}_N$ to maintain consistency with their use between Schr\"{o}dinger wave functions. \\
~}
\resizebox{\textwidth}{!}{
\begin{tabular}{|c|c|c|c|}
\hline
& & &  \\[-.2cm]
 $j$ & ${\cal L}^j_\mathrm{int}$ & Pauli operator reduction  &~~$ \displaystyle\sum_i c_i \CO_i$ \\[0.3cm]
\hline
& & &  \\
1&$ \displaystyle{ \bar{\chi}_e \chi_\mu~\bar{N} N}$ & $ 1_L~ 1_N $&$ \CO_1$ \\[0.25cm]
2&$ \displaystyle{ \bar{\chi}_e \chi_\mu ~\bar{N}i  \gamma^5 N}$ &$  1_L~ \left( i \displaystyle{{\vec{q} \over 2 m_N} \cdot \vec{\sigma}_N} \right)$ & $\displaystyle{{q \over 2 m_N}} \CO_{10}$ \\[0.25cm]
3&$  \displaystyle{ \bar{\chi}_e i \gamma^5 \chi _\mu~\bar{N} N}$ & $ \displaystyle{ \left( -i \hat{q} \cdot \vec{\sigma}_L \right)~1_N} $ & $-\displaystyle{\CO_{11}}$ \\[0.25cm]
4&$  \displaystyle{\bar{\chi}_e  i \gamma^5 \chi_\mu \bar{N}  i \gamma^5 N} $&$ \displaystyle{ \left( - i \hat{q} \cdot \vec{\sigma}_L  \right) ~\left( i {\vec{q} \over 2 m_N} \cdot \vec{\sigma}_N \right)}$  &$\displaystyle{-{q \over 2 m_N} \CO_6}$  \\[0.25cm]
5&$  \displaystyle{\bar{\chi}_e \gamma^\mu \chi_\mu  \bar{N} \gamma_\mu N }$& $ \displaystyle{1_L 1 _N }$&$\displaystyle{ \CO_1 }$  \\[0.0cm]
~& & $\displaystyle{- \left( \hat{q} 1_L-i \hat{q} \times \vec{\sigma}_L  \right) \cdot \left( \vec{v}_N +i {\vec{q} \over 2m_N} \times \vec{\sigma}_N \right)}$ &$\displaystyle{+i \CO_2^\prime-\CO_5-{q \over 2m_N} (\CO_4+\CO_6) }$   \\[0.25cm]
6&$  \displaystyle{\bar{\chi}_e \gamma^\mu \chi _\mu \bar{N} i \sigma_{\mu \alpha} {q^\alpha \over m_N} N}$ &$  \displaystyle{- \left( \hat{q}1_L -i \hat{q} \times \vec{\sigma}_L  \right) \cdot \left( -i{\vec{q} \over m_N} \times \vec{\sigma}_N \right)}    $&$ \displaystyle{{q \over m_N} \left( \CO_4 + \CO_6 \right) } $  \\[0.25cm]
7&$  \displaystyle{\bar{\chi}_e \gamma^\mu \chi_\mu \bar{N} \gamma_\mu \gamma^5 N}$ & $ \displaystyle{ 1_L \left( \vec{v}_N \cdot \vec{\sigma}_N \right)  - \left( \hat{q}1_L -i \hat{q} \times \vec{\sigma}_L \right) \cdot \vec{\sigma}_N }$ & $ \CO_7 +i \CO_{10}- \CO_9 $   \\[0.25cm]
8& $  \displaystyle{ \bar{\chi}_e \gamma^\mu \chi_\mu  \bar{N} \sigma_{\mu \alpha} \frac{q^\alpha}{m_N}\gamma^5 N}$ & $ \displaystyle{ 1_L \left(-i {\vec{q} \over m_N} \cdot \vec{\sigma}_N \right) }$ & $ \displaystyle{ -{q \over m_N}\CO_{10}} $  \\[0.25cm]
9& $  \displaystyle{\bar{\chi}_e i \sigma^{\mu\nu} \displaystyle{q_\nu \over m_L} \chi_\mu  \bar{N} \gamma_\mu N} $ &$\displaystyle{  -{q \over m_L} 1_L~1_N }$&$ \displaystyle{ -{q \over m_L}  \CO_1} $ \\[0.0cm]
~& & $-\displaystyle{ \left( -i {\vec{q} \over m_L}\times \vec{\sigma}_L \right)  \cdot \left(\vec{v}_N+i {\vec{q} \over 2m_N} \times \vec{\sigma}_N \right)}$ &$\displaystyle{-{q \over m_L}  \left( \CO_5 +{q \over 2m_N} \left( \CO_4+\CO_6 \right) \right)}$  \\[0.25cm]
10& $ \displaystyle{\bar{\chi}_e i \sigma^{\mu\nu} \displaystyle{q_\nu \over m_L} \chi_\mu \bar{N} i \sigma_{\mu\alpha} \displaystyle{q^\alpha \over m_N} N}$ & $\displaystyle{-\left(  -i{\vec{q} \over m_L} \times \vec{\sigma}_L  \right)\cdot  \left(- i {\vec{q} \over m_N} \times \vec{\sigma}_N \right)  } $ & $\displaystyle{ {q \over m_L} {q \over m_N} \left( \CO_4+\CO_6 \right) } $  \\[0.25cm]
11& $ \displaystyle{\bar{\chi}_e i \sigma^{\mu\nu} \displaystyle{q_\nu \over m_L} \chi_\mu \bar{N} \gamma_\mu \gamma^5 N} $& $ \displaystyle{ \left(-{q \over m_L} 1_L \right) \vec{v}_N \cdot \vec{\sigma}_N  -\left(-i {\vec{q} \over m_L} \times \vec{\sigma}
_L \right) \cdot \vec{\sigma}_N}$ & $ \displaystyle{-{q \over m_L} \left( \CO_7+\CO_9 \right)  }$ \\[0.25cm]
12 & $  \displaystyle{\bar{\chi}_e i \sigma^{\mu\nu} \displaystyle{q_\nu \over m_L} \chi_\mu  \bar{N} \sigma_{\mu \alpha} \frac{q^\alpha}{m_N} \gamma^5 N}$ & $\displaystyle{\left( -{q \over m_L} 1_L \right) \left( -i {\vec{q} \over m_N} \cdot \vec{\sigma}_N \right)}$  &$ \displaystyle{ {q \over m_L} {q \over m_N} \CO_{10} } $ \\[0.25cm]
13 & $ \displaystyle{\bar{\chi}_e \gamma^\mu \gamma^5 \chi _\mu \bar{N}\gamma_\mu N }$&$\displaystyle{ \left(\hat{q} \cdot \vec{\sigma}_L \right) 1_N -\vec{\sigma}_L \cdot \left(\vec{v}_N + i {\vec{q} \over 2m_N} \times \vec{\sigma}_N \right)}$ &$ \displaystyle{-i \CO_{11} -\CO_8 -{q \over 2m_N} \CO_9 }$ \\[0.3cm]
14 &$  \displaystyle{\bar{\chi}_e \gamma^\mu \gamma^5 \chi_\mu \bar{N} i \sigma_{\mu\alpha} \displaystyle{q^\alpha \over m_N} N }$&$ \displaystyle{ -\vec{\sigma}_L  \cdot \left( -i {\vec{q} \over m_N} \times \vec{\sigma}_N \right)} $ & $ \displaystyle{q \over m_N}\CO_9$  \\[0.25cm]
15 & $ \displaystyle{\bar{\chi}_e \gamma^\mu \gamma^5 \chi_\mu \bar{N} \gamma_\mu \gamma^5 N}$ &$\displaystyle{\left( \hat{q} \cdot \vec{\sigma}_L \right) \left(\vec{v}_N  \cdot \vec{\sigma}_N \right) }-\vec{\sigma}_L \cdot \vec{\sigma}_N $ & $-i \CO_{14}- \CO_4$   \\[0.25cm]
16 & $ \displaystyle{\bar{\chi}_e \gamma^\mu \gamma^5 \chi_\mu  \bar{N} \sigma_{\mu \alpha} \frac{q^\alpha}{m_N}\gamma^5 N}$ & $ \displaystyle{\left(  \hat{q} \cdot \vec{\sigma}_L \right)  \left( -i {\vec{q} \over m_N} \cdot \vec{\sigma}_N \right)}$  &$ \displaystyle{i {q \over m_N} \CO_6}$  \\[0.25cm]
17 & $ \displaystyle{  \bar{\chi}_e \sigma^{\mu \nu} \frac{q_\nu}{m_L} \gamma^5 \chi_\mu \bar{N}\gamma_\mu  N} $& $ \displaystyle{\left( -i {\vec{q} \over m_L} \cdot \vec{\sigma}_L \right) 1_N }$ & $ \displaystyle{-{q \over m_L}  \CO_{11}}$   \\[0.0cm]
~& & $ \displaystyle{-  i{q \over m_L} \left(\vec{\sigma}_L -\hat{q} \hat{q} \cdot \vec{\sigma}_L  \right) \cdot \left( \vec{v}_N +i {\vec{q} \over 2m_N} \times \vec{\sigma}_N \right) }$ &  $ \displaystyle{-{q \over m_L} \left( i \CO_8+i {q \over 2m_N}\CO_9 +i\CO_{16}^\prime \right)}$   \\[0.25cm]
18 & $ \displaystyle{ \bar{\chi}_e  \sigma^{\mu \nu} \frac{q_\nu}{m_L} \gamma^5 \chi_\mu \bar{N} i \sigma_{\mu \alpha} \displaystyle{{q^\alpha \over m_N}} N} $ & $ \displaystyle{  - i{q \over m_L}  \left(\vec{\sigma}_L - \hat{q} \hat{q} \cdot \vec{\sigma}_L  \right) \cdot \left(-i {\vec{q} \over m_N} \times \vec{\sigma}_N \right) }$& $ \displaystyle{i {q \over m_L} {q \over m_N} \CO_9} $   \\[0.25cm]
19& $  \displaystyle{ \bar{\chi}_e  \sigma^{\mu \nu} \frac{q_\nu}{m_L} \gamma^5 \chi_\mu \bar{N} \gamma_\mu \gamma^5 N} $& $ \displaystyle{\left( -i {\vec{q} \over m_L} \cdot \vec{\sigma}_L \right) \left(\vec{v}_N \cdot \vec{\sigma}_N \right)}$ & $ \displaystyle{-{q \over m_L}  \CO_{14} }$  \\[0.0cm]
&  & $ \displaystyle{-  i{q \over m_L} \left(\vec{\sigma}_L - \hat{q} \hat{q} \cdot \vec{\sigma}_L  \right) \cdot \vec{\sigma}_N }$ & $ \displaystyle{-{q \over m_L} \left( i \CO_4 +i \CO_6 \right)}$  \\[0.25cm]
20 & $  \displaystyle{ \bar{\chi}_e  \sigma^{\mu \nu} \frac{q_\nu}{m_L} \gamma^5 \chi _\mu \bar{N} \sigma_{\mu \alpha} \frac{q^\alpha}{m_N} \gamma^5 N}$   & $ \displaystyle{\left( -i {\vec{q} \over m_L} \cdot \vec{\sigma}_L \right) \left( -i {\vec{q} \over m_N} \cdot \vec{\sigma}_N \right)}$ &$ \displaystyle{{q \over m_L} {q \over m_N} \CO_6}$  \\
 & &  & \\
 \hline
\end{tabular}}
\end{table*}

Motivated by the dimension $d\leq 7$ light-quark interactions described in Sec. \ref{sec:LEFT}, we add to the covariant interaction in Eq. \eqref{eq:d:Lagr} an additional 12 operators that take the form of products of tensor currents. The additional ${\cal L}_{{\rm int}}^{j,N}$ interactions are listed in the first column in Table \ref{tab:tensor_upper}, which also gives the results of the nonrelativistic reduction, working at ${\mathcal O}(v_N)$. The four NRET operators previously missing now appear. When $\vec{v}_\mu$ corrections are included, this generates the additional NRET contributions listed in Table \ref{tab:tensor_lower}.   The extended nucleon-level Dirac basis, consisting of 20+12=32 operators, is employed in the updated script, \mutoenretvtwo.

\begin{table*}[!]
\caption{Nonrelativistic reduction of tensor currents and their correspondence to the upper-component operators $\mathcal{O}_i$.}
\label{tab:tensor_upper}
\resizebox{\textwidth}{!}{
\begin{tabular}{|c|c|c|c|}
\hline
& & &  \\
 $j$ & $\mathcal{L}^j_\mathrm{int}$ & Pauli Operator Reduction  &~~$ \sum_i c_i \mathcal{O}_i$   \\[0.4cm]
\hline
& & &  \\
21 & $\bar{\chi}_e\sigma^{\mu\nu}\chi_\mu \bar{N}\sigma_{\mu\nu}N$ & $-\frac{q}{m_N}1_L1_N -2i\hat{q}\cdot\left(\vec{v}_N\times\vec{\sigma}_N\right)+2\vec{\sigma}_L\cdot\vec{\sigma}_N + 2\vec{\sigma}_L\cdot\left[\hat{q}\times\left(\vec{v}_N\times\vec{\sigma}_N\right)\right]$ &  $-\frac{q}{m_N}\mathcal{O}_1-2 \mathcal{O}_3+2\mathcal{O}_4-2i\mathcal{O}_{13}'$ \\[0.3cm]
22 & $\bar{\chi}_e\sigma^{\mu\nu}\chi_\mu \frac{i}{2m_N}\bar{N}\gamma_{[\mu}q_{\nu]}N$ & $-\frac{q}{m_N}1_L1_N$ &  $-\frac{q}{m_N}\mathcal{O}_1$ \\[0.3cm]
23 & $\bar{\chi}_e\sigma^{\mu\nu}\chi_\mu \frac{i}{m_N^2}\bar{N}q_{[\mu}k_{12,\nu]}N$ & $4\frac{q}{m_N}1_L1_N$ & $4\frac{q}{m_N}\mathcal{O}_1$ \\[0.3cm]
24 & $\bar{\chi}_e\sigma^{\mu\nu}\chi_\mu \frac{1}{m_N}\bar{N}\gamma_{[\mu}\slashed{q}\gamma_{\nu]}N$ & $-4i\frac{q}{m_N}\left(\hat{q}\times\vec{\sigma}_L\right)\cdot\left(\hat{q}\times\vec{\sigma}_N\right)$ & $-4i\frac{q}{m_N}\left(\mathcal{O}_4+\mathcal{O}_6\right)$ \\[0.3cm]
25 & $\bar{\chi}_ei\sigma^{\mu\nu}\gamma_5\chi_\mu \bar{N}\sigma_{\mu\nu}N$ & $ 2\vec{\sigma}_L\cdot\left(\hat{q}\times\vec{\sigma}_N\right)-2i\hat{q}\cdot\vec{\sigma}_N+\frac{q}{m_N}i\hat{q}\cdot\vec{\sigma}_L-2\vec{\sigma}_L\cdot\left(\vec{v}_N\times\vec{\sigma}_N\right)  $ &  $-2i\mathcal{O}_{9}-2 \mathcal{O}_{10}+\frac{q}{m_N}\mathcal{O}_{11}-2\mathcal{O}_{12}$ \\[0.3cm]
26 & $\bar{\chi}_ei\sigma^{\mu\nu}\gamma_5\chi_\mu \frac{i}{2m_N}\bar{N}\gamma_{[\mu}q_{\nu]}N$ & $\frac{q}{m_N}i\hat{q}\cdot\vec{\sigma}_L1_N$ & $\frac{q}{m_N}\mathcal{O}_{11}$ \\[0.3cm]
27 & $\bar{\chi}_ei\sigma^{\mu\nu}\gamma_5\chi_\mu \frac{i}{m_N^2}\bar{N}q_{[\mu}k_{12,\nu]}N$ & $-4\frac{q}{m_N}i\hat{q}\cdot\vec{\sigma}_L1_N$  & $-4\frac{q}{m_N}\mathcal{O}_{11}$ \\[0.3cm]
28 & $\bar{\chi}_ei\sigma^{\mu\nu}\gamma_5\chi_\mu \frac{1}{m_N}\bar{N}\gamma_{[\mu}\slashed{q}\gamma_{\nu]}N$ & $-4\frac{q}{m_N}\vec{\sigma}_L\cdot\left(i\hat{q}\times\vec{\sigma}_N\right)$  & $-4\frac{q}{m_N}\mathcal{O}_9$  \\[0.3cm]
29 & $\frac{i}{2m_L}\bar{\chi}_e\gamma^{[\mu}q^{\nu]}\chi_\mu \bar{N}\sigma_{\mu\nu}N$ & $\frac{q}{m_L} \left(\frac{q}{2m_N}1_L1_N+i\hat{q}\cdot\left(\vec{v}_N\times\vec{\sigma}_N\right)+\left(\hat{q}\times\vec{\sigma}_L\right)\cdot\left(\hat{q}\times\vec{\sigma}_N\right)  \right)$ & $\frac{q}{m_L}\left(\frac{q}{2m_N}\mathcal{O}_1+\mathcal{O}_3+\mathcal{O}_4+\mathcal{O}_6\right)$\\[0.3cm]
30 & $\frac{i}{2m_L}\bar{\chi}_e\gamma^{[\mu}q^{\nu]}\gamma_5\chi_{\mu}\bar{N}\sigma_{\mu\nu}N$ & $\frac{q}{m_L} \left( \vec{\sigma}_L\cdot\left(i\hat{q}\times\vec{\sigma}_N\right)+\frac{q}{2m_N}\hat{q}\cdot\vec{\sigma}_L 1_N + \hat{q}\cdot\vec{\sigma}_Li\hat{q}\cdot\left(\vec{v}_N\times\vec{\sigma}_N\right)\right)$ & $\frac{q}{m_L} \left(\mathcal{O}_9-i\frac{q}{2m_N}\mathcal{O}_{11}-i\mathcal{O}_{15} \right)$ \\[0.3cm]
31 & $\frac{1}{m_L}\bar{\chi}_e \gamma^{[\mu}\slashed{q}\gamma^{\nu]}\chi_\mu \bar{N}\sigma_{\mu\nu}N$ & $4\frac{q}{m_L}\Big(-i\left(\hat{q}\cdot\vec{\sigma}_L\right)\left(\hat{q}\cdot\vec{\sigma}_N\right)+\vec{\sigma}_L\cdot\left[i\hat{q}\times\left(\vec{v}_N\times\vec{\sigma}_N\right)\right]\Big)$ & $4\frac{q}{m_L}\left(i\mathcal{O}_6+\mathcal{O}_{13}'\right)$ \\[0.3cm]
32 & $\frac{1}{m_L}\bar{\chi}_e\gamma^{[\mu}\slashed{q}\gamma^{\nu]}\gamma_5\chi_{\mu}\bar{N}\sigma_{\mu\nu}N$ & $4\frac{q}{m_L}\Big(-i\hat{q}\cdot\vec{\sigma}_N+\vec{\sigma}_L\cdot\left(\vec{v}_N\times\vec{\sigma}_N\right)-\hat{q}\cdot\vec{\sigma}_L\hat{q}\cdot\left(\vec{v}_N\times\vec{\sigma}_N\right) \Big)$  & $4\frac{q}{m_L}\left(-\mathcal{O}_{10}+\mathcal{O}_{12}+\mathcal{O}_{15}\right)$ \\[0.3cm]
 \hline
\end{tabular}}
\end{table*}

\begin{table*}[!]
\caption{ \label{tab:LWL2} As in Table \ref{tab:LWL}, but listing the additional terms generated for scalar- and vector-mediated interactions when the linear expansion in velocities includes 
$\vec{v}_\mu$, so that
$\chi_\mu=\left( \begin{array}{c} \xi_s \\ {\vec{\sigma}_L \cdot \vec{v}_\mu \over 2} \xi_s  \end{array} \right)$.  \\
~}
\resizebox{\textwidth}{!}{
\begin{tabular}{|c|c|c|c|}
\hline
& & &  \\[-.2cm]
 $j$ & $\mathcal{L}^j_\mathrm{int}$ & Pauli operator reduction  &~~$ \displaystyle\sum_i b_i \CO^f_i$   \\[0.3cm]
\hline
& & &  \\
1&$ \displaystyle{ \bar{\chi}_e \chi_\mu~\bar{N} N}$ & $ -\displaystyle{1 \over 2} \hat{q} \cdot \vec{v}_\mu~ 1_N -\displaystyle{i \over 2} \hat{q} \cdot [\vec{v}_\mu \times \vec{\sigma}_L] ~ 1_N$&$ \displaystyle{i } \CO_2^{f \, \prime} - \CO_3^f $  \\[0.25cm]
3&$  \displaystyle{ \bar{\chi}_e i \gamma^5 \chi _\mu~\bar{N} N}$ & $ \displaystyle{ i \over 2} \vec{v}_\mu  \cdot \vec{\sigma}_L  1_N$ & $ \displaystyle{i} \CO_7^f$ \\[0.25cm]
5&$  \displaystyle{\bar{\chi}_e \gamma^\mu \chi_\mu  \bar{N} \gamma_\mu N }$& $\displaystyle{1 \over 2} \hat{q} \cdot \vec{v}_\mu ~1_N+\displaystyle{i \over 2} \hat{q} \cdot [\vec{v}_\mu \times \vec{\sigma}_L] ~1_N$&  $ -\displaystyle{i} \CO_2^{f \, \prime} +\CO_3^f $    \\[0.25cm]
7&$  \displaystyle{\bar{\chi}_e \gamma^\mu \chi_\mu \bar{N} \gamma_\mu \gamma^5 N}$ & $-\displaystyle{1 \over 2} \vec{v}_\mu \cdot \vec{\sigma}_N -\displaystyle{i \over 2} [ \vec{v}_\mu \times \vec{\sigma}_L] \cdot \vec{\sigma}_N$ & $-  \CO_8^f - \displaystyle{i }  \CO^f_{12} $   \\[0.3cm]
9& $  \displaystyle{\bar{\chi}_e i \sigma^{\mu\nu} \displaystyle{q_\nu \over m_L} \chi_\mu  \bar{N} \gamma_\mu N} $ & $\displaystyle{q \over 2 m_L} \left(  \hat{q} \cdot \vec{v}_\mu ~1_N +i \hat{q} \cdot [\vec{v}_\mu \times \vec{\sigma}_L] ~1_N \right)$&$ \displaystyle{ q \over m_L}  \left( -i  \CO_2^{f \, \prime} + \CO_3^f \right)  $\\[0.25cm]
11& $ \displaystyle{\bar{\chi}_e i \sigma^{\mu\nu} \displaystyle{q_\nu \over m_L} \chi_\mu \bar{N} \gamma_\mu \gamma^5 N} $& $ \displaystyle{q \over 2 m_L} \left( \vec{v}_\mu \cdot \vec{\sigma}_N +i [\vec{v}_\mu \times \vec{\sigma}_L] \cdot \vec{\sigma}_N \right. $ & $ \displaystyle{{q \over  m_L} \left( \CO_8^f+i \CO^f_{12} \right.~~~~  }$ \\[0.0cm]
 & & $~\left.  -i \hat{q} \cdot [\vec{v}_\mu \times \vec{\sigma}_L ] \hat{q} \cdot \vec{\sigma}_N-\hat{q} \cdot \vec{v}_\mu \hat{q} \cdot \vec{\sigma}_N  \right)$  & $~~~~~~\left.  +i  \CO_{15}^f + \CO_{16}^{f \, \prime}\right) $  \\[0.25cm]
13 & $ \displaystyle{\bar{\chi}_e \gamma^\mu \gamma^5 \chi _\mu \bar{N}\gamma_\mu N }$&$ \displaystyle{1 \over 2} \vec{v}_\mu \cdot \vec{\sigma}_L ~ 1_N$ & $ \CO_7^f$  \\[0.25cm]
15 & $ \displaystyle{\bar{\chi}_e \gamma^\mu \gamma^5 \chi_\mu \bar{N} \gamma_\mu \gamma^5 N}$ & $\displaystyle{i \over 2} [ \hat{q} \times \vec{v}_\mu] \cdot \vec{\sigma}_N-\displaystyle{1 \over 2} ( \hat{q} \times [\vec{v}_\mu \times \vec{\sigma}_L]) \cdot \vec{\sigma}_N  $ & $  \CO^f_{5}+\displaystyle{i }  \CO_{13}^{f \, \prime}$   \\[0.0cm]
 & & $-\displaystyle{1 \over 2} \vec{v}_\mu \cdot \vec{\sigma}_L ~\hat{q} \cdot \vec{\sigma}_N $& $+ \displaystyle{i } \CO_{14}^f $   \\[0.25cm]
17 & $ \displaystyle{  \bar{\chi}_e \sigma^{\mu \nu} \frac{q_\nu}{m_L} \gamma^5 \chi_\mu \bar{N}\gamma_\mu  N} $ & $ \displaystyle{ i q \over 2 m_L} \vec{v}_\mu \cdot \vec{\sigma}_L ~1_N $ & $ \displaystyle{i q \over  m_L}  \CO^f_7 $   \\[0.25cm]
19& $  \displaystyle{ \bar{\chi}_e  \sigma^{\mu \nu} \frac{q_\nu}{m_L} \gamma^5 \chi_\mu \bar{N} \gamma_\mu \gamma^5 N} $& $ \displaystyle{ q \over 2 m_L} ( [\hat{q} \times \vec{v}_\mu] \cdot \vec{\sigma}_N+(i \hat{q} \times [\vec{v}_\mu \times \vec{\sigma}_L]) \cdot \vec{\sigma}_N) $ & $ \displaystyle{q \over  m_L}  (-i \CO_5^f +\CO_{13}^{f \, \prime}) $  \\
 & &  &  \\
 \hline
\end{tabular}}
\end{table*} 

\mutoenretvtwo{} thus includes 32 Dirac interactions associated with scalar, vector, or tensor exchanges.  The NRET reduction used in the new script also includes the lower-component contributions of $\vec{v}_\mu$, extending the NRET basis used in \mutoenretvone{} by an additional 10 operators, for a total of 26.
When the nuclear physics multipole expansion is performed for the additional operators, new nuclear multipoles arise.  These appear in Eq. (B7) of \cite{Haxton:2022piv}.  Matrix elements of the new multipole operators can still be evaluated analytically, if the Slater determinants used in the shell model are constructed in a harmonic oscillator single-particle basis.  But unlike the original 16 NRET operators, the results are no longer expressible in the form of $e^{-2y} p(y)$ where $p(y)$ is a finite polynomial in $y$.

\begin{table*}[!]
\caption{As in Table \ref{tab:tensor_upper}, but listing the additional terms generated for tensor-mediated interactions when the linear expansion in velocities includes 
$\vec{v}_\mu$.}
\label{tab:tensor_lower}
\resizebox{\textwidth}{!}{
\begin{tabular}{|c|c|c|c|}
\hline
& & &  \\
 $j$ & $\mathcal{L}^j_\mathrm{int}$ & Pauli Operator Reduction  &~~$ \sum_i b_i \mathcal{O}^f_i$   \\[0.4cm]
\hline
& & &  \\
21 & $\bar{\chi}_e\sigma^{\mu\nu}\chi_\mu \bar{N}\sigma_{\mu\nu}N$ & $\left(i\hat{q}\times\vec{v}_{\mu}\right)\cdot\vec{\sigma}_N-\left[\hat{q}\times\left(\vec{v}_{\mu}\times\vec{\sigma}_L\right)\right]\cdot\vec{\sigma}_N-\left(\vec{v}_{\mu}\cdot\vec{\sigma}_L\right)\left(\hat{q}\cdot\vec{\sigma}_N\right)$  & $2\mathcal{O}_5^f+2i\mathcal{O}_{13}^{f\prime}+2i\mathcal{O}_{14}^{f}$ \\[0.3cm]
25 & $\bar{\chi}_ei\sigma^{\mu\nu}\gamma_5\chi_\mu \bar{N}\sigma_{\mu\nu}N$ & $i\vec{v}_{\mu}\cdot\vec{\sigma}_N-\left(\vec{v}_{\mu}\times\vec{\sigma}_L\right)\cdot\vec{\sigma}_N$  & $2i\mathcal{O}_{8}^{f}-2\mathcal{O}_{12}^f$ \\[0.3cm]
29 & $\frac{i}{2m_L}\bar{\chi}_e\gamma^{[\mu}q^{\nu]}\chi_\mu \bar{N}\sigma_{\mu\nu}N$ & $-\frac{q}{m_L}\Big(\left(i\hat{q}\times \frac{\vec{v}_{\mu}}{2}\right)\cdot\vec{\sigma}_N-\left[\hat{q}\times\left(\frac{\vec{v}_{\mu}}{2}\times\vec{\sigma}_L\right)\right]\cdot\vec{\sigma}_N \Big)$ & $-\frac{q}{m_L}\left(\mathcal{O}_5^f+i\mathcal{O}_{13}^{f\prime}\right)$\\[0.3cm]
30 & $\frac{i}{2m_L}\bar{\chi}_e\gamma^{[\mu}q^{\nu]}\gamma_5\chi_\mu \bar{N}\sigma_{\mu\nu}N$ & $\frac{q}{m_L} \Big( \frac{\vec{v}_{\mu}}{2} \cdot \vec{\sigma}_N+i \left(\frac{\vec{v}_\mu}{2} \times \vec{\sigma}_L \right) \cdot \vec{\sigma}_N-i \hat{q}\cdot\left(\frac{\vec{v}_{\mu}}{2}\times\vec{\sigma}_L\right)\hat{q}\cdot\vec{\sigma}_N-\hat{q}\cdot \frac{\vec{v}_{\mu}}{2}\hat{q}\cdot\vec{\sigma}_N \Big)$ & $\frac{q}{m_L}\left(\mathcal{O}_8^f+i \mathcal{O}_{12}^f+i\mathcal{O}_{15}^f+\mathcal{O}_{16}^{f\prime}\right)$\\[0.3cm]
31 & $\frac{1}{m_L}\bar{\chi}_e \gamma^{[\mu}\slashed{q}\gamma^{\nu]}\chi_\mu \bar{N}\sigma_{\mu\nu}N$ & $-2i\frac{q}{m_L}\vec{v}_{\mu}\cdot\vec{\sigma}_L\hat{q}\cdot\vec{\sigma}_N$ & $-4\frac{q}{m_L}\mathcal{O}_{14}^f$ \\[0.3cm]
32 & $\frac{1}{m_L}\bar{\chi}_e \gamma^{[\mu}\slashed{q}\gamma^{\nu]}\gamma_5\chi_\mu \bar{N}\sigma_{\mu\nu}N$ & $2\frac{q}{m_L} \Big(\hat{q}\cdot\left(\vec{v}_{\mu}\times\vec{\sigma}_L\right)\hat{q}\cdot\vec{\sigma}_N -i\hat{q}\cdot\vec{v}_{\mu}\hat{q}\cdot\vec{\sigma}_N\Big)$ & $4\frac{q}{m_L}\left(-\mathcal{O}_{15}^f+i\mathcal{O}_{16}^{f \prime}\right)$ \\[0.3cm]
 \hline
\end{tabular}}
\end{table*}

\section{Time-reversal-odd nuclear currents}
\label{app:AFF}
In addition to the terms displayed on the right-hand sides
of Eqs. (\ref{vec:form:factor}), (\ref{axial:form:factor}),
and (\ref{tensor:form:factor}), other terms can be constructed that satisfy the conditions of Lorentz covariance and hermiticity 
\begin{align}
\label{vec:form:factor1}
\langle N'|\bar q \gamma^\mu q|N\rangle &= \cdots ~+~\bar u_N'\Big[iF^{q/N}_3(q_\mathrm{rel}^2) \frac{q^\mu}{m_N}\Big]u_N, \tag{4.1a} \\
\label{axial:form:factor1}
\langle N'|\bar q \gamma^\mu \gamma_5 q|N\rangle &= \cdots~+~\bar u_N'\Big[\gamma_5\frac{\sigma^{\mu\nu}q_\nu}{m_N} F_{A,3}^{q/N}(q_\mathrm{rel}^2)\Big]u_N, \tag{4.2a} \\
\label{tensor:form:factor1}
\langle N'|\bar q \sigma^{\mu\nu} q |N\rangle &= \cdots~+~\bar u_N'\Big[-\frac{1}{m_N}\gamma^{[\mu}\slashed q \gamma^{\nu]}\hat{F}_{T,3}^{q/N}(q_\mathrm{rel}^2) \Big] u_N, \tag{4.7a}
\end{align}
where the ellipses denote the terms already displayed in Eqs. (\ref{vec:form:factor}), (\ref{axial:form:factor}),
and (\ref{tensor:form:factor}). The single-nucleon currents displayed on the RHS of Eqs.~\eqref{vec:form:factor}--\eqref{tensor:form:factor} transform under time reversal in the same way as the corresponding bare quark operators on the LHS, from which the nucleon
currents are generated.  In contrast, the single-nucleon currents on the RHS of Eqs.~\eqref{vec:form:factor1}--\eqref{tensor:form:factor1} have the opposite behavior under $T$. Since QCD is invariant under $T$ (ignoring for now the experimentally well constrained QCD $\theta$ term), the form factors on the RHS of Eqs.~\eqref{vec:form:factor1}--\eqref{tensor:form:factor1} cannot be generated in QCD.

If all new physics is heavy, it can be integrated out and described by the WET Lagrangian in Sec.~\ref{sec:LEFT}. At low energies, we then only have QCD and QED as propagating degrees of freedom, both of which are invariant under $T$. In this case, the form factors $F_3^{q/N}$, $F_{A,3}^{q/N}$, $F_{T,3}^{q/N}$ are zero and do not contribute to $\mu\to e$ conversion. CP-violating light new physics, on the other hand, could give rise to nonzero $T$-odd nuclear currents on the RHS of Eqs.~\eqref{vec:form:factor1}--\eqref{tensor:form:factor1}. Since such light NP couples to nucleons, it is necessarily weakly coupled, otherwise it would have been observed already, and thus will only give rise to highly suppressed contributions: $F^{q/N}_3$, $F_{A,3}^{q/N}$, $F_{T,3}^{q/N} \ll 1$. (The other option is that these are generated from the small but nonzero QCD $\theta$ term or from the CP-violating part of the SM weak interactions, again leading to highly suppressed contributions.)

While we expect such symmetry-odd operators to be highly suppressed, nevertheless we include them here for completeness. The modifications to the $d_j^{(N)}$ coefficients are 
\begin{align}
\begin{split}
d_1^{N}&=\cdots - i\frac{m_-}{m_N}\sum_q\hat{\cal{C}}_{1,q}^{(6)}F_3^{q/N}-i\left(m_\mu^2-m_e^2\right)\sum_q \hat{\cal{C}}_{9,q}^{(7)}F_3^{q/N},
\end{split}\tag{B.10a}
\\
\begin{split}
d_2^{N}&=\cdots +\frac{m_{+}}{m_N}\sum_q \hat{\cal{C}}_{2,q}^{(6)}F_3^{q/N}-4i m_-\sum_q \C^{(7)}_{15,q}\hat F_{T,3}^{q/N}
\end{split}\tag{B.11a}
\\
\begin{split}
d_3^{N}&=\cdots -i\left(m_\mu^2-m_e^2\right)\sum_q \hat{\cal{C}}_{10,q}^{(7)}F_3^{q/N},
\end{split}\tag{B.12a}
\\
\begin{split}
d_4^{N}&=\cdots + 
4 m_+ \sum_q \C^{(7)}_{16,q}\hat F_{T,3}^{q/N},
\end{split}\tag{B.13a}
\\
d_7^{N}&=\cdots +2\frac{q_{\text{rel.}}^2}{m_N } \sum_q
\C^{(7)}_{15,q}\hat F_{T,3}^{q/N},\tag{B.16a}
\\
d_8^{N}&=\cdots + \sum_q\hat{\cal{C}}_{3,q}^{(6)}F_{A,3}^{q/N}+m_+\sum_q\hat{\cal{C}}_{11,q}^{(7)}F_{A,3}^{q/N},\tag{B.17a}
\\
d_{12}^N&=\cdots -m_L\sum_q \hat{\cal{C}}_{11,q}^{(7)}F_{A,3}^{q/N},\tag{B.21a}
\\
d_{15}^{N}&=\cdots +2\frac{q_{\text{rel.}}^2}{m_N}\sum_q \C^{(7)}_{16,q}\hat F_{T,3}^{q/N},\tag{B.24a}
\\
d_{16}^{N}&=\cdots +\sum_q \hat{\cal{C}}^{(6)}_{4,q}F_{A,3}^{q/N}-im_-\sum_q \hat{\cal{C}}_{12,q}^{(7)}F_{A,3}^{q/N},\tag{B.25a}
\\
d_{20}^{N}&=\cdots + m_L\sum_q \hat{\cal{C}}_{12,q}^{(7)}F_{A,3}^{q/N},\tag{B.29a}
\\
d_{24}^{N}&=\cdots -\sum_q  \hat {\cal C}_{9,q}^{(6)} \hat F_{T,3}^{q/N},\tag{B.33a}
\\
d_{28}^{N}&=\cdots -\sum_q  \hat {\cal C}_{10,q}^{(6)} \hat F_{T,3}^{q/N}.\tag{B.37a}
\end{align}
With this extension, all $d_i^N$ coefficients are nonzero. 

\section{Numerical values of nucleon form factors}
\label{app:nucleon:ff:values}

In the numerical results of Sec.~\ref{sec:NewPhysicsModels}, we evaluated the nucleon form factors in Eqs.~\eqref{vec:form:factor}-\eqref{CPodd:photon:form:factor} at $q_{\rm rel.}^2=-q_{\rm eff}^2$, where $q_{\rm eff}=110.81$\,MeV for ${}^{27}$Al. Below, we provide the expressions that were used in the numerical evaluations. The expressions are given in a form where it is straightforward to re-use them for any other target element and the corresponding $q_{\rm eff}$.
To do so, we Taylor expand the nucleon form factors around $q_{\rm rel.}^2=0$,
\begin{equation}
\label{eq:Fi}
F_{i}^{q/N}(q_{\rm rel.}^2)=F_{i}^{q/N}(0)+F_{i}^{\prime\, q/N}(0) q_{\rm rel.}^2+ \cdots.
\end{equation}
The axial, pseudoscalar, and CP-odd gluonic current form factors include light-meson poles, so that for these we have  (for $i=\tilde G$ the superscript $q/N\to N$)
\begin{align}
\label{eq:F_PP'}
F_{i}^{q/N}(q_{\rm rel.}^2)&=\frac{m_N^2}{m_\pi^2-q_{\rm rel.}^2} a_{i,\pi}^{q/N}+\frac{m_N^2}{m_\eta^2-q_{\rm rel.}^2} a_{i,\eta}^{q/N}
+b_{i}^{q/N}+\cdots\,, \quad i=P,P', \tilde G\,.
\end{align}
Throughout this work, we use the $\pi^0$ mass for $m_\pi$. The ellipses denote terms of the form $(q_{\rm rel.}^2/m_N^2)^n$, which we do not include.

Unless specified otherwise, we work in the isospin limit, so that the nucleon form factors for neutrons can be obtained from the ones for protons, e.g.,
\beq
F_{1,2}^{u(d,s)/p}(q_{\rm rel.}^2)= F_{1,2}^{d(u,s)/n}(q_{\rm rel.}^2).
\eeq
For the average nucleon mass, we use $m_N=(m_p+m_n)/2$.
We use the FLAG
quality requirements~\cite{FlavourLatticeAveragingGroupFLAG:2021npn} to decide which lattice QCD results to include in our determinations of the hadronic input parameters. In many respects, our results are an update of the global averaging for the values of nucleon form factors for dark matter direct detection in Refs. \cite{Bishara:2017pfq,Bishara:2017nnn}.

For certain quark currents, two-nucleon terms contribute at the same order in chiral power counting as ``form-factor effects'' arising from the momentum dependence of single-nucleon form factors. Recent calculations of scalar-mediated scattering of dark matter on $^4$He suggest that the na\"ive power-counting scheme may be flawed in two ways:  (1) The calculated two-nucleon contributions are significantly smaller than expected, and (2) they may require new short-distance operators at NLO with currently unknown LECs in order to restore regulator independence \cite{deVries:2023hin}. Given the current uncertainty regarding two-nucleon effects, our approach is to retain the relatively well understood NLO physics associated with the momentum dependence of single-nucleon form factors while leaving a complete evaluation of subleading effects to future investigations.

\subsection{Vector currents}
\label{app:sec:vector:current}
The hadronic matrix elements of quark currents in Eq.~\eqref{vec:form:factor} depend on two sets of form factors. The Dirac form factors for the proton, at zero recoil, are  given by
\begin{equation}\label{eq:F1:num}
F_1^{u/p}(0)=2, \qquad F_1^{d/p}(0)=1, \qquad F_1^{s/p}(0)=0,
\end{equation}
while the derivatives at $q^2=0$ are (see, e.g., Ref.~\cite{Hill:2014yxa}, with $a_p=\mu_p-1$, $a_n=\mu_n$)
\begin{align}
\begin{split}
F_1^{\prime \, u/p}(0)& = \frac{1}{6}\big(2 \big[r_E^{p} \big]^2 + 
  \big[r_E^{n} \big]^2 +\big[r_E^s\big]^2\big)
  \\
  &\quad-\frac{1}{4m_N^2}\big(2
  \mu_p + \mu_n+\mu_s-2)=5.07(1){\rm~GeV}^{-2}\,, 
\end{split}  
  \\
  \begin{split}
F_1^{\prime \, d/p}(0) & = \frac{1}{6}\big(\big[r_E^{p} \big]^2 +
  2 \big[r_E^{n} \big]^2 +\big[r_E^s\big]^2\big)
  \\
 &\quad -\frac{1}{4m_N^2}\big(
  \mu_p + 2\mu_n+\mu_s-1)=2.61(2){\rm~GeV}^{-2}\,, \label{eq:F1pdp0:num}
  \end{split}
  \\
  F_1^{\prime \, s/p}(0) & = \frac{1}{6} \big[r_E^s\big]^2-\frac{\mu_s}{4m_N^2}=-8.9(8.4)\cdot 10^{-3}{\rm~GeV}^{-2}\,.
\end{align}
The Pauli form factors at zero momentum are
\begin{align}
F_2^{u/p}(0)&=2
(\mu_p-1)+\mu_n+\mu_s=1.64(2),
\\
 F_2^{d/p}(0)&=2
\mu_n+(\mu_p-1)+\mu_s=-2.07(2),
\\
\label{eq:mus:value}
F_2^{s/p}(0)&=\mu_s=
-0.036(21),
\end{align}
and the first derivatives are
\begin{align}
\begin{split}
F_2^{\prime \, u/p}(0)& = \frac{1}{6}\Big( 2\big( \big[r_M^{p} \big]^2- \big[r_E^{p} \big]^2\big) + 
  \big[r_M^{n} \big]^2 -\big[r_E^{n} \big]^2 +\big[r_M^{s} \big]^2 -\big[r_E^{s} \big]^2 \Big)
  \\
  &+\frac{1}{4m_N^2}\big(2
  \mu_p + \mu_n+\mu_s-2)=4.3(4){\rm~GeV}^{-2}\,, 
  \end{split}
  \\
  \begin{split}
F_2^{\prime \, d/p}(0) & = \frac{1}{6}\Big(  \big[r_M^{p} \big]^2- \big[r_E^{p} \big]^2 + 2\big(
  \big[r_M^{n} \big]^2 -\big[r_E^{n} \big]^2\big) +\big[r_M^{s} \big]^2 -\big[r_E^{s} \big]^2 \Big)
  \\
  &+\frac{1}{4m_N^2}\big(
  \mu_p + 2\mu_n+\mu_s-1)=6.8(2){\rm~GeV}^{-2}\,,
   \label{eq:F2pdp0:num}
  \end{split}
  \\
  F_2^{\prime \, s/p}(0) & = \frac{1}{6} \big([r_M^s]^2-[r_E^s]^2\big)+\frac{\mu_s}{4m_N^2}=-0.03(5){\rm~GeV}^{-2}\,.
\end{align}

In the numerical evaluations we used  $\mu_p=2.792847$, $\mu_n=-1.91304$ for the proton and
neutron magnetic moments in units of nuclear magnetons $\hat\mu_N
 =e/(2m_N)$ (the errors are negligibly small)~\cite{ParticleDataGroup:2022pth}. The value for $\mu^s$ in \eqref{eq:mus:value} is the average of results from Refs.~\cite{Djukanovic:2019jtp, Sufian:2016pex}, inflating the
errors according to the PDG prescription (as before, we do not include~\cite{Alexandrou:2019olr} in the average).
For the values of electric and magnetic charge radii, we use 
$\big[ r_E^{p} \big]^2 =0.7071(7)\,$fm$^2$~\cite{ParticleDataGroup:2022pth}, 
$\big[ r_M^{p} \big]^2 =0.724(45)\,$fm$^2$~\cite{ParticleDataGroup:2022pth}, 
$\big[ r_E^{n} \big]^2 = -0.1155(17)\,$fm$^2$~\cite{ParticleDataGroup:2022pth}, 
$\big[ r_M^{n} \big]^2 = 0.743(16)\,$fm$^2$~\cite{ParticleDataGroup:2022pth},
and $\big[r_E^s\big]^2= -0.0045(14)\,$fm$^2$~\cite{Sufian:2016pex,Djukanovic:2019jtp}, 
$\big[r_M^s\big]^2= -0.010(11)\,$fm$^2$~\cite{Djukanovic:2019jtp}.  (We do not use the
$N_f=2+1+1$ results for $\big[r_{E}^s\big]^2, \big[r_{M}^s\big]^2$ from Ref.~\cite{Alexandrou:2019olr} due to lack of continuum extrapolation.)

For $\mu\to e$ conversion on ${}^{27}$Al the values of the Dirac and Pauli form factors are thus,
\begin{align}\label{eq:F1:Al}
F_1^{u/p}\big|_{-q_{\rm eff}^2}&=1.9378(4), \quad F_1^{d/p}\big|_{-q_{\rm eff}^2}=0.9681(3), \quad F_1^{s/p}\big|_{-q_{\rm eff}^2}=1.1(1.0)\cdot 10^{-4},
\\
F_2^{u/p}\big|_{-q_{\rm eff}^2}&=1.58(2), \quad F_2^{d/p}\big|_{-q_{\rm eff}^2}=-2.15(2), \quad F_2^{s/p}\big|_{-q_{\rm eff}^2}=-3.6(2.2)\cdot 10^{-2}.
\end{align}
The errors on $F_1^{u/p}$ and $F_1^{d/p}$ are dominated by the neglected ${\mathcal O}(q^4)$ contributions, which we estimate to be equal to $F_i(0)\times (q_{\rm eff}/m_N)^4$. 

In the derivation of expressions for the zero recoil values of the $F_{1,2}^{q/N}$ form factors and their derivatives in terms of the conventionally defined observables we used 
\begin{equation}
G_{E}^N(q_{\rm rel.}^2)=G_{E}^N(0)+\frac{1}{6}\big[r_{E}^N]^2 q_{\rm rel.}^2+\cdots\,, \qquad G_{M}^N(q_{\rm rel.}^2)=G_{M}^N(0)+\frac{1}{6}\big[r_{M}^N]^2 q_{\rm rel.}^2+\cdots,
\end{equation}
where the Sachs electric and magnetic form
factors are~\cite{Ernst:1960zza} (see also, e.g.,
\cite{Hill:2010yb})
\begin{equation}
G_E^N(q_{\rm rel.}^2)=F_1^N(q_{\rm rel.}^2)+\frac{q_{\rm rel.}^2}{4 m_N^2} F_2^N(q_{\rm rel.}^2)\,, \quad {\rm and}\quad
G_M^N(q_{\rm rel.}^2)=F_1^N(q_{\rm rel.}^2)+F_2^N(q_{\rm rel.}^2)\,.
\end{equation}

\subsection{Axial vector currents} 
The axial vector form factor at zero recoil is given by
\begin{equation}
\label{eq:Deltaq}
F_A^{q/p}(0)=\Delta q, 
\end{equation}
where numerically, 
\begin{align}
\label{eq:Deltau-d}
\Delta u-\Delta d&=g_A=1.2754(13),
\\
\label{eq:Deltau+d}
\Delta u+\Delta d &\equiv \Delta \Sigma_{ud}=0.397(40),
\\
\label{eq:Deltas}
\Delta s&=-0.045(9).
\end{align}
The isovector combination $\Delta u-\Delta d$ is determined very precisely from nuclear $\beta$ decay~\cite{ParticleDataGroup:2022pth}
 (above we set $g_V=1$ in $\lambda=g_A/g_V$, i.e., we ignored corrections of second order in isospin breaking \cite{Ademollo:1964sr,Donoghue:1990ti}, and used positive sign convention for $g_A$). The value for $\Delta u+\Delta d$ (for $\Delta s$) follows from averages of the lattice QCD results for $\Delta u$ and $\Delta d$ (for $\Delta s$), summing errors in quadrature~\cite{Lin:2018obj,Liang:2018pis}, and rescaling the errors for $\Delta u$ and $\Delta s$ according to the PDG description.

For derivatives of the axial form factor at zero recoil we can write
\beq
F_A^{q/p}(q_{\rm rel.}^2)=F_A^{q/p}(0)\Big(1+\frac{\langle r_A^2\rangle_q}{6} q_{\rm rel.}^2+\cdots\Big).
\eeq
For the $u-d$ current the axial charge radius is well known,  $\langle r_A^2\rangle_{u-d}=0.392(28)\,\text{fm}^2$, averaging over lattice QCD determinations  ~\cite{Jang:2023zts,Bali:2023sdi,Park:2021ypf,Djukanovic:2022wru,Alexandrou:2023qbg} (see also \cite{Alexandrou:2020okk,Tsuji:2023llh}). For $u+d$ and $s$ quark currents we use the results from \cite{Alexandrou:2021wzv} $\langle r_A^2\rangle_{u+d}=0.49(31)\,\text{fm}^2$,  $\langle r_A^2\rangle_{s}=0.48(48)\,\text{fm}^2$, with the caveat that this determination still lacks proper continuum extrapolation.

In terms of the above input quantities the zero-recoil form factors are
\begin{align}
F_A^{u/p}(0)&=\Delta u=\frac{1}{2}\big(g_A+\Delta \Sigma_{ud}\big)=0.836(20), 
\\
F_A^{d/p}(0)&=\Delta d= \frac{1}{2}\big(-g_A+\Delta \Sigma_{ud}\big)=-0.439(20), 
\\
 F_A^{s/p}(0)&=\Delta s=-0.045(9),
\end{align}
with the derivatives given by
\begin{align}
F_A^{u/p\prime}(0)&=\frac{1}{12}\big(g_A \langle r_A^2\rangle_{u-d}+\Delta \Sigma_{ud} \langle r_A^2\rangle_{u+d}\big)=1.49(28)\,\text{GeV}^{-2}, 
\\
F_A^{d/p\prime}(0)&=\frac{1}{12}\big(-g_A \langle r_A^2\rangle_{u-d}+\Delta \Sigma_{ud} \langle r_A^2\rangle_{u+d}\big)=-0.653(28)\,\text{GeV}^{-2}, 
\\
F_A^{s/p\prime}(0)&=\frac{1}{6}\Delta s \langle r_A^2\rangle_s=-0.09(9)\,\text{GeV}^{-2}.
\end{align}
Note that the errors on $F_A^{u/p}(0)$ and $F_A^{d/p}(0)$ are fully correlated (and similarly on $F_A^{u/p\prime}(0)$ and $F_A^{d/p\prime}(0)$).

For the induced pseudoscalar form factors $F_{P'}^{q/N}$ we use the expansion in \eqref{eq:F_PP'}.
At LO in Heavy Baryon Chiral Perturbation Theory (HBChPT) the residues of the pion- and eta-pole contributions to $F_{P'}^{q/N}$ are given by \cite{Bishara:2016hek}
\begin{align}
a_{P',\pi}^{u/p} &= -a_{P',\pi}^{d/p} = 2 g_A, \qquad a_{P',\pi}^{s/p} =
0,
\\
\label{eq:P':eta}
a_{P',\eta}^{u/p} &= a_{P',\eta}^{d/p} = -\frac{1}{2}a_{P',\eta}^{s/p}= \frac{2}{3}\big(\Delta \Sigma_{ud}-2 \Delta s\big)(1+\Delta_{\rm GT}^8),
\end{align}
where for the coefficients of the $\eta$ pole we also include the correction that was found in the lattice QCD study for the $u+d-2s$ current, $\Delta_{\rm GT}^8=0.50(14)$ \cite{Alexandrou:2021wzv}, and apply it as a common factor for each quark flavor, so that still only the octet current contributes to the $\eta$ pole (we use the $z$ expansion value, determination from a dipole ansatz agrees with it). For constant terms we write
\beq
b_{P'}^{q/N}=F_{P'}^{q/N}(0)-m_N^2\biggr(\frac{a_{P',\pi}^{q/N}}{m_\pi^2}+\frac{a_{P',\eta}^{q/N}}{m_\eta^2}\biggr),
\eeq
where $F_{P'}^{u/p}(0)= 119(7)$, $F_{P'}^{d/p}(0)= -130(17)$,  $F_{P'}^{s/p}(0)= -1.6(1.0)$ \cite{Alexandrou:2021wzv} (note that the continuum extrapolation does not satisfy FLAG criteria, so these results should still be treated as preliminary). 

For $\mu\to e$ conversion on ${}^{27}$Al we thus have 
\begin{align}\label{eq:FA:Al}
F_A^{u/p}\big|_{-q_{\rm eff}^2}&=0.818(20), & F_A^{d/p}\big|_{-q_{\rm eff}^2}&=-0.431(20),& F_A^{s/p}\big|_{-q_{\rm eff}^2}&=-4.4(9)\cdot 10^{-2},&
\\
F_{P'}^{u/p}\big|_{-q_{\rm eff}^2}&=69(7), &F_{P'}^{d/p}\big|_{-q_{\rm eff}^2}&=-80(17),& F_{P'}^{s/p}\big|_{-q_{\rm eff}^2}&=-1.5(1.1).&
\end{align}

\subsection{Scalar currents}
At zero recoil, the scalar form factors defined by Eq. \eqref{scalar:form:factor} are given by the nuclear sigma terms
\begin{equation}\label{eq:FS:def}
F_S^{q/N}(0)= \sigma_q^N\,.
\end{equation}
For $u$ and $d$ quarks their values follow from the pion-nucleon sigma term, $\sigma_{\pi N} = \langle N | \bar
m (\bar u u + \bar d d) | N \rangle$~\cite{Hoferichter:2023ptl}
\beq
\sigma_u^{p/n}=\frac{1}{2}\tilde \sigma_{\pi N} (1-\xi)\pm\hat c_5(1-1/\xi), \quad \sigma_d^{p/n}=\frac{1}{2}\tilde \sigma_{\pi N} (1+\xi)\pm\hat c_5(1+1/\xi),
\eeq
where $\tilde \sigma_{\pi N}=\sigma_{\pi N}+\delta\sigma_{\pi N}$, with $\delta\sigma_{\pi N}$ an isospin breaking correction, while $\xi=
(1-r_{ud})/(1+r_{ud})=0.357(6)$~\cite{ParticleDataGroup:2022pth}, where $r_{ud}=m_u/m_d$ is given in Eq. \eqref{eq:mass:ratios} and we shortened the product of low energy constants to $\hat c_5\equiv B c_5 (m_d-m_u)=-0.51(8)$\,MeV~\cite{Hoferichter:2023ptl}.  The isospin corrections $\delta\sigma_{\pi N}$ depend on what part of isospin breaking has been included in the extraction of $\sigma_{\pi N}$. The average of lattice results gives $\sigma_{\pi N}=44.2(2.6)$\,MeV \cite{Durr:2011mp,Alexandrou:2014sha,Durr:2015dna,Yang:2015uis,Agadjanov:2023efe,RQCD:2022xux,Gupta:2021ahb}, with $\delta\sigma_{\pi N}=-0.5(5)$\,MeV~\cite{Hoferichter:2023ptl}, which is significantly lower than the pion-atom based value $\sigma_{\pi N}=59.0(3.5)$\,MeV [for which  $\delta\sigma_{\pi N}=-3.6(2)$ MeV]~\cite{Hoferichter:2023ptl}. Averaging the two determinations, and inflating the errors according to the PDG prescription, gives
\beq
\tilde \sigma_{\pi N}=48(6)\,\text{MeV},
\eeq
from which
\begin{equation}
\label{eq:values:sigmaqN}
\begin{split}
\sigma_u^p=16.3(2.5)\,\text{MeV}\,, \qquad \sigma_d^p=30.6(4.2)\,\text{MeV}\,,
\\
\sigma_u^n=14.5(2.2)\,\text{MeV}\,, \qquad \sigma_d^n=34.5(4.0)\,\text{MeV}\,.
\end{split}
\end{equation}
 
For the strange quark nuclear sigma term the average of lattice QCD determinations is~\cite{RQCD:2022xux,Agadjanov:2023efe,Durr:2015dna,Junnarkar:2013ac,Durr:2011mp,Freeman:2012ry,Yang:2015uis} 
  \begin{equation}
  \label{eq:values:sigmas}
  \sigma_s^p=\sigma_s^n=(43.3\pm 4.8){\rm~MeV}\,.
  \end{equation}

The derivatives of the scalar form factors at zero recoil can be related to the quark contributions to the $A$ and $D$ gravitational form factors (we use the notation from \cite{Hackett:2023rif})
\beq
\label{eq:Fsprime}
F_S^{q/p\prime}(0)=m_N A_q'(0)-\frac{1}{4m_N}\Big[3 D_q(0)+A_q(0)-2 J_q(0)\Big], \qquad q=u,d,s.
\eeq

Using the dipole fit to the $q_{\rm rel.}^2$-dependent form factors from Ref. \cite{Hackett:2023rif} we then obtain
\beq
F_S^{u/p\prime}(0)=0.72(14)\,\text{GeV}^{-1}, \quad F_S^{d/p\prime}(0)=0.59(14)\,\text{GeV}^{-1}, \quad F_S^{s/p\prime}(0)=0.17(14)\,\text{GeV}^{-1}.
\eeq

This then translates to the following values of scalar form factors for $\mu\to e$ conversion on ${}^{27}$Al, 
\begin{equation}
\label{eq:values:sigmaqN:Al}
\begin{split}
F_S^{u/p}\big|_{-q_{\rm eff}^2}&=7.5(3.0)\,\text{MeV}, \qquad F_S^{d/p}\big|_{-q_{\rm eff}^2}=23.4(4.5)\,\text{MeV}, 
\\
F_S^{u/n}\big|_{-q_{\rm eff}^2}&=7.3(2.7)\,\text{MeV}, \qquad F_S^{d/n}\big|_{-q_{\rm eff}^2}=25.6(4.3)\,\text{MeV},
\end{split}
\end{equation}
and
\beq
F_S^{s/N}\big|_{-q_{\rm eff}^2}=41(5)\,\text{MeV}\,.
\eeq

\subsection{Pseudoscalar currents}
For the pseudoscalar form factors $F_P^{q/N}$ defined by Eq. \eqref{pseudoscalar:form:factor}, the LO HBChPT expressions for the residues of the pole are given by
\begin{align}
\label{eq:aPpi}
\frac{a_{P,\pi}^{u/p}}{m_u}&=-\frac{a_{P,\pi}^{d/p}}{m_d}=\frac{B_0}{m_N} g_A\,, \qquad \frac{a_{P,\pi}^{s/p}}{m_s}=0\,,
\\
\label{eq:aPeta}
\frac{a_{P,\eta}^{u/p}}{m_u}&=\frac{a_{P,\eta}^{d/p}}{m_d}=-\frac{1}{2}\frac{a_{P,\eta}^{s/p}}{m_s}=\frac{B_0}{3m_N}\big(\Delta u+\Delta d-2 \Delta s\big)\,,
\end{align}
where $B_0$ is
a ChPT constant related to the quark condensate given, up to
corrections of ${\mathcal O}(m_q)$, by $\langle \bar q q\rangle \simeq
-f^2 B_0$. In order to satisfy the PCAC relations \eqref{eq:PCAC1}, \eqref{eq:PCAC2}, we use the SU(3) flavor symmetric meson mass relations
\beq
m_\pi^2=B_0(m_u+m_d), \qquad m_\eta^2=\frac{B_0}{3}(m_u+m_d+4m_s),
\eeq
for the residues of the corresponding meson poles, and include the $\Delta_\mathrm{GT}^8$ correction factor, Eq. \eqref{eq:P':eta}, which then gives,
\begin{align}
a_{P,\pi}^{u/p}&=\frac{m_\pi^2}{m_N}\frac{1}{1+1/r_{ud}}g_A=7.9(7)\cdot 10^{-3} \,\text{GeV},
\\
a_{P,\pi}^{d/p}&=-\frac{m_\pi^2}{m_N}\frac{1}{1+r_{ud}}g_A=-16.8(7)\cdot 10^{-3}\,\text{GeV},
\\
a_{P,\pi}^{s/p}&=0,
\end{align}
and 
\begin{align}
a_{P,\eta}^{u/p}&=\frac{m_\eta^2}{m_N}\frac{1}{1+1/r_{ud}}\frac{1}{1+2r_s} \big(\Delta \Sigma_{ud} -2 \Delta s\big)\big(1+\Delta_{\rm GT}^8\big)=1.4(2)\cdot 10^{-4} \,\text{GeV},
\\
a_{P,\eta}^{d/p}&=\frac{m_\eta^2}{m_N}\frac{1}{1+r_{ud}}\frac{1}{1+2r_s} \big(\Delta \Sigma_{ud} -2 \Delta s\big)\big(1+\Delta_{\rm GT}^8\big)=2.9(4)\cdot 10^{-3} \,\text{GeV},
\\
a_{P,\eta}^{s/p}&=-\frac{m_\eta^2}{m_N}\frac{1}{2+1/r_s} \big(\Delta \Sigma_{ud} -2 \Delta s\big)\big(1+\Delta_{\rm GT}^8\big)=-0.11(1) \,\text{GeV},
\end{align}
where the ratios of the quark masses are \cite{ParticleDataGroup:2022pth}
\beq
\label{eq:mass:ratios}
r_{ud}=\frac{m_u}{m_d}=0.474(65), \qquad r_s=\frac{2 m_s}{m_u+m_d}=27.33(72).
\eeq

Using the PCAC expressions in Sec. \ref{app:PCAC} we obtain for the constant terms 
\begin{align}
\begin{split}
\label{eq:bPu}
b_P^{u/N}&=b_P^{d/N}=\frac{m_N}{2 r_s+1}\Big[\big(r_s-\frac{1}{2}\Delta_{\rm GT}^8\big)\Delta\Sigma_{ud}+\big(1+\Delta_{\rm GT}^8\big)\Delta s\Big]
\\
&\qquad\qquad+\frac{m_N}{3}\frac{(1-r_{ud})}{(1+r_{ud})}g_A-m_N\tilde m\sum_{q=u,d,s}\frac{\Delta q}{m_q},
\end{split}
\\
\label{eq:bPs}
b_P^{s/N}&=b_P^{u/N}+\frac{m_N}{2}\Delta_{\rm GT}^8\big(\Delta u+\Delta d-2 \Delta s\big),
\end{align}

where $1/\tilde m=(1/m_u+1/m_d+1/m_s)$.
Numerically, $b_P^{u/N}=b_P^{d/N}=-0.070(12)\,$GeV and $b_P^{s/N}=0.044(35)\,$GeV.

For $\mu\to e$ conversion on ${}^{27}$Al thus 
\beq
F_P^{u/p}\big|_{-q_{\rm eff}^2}=0.16(3)\,\text{GeV}, \quad F_P^{d/p}\big|_{-q_{\rm eff}^2}=-0.55(3)\,\text{GeV}, \quad F_P^{s/p}\big|_{-q_{\rm eff}^2}=-0.28(2)\,\text{GeV}.
\eeq

\subsection{CP-even gluonic current}

For evaluating the matrix element of the CP-even gluonic current in Eq. \eqref{CPeven:gluonic:form:factor} we can use the relation with the trace of the stress-energy tensor $T_\mu^\mu=(\beta/2 \alpha_s) G_{\mu \nu}^a
G^{a \mu\nu}+\sum_{u,d,s}(1+2\gamma_m) m_q\bar q q$ \cite{Shifman:1978zn,Collins:1976yq} (see also discussions in \cite{Ji:1994av,Ji:2021mtz,Wang:2024lrm,Liu:2021gco}), where $\beta=-(b_0 \alpha_s^2+b_1 \alpha_s^3+\cdots)$, $\gamma_m=\gamma_1 (\alpha_s/4\pi)+\gamma_2  (\alpha_s/4\pi)^2+\cdots$, where $b_0=(33-2n_f)/12\pi=27/12\pi$, 
$b_1=(153-19n_f)/(24\pi^2)$, $\gamma_1=4$, $\gamma_2=202/3-20 n_f/9$, and the results for higher orders can be found in \cite{ParticleDataGroup:2022pth}. The nucleon matrix element of the trace of the stress-energy tensor at zero momentum-transfer is $\langle N|T_\mu^\mu|N\rangle=m_N\bar u_Nu_N$, which then gives 
\begin{equation}\label{eq:FG:def}
F_G^{N}(0)=-\frac{2m_G}{27}\frac{1}{1+(b_1\alpha_s+\cdots)/b_0}=-50.4(6)\,\text{MeV},
\end{equation}
where $m_G=m_N-(1+2\gamma_m)\sum_q \sigma_q^N=823(10)$\,MeV, where we used $\alpha_s(2 \,\text{GeV})=0.297(6)$ and averaged over values for $N=n,p$ since the difference is much smaller than the error.

In terms of the gravitational form factors 
\begin{equation}
\begin{split}
    F_G^{\prime N}(0)=-\frac{2}{27}\frac{m_N}{1+(b_1\alpha_s+\cdots )/b_0}\biggr[&A'_g(0)+A'_q(0) -\frac{3}{4m_N^2}\big(D_g(0)+D_q(0)\big) 
    \\
    &-\frac{(1+2\gamma_m)}{m_N}\sum_q F_S^{\prime q/N}(0) \biggr]
    \\
    &=-0.14(5)\text{GeV}^{-1}.
\end{split}
\end{equation}

This then gives 
\beq
\quad F_G^{N}\big|_{-q_{\rm eff}^2}=-48.7(0.9)\,\text{MeV}\,.
\eeq

\subsection{CP-odd gluonic current}
For the hadronic matrix element of the CP-odd gluonic current in Eq. \eqref{CPodd:gluonic:form:factor}, we use the leading-order HBChPT expression \cite{Bishara:2017pfq},
\begin{equation}
\begin{split}\label{FGtilde:LO}
F_{\tilde G}^{N}(q_{\rm rel.}^2)=-\tilde m m_N\Big[&\frac{\Delta u}{m_u}+\frac{\Delta d}{m_d}+\frac{\Delta s}{m_s}+\frac{g_A}{2}\Big(\frac{1}{m_u}-\frac{1}{m_d}\Big)\frac{q_{\rm rel.}^2}{m_\pi^2-q_{\rm rel.}^2}
\\
&+\frac{1}{6}\big(\Delta u+\Delta d-2 \Delta s)\Big(\frac{1}{m_u}+\frac{1}{m_d}-\frac{2}{m_s}\Big)\frac{q_{\rm rel.}^2}{m_\eta^2-q_{\rm rel.}^2}\Big],
\end{split}
\end{equation}
where $1/\tilde m=(1/m_u+1/m_d+1/m_s)$. Using the values and expression for $\Delta q$ in  \eqref{eq:Deltau-d}-\eqref{eq:Deltas}, along with ratios of masses in \eqref{eq:mass:ratios}, gives 
\beq
\quad F_{\tilde G}^{N}\big|_{-q_{\rm eff}^2}=-0.306(28)\,\text{GeV}\,.
\eeq

\subsection{Tensor current}
The matrix elements of the tensor current in Eq. \eqref{tensor:form:factor} are described by three sets of form factors,  related to the
generalized tensor form factors~\cite{Gockeler:2005cj,Diehl:2001pm,Bishara:2017pfq}. At zero recoil we have $
\hat F_{T,0}^{q/p}(0)=  g_{T}^{q} $, 
with  the lattice QCD results \cite{Gupta:2018lvp,FlavourLatticeAveragingGroupFLAG:2021npn} (see also \cite{Park:2020axe})
\beq
\label{eq:gTq}
g_T^u=0.784(30),\qquad g_T^d=-0.204(15), \qquad g_T^s=-2.7(1.6)\cdot 10^{-3}.
\eeq
In Ref. \cite{Hoferichter:2018zwu} these results were used, combined with the assumption of pole dominance, as well as constraints from analyticity and unitarity,  to obtain the zero-recoil values of the remaining two tensor form factors, as well as the derivatives at $q_{\rm rel.}^2=0$. We use the values in \eqref{eq:gTq}, with the remaining inputs given in Table I of \cite{Hoferichter:2018zwu}, together with the translation to our notation, $\hat F_{T,0}^{q/p}=F_{1,T}^{q}$, $\hat F_{T,1}^{q/p}/2=F_{2,T}^{q}$, $\hat F_{T,2}^{q/p}=-F_{3,T}^{q}$, to obtain for  $\hat F_{T,i}^{q/N}\big|_{-q_{\rm eff}^2}$
\begin{align}
\hat F_{T,0}^{u/p} &=0.777(30),& \hat F_{T,0}^{d/p}&=-0.203(15),& \hat F_{T,0}^{s/N}&=-2.7(1.6)\cdot10^{-3},&
\\
\hat F_{T,1}^{u/p}&=-2.8(2.0),&\hat F_{T,1}^{d/p}&=0.9(6),& \hat F_{T,1}^{s/N}&=1.7(1.0)\cdot10^{-2},&
\\
\hat F_{T,2}^{u/p}&=-0.08(20),&\hat F_{T,2}^{d/p}&=0.57(30),& \hat F_{T,2}^{s/N}&=3.8(3.0)\cdot10^{-3}.&
\end{align}

\begin{table}
\begin{center}
\small
  {\renewcommand{\arraystretch}{1.35}
\begin{tabular}{cccccc}
\hline\hline
parameter & value & parameter & value & parameter & value
 \\
\hline
$\mu_p$			 &  $2.792847$ 	          & $\mu_n$ 		          & $-1.91304$                   & $\mu_s$                 & $-0.036(21)$
\\
$\big[ r_E^{p} \big]^2$ & $0.7071(7)\,\text{fm}^2$ & $\big[ r_E^{n} \big]^2$ & $-0.1155(17)\,$fm$^2$ & $\big[r_E^s\big]^2$ & $ -0.0045(14)\,$fm$^2$
\\
$\big[ r_M^{p} \big]^2$  &$0.724(45)\,$fm$^2$  & $\big[ r_M^{n} \big]^2 $ & $ 0.743(16)\,$fm$^2$ & $\big[r_M^s\big]^2$ & $-0.010(11)\,$fm$^2$
\\[1mm]
\hline
$g_A$			&  $1.2754(13)$		& $ \Delta \Sigma_{ud}$     & $0.397(40)$			& $\Delta s$ 		& $-0.045(9)$
\\
$\langle r_A^2\rangle_{u-d}$ & $0.392(28)\,\text{fm}^2$, &  $\langle r_A^2\rangle_{u+d}$ & $0.49(31)\,\text{fm}^2$ &  $\langle r_A^2\rangle_{s}$ & $0.48(48)\,\text{fm}^2$
\\
$F_{P'}^{u/p}(0)$  & $119(7)$ & $F_{P'}^{d/p}(0)$ & $ -130(17)$ & $F_{P'}^{s/p}(0)$  & $-1.6(1.0)$
\\
$\Delta_{\rm GT}^8$ & $0.50(14)$ & & &
\\
\hline
$\tilde \sigma_{\pi N}$ & $ 48(6)\,$MeV &  $\hat c_5$ & $-0.51(8)$ & $ \sigma_s^N$ & $43.3(4.8){\rm~MeV}$
\\
$F_S^{u/p\prime}(0)$ & $0.72(14)\,\text{GeV}^{-1}$ & $F_S^{d/p\prime}(0)$ & $0.59(14)\,\text{GeV}^{-1}$ &  $F_S^{s/p\prime}(0)$ & $0.17(14)\,\text{GeV}^{-1}$
\\
$F_G^N(0)$ & $-50.4(6)\,\text{MeV}$ & $F_G^{\prime N}(0)$ & $-0.14(5)\,\text{GeV}^{-1}$ & $\alpha_s(2 \,\text{GeV}) $ & $0.297(6)$ 
\\
\hline
$g_T^u$ & $0.784(30)$ & $ g_T^d$ & $-0.204(15)$ & $g_T^s$ & $-2.7(1.6)\cdot 10^{-3}$
\\
$\hat F_{T,0}^{'u/p}(0)$  & $0.54(11)\,\text{GeV}^{-2}$ & $\hat F_{T,0}^{'d/p}(0)$ & $-0.11(2)\,\text{GeV}^{-2}$ & $\hat F_{T,0}^{'s/N}(0)$ & $-0.0014(9) \,\text{GeV}^{-2} $ 
\\
$\hat F_{T,1}^{u/p}(0)$ & $-3.0(2.0) $ & $\hat F_{T,1}^{d/p}(0)$ & $1.0(6)$ & $\hat F_{T,1}^{s/N}(0)$ & $0.018(10)$
\\
$\hat F_{T,1}^{'u/p}(0)$ & $-14.0(1.6)\,\text{GeV}^{-2} $ & $\hat F_{T,1}^{'d/p}(0)$ & $5.0(6)\,\text{GeV}^{-2}$ & $\hat F_{T,1}^{'s/N}(0)$ & $0.082(52)\,\text{GeV}^{-2}$
\\
$\hat F_{T,2}^{u/p}(0)$ & $-0.1(2) $ & $\hat F_{T,2}^{d/p}(0)$ & $0.6(3)$ & $\hat F_{T,2}^{s/N}(0)$ & $0.004(3)$
\\
$\hat F_{T,2}^{'u/p}(0)$ & $-1.8(2)\,\text{GeV}^{-2} $ & $\hat F_{T,2}^{'d/p}(0)$ & $2.1(2)\,\text{GeV}^{-2}$ & $\hat F_{T,2}^{'s/N}(0)$ & $0.015(13)\,\text{GeV}^{-2}$
\\
\hline
$r_{ud}$ &$0.474(65)$ & $ r_s$ & $27.33(72)$ &  &
\\
\hline 
$F_\gamma^p(0)$ & {$4.7(2.6) \cdot 10^{-7}$ GeV} & {$F_\gamma^n(0)$} & {$1.5(0.5) \cdot 10^{-6}$ GeV} &  &  \\
{$F_{\tilde{\gamma}}^p(0)$} &  $3.83(3) \cdot 10^{-6}~\mathrm{GeV}$ & {$F_{\tilde{\gamma}}^n(0)$} &  $-3.9(7) \cdot 10^{-7}~\mathrm{GeV}$ & & 
\\
\hline
\hline
\end{tabular}
}
\end{center}
\caption{ \label{tab:inputs} Numerical values of input parameters that enter the expressions for matrix elements of vector, axial, scalar, pseudoscalar, and tensor quark currents, CP-even and CP-odd gluonic currents, as well as the CP-even and CP-odd Rayleigh operators. See the main text for details and references.
}
\end{table}

\subsection{Rayleigh operators} 
The matrix element of the CP-even Rayleigh operator at zero momentum transfer, $F_{\gamma}^N(0)$ in Eq. \eqref{CPeven:photon:form:factor}, can be obtained from the spin-averaged forward nucleon matrix element (we use the same normalization as \cite{Walker-Loud:2012ift})
\beq
\label{eq:TN:munu}
T_{N}^{\mu\nu} \bar u_N u_N=\frac{i}{2}\sum_s\int d^4 x~ e^{i q\cdot x}\langle N(k,s)| T\{J^{\mu}_{\rm e.m.}(x), J^\nu_{\rm e.m.}(0)\big\}|N(k,s)\rangle,
\eeq
where $N=p,n$, while $J^{\mu}_{\rm e.m.} =\sum_{q=u,d,s} Q_q \bar q \gamma^\mu q$, with $Q_q=\{2/3, -1/3, -1/3\}$ for $q=\{u,d,s\}$. The above matrix element can be decomposed
as
\beq
\label{eq:WNmunu}
\begin{split}
T_{N}^{\mu\nu}=\biggr\{&\biggr(-g^{\mu\nu}+\frac{q^\mu q^\nu}{q^2}\biggr)T_1^N(\nu, Q^2)
\\
&+\frac{1}{m_N^2}\biggr(k^\mu-\frac{k\cdot q q^\mu}{q^2}\biggr)\biggr(k^\nu -\frac{k\cdot q q^\nu}{q^2}\biggr) T_2^N (\nu, Q^2)\biggr\},
\end{split}
\eeq
with $Q^2=-q^2$ and $\nu =  k\cdot q/m_N$. 

The hadronic matrix element  $F_\gamma^N(0)$ follows from $FF$ operator insertion, attaching the two photon lines  to $T_N^{\mu\nu}$, giving the one-loop expression
\beq
F_\gamma^N(0)=-i \frac{\alpha^2}{6 \pi m_N } \int_R \frac{d^4 q}{(2\pi)^3} \frac{T_N^{\mu\nu} g_{\mu\nu}}{q^2+i\epsilon},
\eeq
where the integral over $d^4q$ is regularized (see below). 
The value of $F_\gamma^N(0)$ is directly proportional to the electromagnetic self-energy of the nucleon $\delta M_N^\gamma$ (see, e.g., Eq. (3) in \cite{Walker-Loud:2012ift}), 
\beq
F_\gamma^N(0)=-\frac{\alpha}{3\pi} \delta M_N^\gamma.
\eeq
For the numerical evaluation of $F_\gamma^N(0)$ we can thus use the appropriately rescaled Cottingham's sum rule for $\delta M_N^\gamma$ 
\beq
F_\gamma^N(0)=-\frac{\alpha}{3\pi} \Big[\delta M_N^{\rm el}+\delta M_N^{\rm inel}+\delta M_N^{\rm sub}+\delta M_N^{\rm ct}\Big],
\eeq
where $\delta M_N^{\rm el}$ is the elastic contribution, $\delta M_N^{\rm inel}$ the inelastic contribution, $\delta M_N^{\rm sub}$ the contribution due to 
$T_1$ requiring a once subtracted dispersion relation, and $\delta M_N^{\rm ct}$ the  counter-terms that can be obtained using operator product expansion (OPE) at large $Q^2$.
The explicit expressions for  $\delta M_N^{\rm el}$, $\delta M_N^{\rm inel}$, and $\delta M_N^{\rm sub}$ are given in Eqs.~(11), (12) and (13) in \cite{Walker-Loud:2012ift}, and depend on a regularization scale $\Lambda_0$ separating the hadronic and the deep inelastic regimes, while OPE results can be found in \cite{Hill:2016bjv}. 
Numerically,
\beq
\label{eq:Fgamma:numerics}
F_\gamma^p(0)=4.7 (2.6) \cdot 10^{-7}\,\text{GeV},\qquad F_\gamma^n(0)=1.5 (0.5) \cdot 10^{-6}\,\text{GeV},
\eeq
where we only included the elastic contribution and the subtractions from elastic and inelastic terms. The errors are squared summed errors due to parameterizations of elastic electric and magnetic form factors (we use the parameterization and numerical values in \cite{Bradford:2006yz}), the magnetic polarizabilities, $\beta_p=2.5(4)\cdot 10^{-4}$\,fm${}^3$, $\beta_n=3.7(1.2)\cdot 10^{-4}$\,fm${}^3$~\cite{ParticleDataGroup:2022pth}, and the variations of the cut-off scale regulating the dispersion integrals, as well as the modeling of $\delta M_N^{\rm sub}$,  where we follow the prescriptions in \cite{Walker-Loud:2012ift}. The omitted inelastic and counter term contributions are expected to be percent level corrections, i.e., much smaller than the above errors on our estimates and can thus be safely neglected.

Since the change in vector form factors $F_{1,2}^N$ going from $Q^2=0$ to $Q^2=-q_{\rm eff}^2$ is at the level of ${\mathcal O}(5\%)$, it is reasonable to expect a similar change also in $F_\gamma^{N}$, which is much smaller than the errors in \eqref{eq:Fgamma:numerics}.
 In the numerics we therefore set $F_\gamma^{N}(-q_{\rm eff}^2)$ to the numerical values for $F_\gamma^{N}(0)$ in \eqref{eq:Fgamma:numerics}.

For the matrix element of the CP-odd Rayleigh operator at zero momentum transfer, $F_{\tilde \gamma}^N(0)$ in Eq.~\eqref{CPodd:photon:form:factor}, we can use the spin-dependent part of the forward double virtual Compton scattering tensor 
\beq
\begin{split}
 \bar u_N M^{\mu\nu}_N u_N 
= i\int d^4 x~ e^{i q\cdot x}\langle N(k,{s'})| T\{J^{\mu}_{\rm e.m.}(x), J^\nu_{\rm e.m.}(0)\big\}|N(k,s)\rangle.
\end{split}
\eeq
The scattering tensor $M^{\mu\nu}_N$ can be decomposed into a symmetric spin-independent piece $T^{\mu\nu}_N$ and an antisymmetric spin-dependent piece $S^{\mu\nu}_N$ as
\begin{equation}
    M^{\mu\nu}_N = T^{\mu\nu}_N + S^{\mu\nu}_N,
\end{equation}
where $T_N^{\mu\nu}$ is given in \eqref{eq:TN:munu}, while the $S^{\mu\nu}_N$ is given by \cite{Lensky:2017dlc} 
\beq \label{eq:antisym_compton}
S_N^{\mu\nu}=\frac{i}{m_N}\epsilon^{\mu\nu \alpha\beta}\Big[q_\alpha s_{N\beta} S_1(\nu, Q^2)+\frac{1}{m_N^2}q_\alpha \big(k\cdot q s_{N\beta}-s_N\cdot q k_\beta\big) S_2(\nu, Q^2)\Big],
\eeq
with $s_N$ being the nucleon spin vector, satisfying $s_N\cdot k=0$, $s_N^2=-1$. Following the same procedure as was done for the CP-even case, the hadronic matrix element $F_{\tilde{\gamma}}^N(0)$ can be obtained from an $F\tilde{F}$ operator insertion, attaching the two photon lines to $M_N^{\mu\nu}$, giving the one-loop amplitude 
\begin{equation}
\begin{split}
    \braket{N(k,s')|\frac{\alpha}{8\pi}F^{\mu\nu}\tilde{F}_{\mu\nu}|N(k,s)}& = -i\frac{\alpha^2}{\pi} \int \frac{d^4q}{(2\pi)^3}\frac{\bar{u}_N \epsilon_{\mu\nu\alpha\beta} q^{\alpha}q'^{\beta} S^{\mu\nu}_N u_N}{q^4},
    \end{split}
\end{equation}
where $q^{\alpha}$ is the loop momentum and $q'^{\beta}=q_\mathrm{rel}^{\beta}=\left(p_e-p_\mu\right)^{\beta}$ is the four-momentum transfer. This is the only place where we retain non-zero momentum transfer. The rest of the loop is evaluated in the forward limit. Plugging in Eq.~(\ref{eq:antisym_compton}), the terms proportional to $S_2$ either vanish identically or are odd under $q^{\alpha} \rightarrow -q^{\alpha}$ and therefore vanish in the integral. Using the non-zero contribution
\begin{equation}
    \epsilon_{\mu\nu\alpha\beta} q^{\alpha}q'^{\beta} S^{\mu\nu}_N \rightarrow \frac{2i}{m_N}\left[q^2 (q'\cdot s_N) - (q \cdot q')(q \cdot s_N) \right]S_1 = \frac{3i}{2m_N} q^2 (q'\cdot s_N) S_1,
\end{equation}
and the relation 
\begin{equation}
    \bar u_N i\gamma_5 u_N \simeq \frac{i}{2 m_N} (q_{\rm rel}\cdot s_N) \bar u_N  u_N,
\end{equation}
we find the following expression for the hadronic matrix element
\begin{equation}
    F_{\tilde{\gamma}}^N(0)=
    \frac{3i\alpha^2}{2\pi}\int \frac{d^4q}{(2\pi)^3}\frac{S_1(\nu,Q^2)}{q^2}.
\end{equation}
Performing the Wick rotation $q^0 \rightarrow i \nu$ and the variable transformation $Q^2 = \vec{q}\,{}^2 + \nu^2$ the matrix element becomes
\begin{equation}
    F_{\tilde{\gamma}}^N(0)=
    \frac{3\alpha^2}{8\pi^3}\int_0^{\infty}dQ^2\int_{-Q}^Qd\nu\frac{\sqrt{Q^2-\nu^2}}{Q^2}S_1(i\nu,Q^2).
\end{equation}
Using the Born-approximation, $S_1$ may be written as \cite{Lensky:2017dlc}
\begin{equation}
    S_1(i\nu,Q^2)=-\frac{1}{2m_N}\left[F_2(Q^2)^2-\frac{Q^2}{\nu^2+\nu_B^2}F_1(Q^2)G_M(Q^2)\right].
\end{equation}
Using this form, the final integral is given by
\begin{equation}
    F_{\tilde{\gamma}}^N(0)=-\frac{3\alpha^2}{16\pi^2m_N}\int_0^{\infty}dQ^2\left[\frac{(G_E-G_M)^2}{2(1+\tau_\mathrm{el})^2}+\left(1-\frac{\sqrt{1+\tau_\mathrm{el}}}{\sqrt{\tau_\mathrm{el}}}\right)\frac{(G_E+\tau_\mathrm{el} G_M)G_M}{1+\tau_\mathrm{el}}\right],
\end{equation}
where $\tau_\mathrm{el}=Q^2/4m_N^2$. This integral is both UV and IR finite, and thus does not require regularization. Using the same form factor parameterization, Ref. \cite{Bradford:2006yz}, as the CP-even calculation,  we find (in the Born approximation)
\begin{equation}
    \begin{split}
        F_{\tilde{\gamma}}^p&= 3.83(3)\times 10^{-6}~\mathrm{GeV},\qquad     F_{\tilde{\gamma}}^n= -3.9(7)\times 10^{-7}~\mathrm{GeV}.
    \end{split}
\end{equation} 
Note that the errors include only uncertainties on $G_{E,M}$ form factors, while we do not attempt to quantify the error due to the neglected inelastic contributions.

\subsection{Partial conservation of axial current}
\label{app:PCAC}
The matrix elements of the axial, Eq.~\eqref{axial:form:factor}, pseudoscalar, Eq.~\eqref{pseudoscalar:form:factor}, and CP-odd gluonic current, Eq.~\eqref{CPodd:gluonic:form:factor}, are related to each other through partial conservation of axial current (PCAC) and through the QCD chiral anomaly. For zero momentum exchange these relations can be derived by performing a chiral rotation of the quark fields, $q\to
\exp(i\beta\gamma_5 )q$, where $\beta$ is a diagonal $3\times3$ matrix, which results in a shift in  the QCD Lagrangian, 
\beq
\label{eq:chiral:shift}
{\cal L}_{\rm QCD}\to {\cal L}_{\rm QCD}+2\Tr\beta\, \frac{\alpha_s}{8\pi} G_{\mu\nu}^a \tilde G^{a\mu\nu} - \partial_\mu\big(\bar q \gamma^\mu \gamma_5 \beta q\big)+2 \bar q i\gamma_5\beta {\cal M}_q q,
\eeq
where ${\cal M}_q=\text{diag}\,(m_u,m_d,m_s)$ and $q=(u,d,s)$ is a vector of light quarks.
Requiring that the results are independent of this phase shift implies an operator identity: the sum of the three last terms in \eqref{eq:chiral:shift} vanishes for any $\beta$, and thus
\begin{equation}\label{eq:beta}
-\Tr\beta\,\langle  N'|\frac{\alpha_s}{8 \pi}G_{\mu\nu}^a
\tilde G^{a\mu\nu}|N\rangle= \langle N'| \bar q i\gamma_5 \beta {\cal M}_q
q|N\rangle - \frac{1}{2} \partial_\mu \langle N'|  \bar q
\gamma^\mu \gamma_5 \beta q|N\rangle \,.
\end{equation}
We can then write the following relation among form factors
\beq
\Tr(\beta)F_{\tilde G}^N=\Tr\big(\beta F_P^{q/N}\big)-m_N \Tr(\beta F_A^{q/N}\big)-\frac{q_{\rm rel.}^2}{4 m_N}\Tr\big(\beta F_{P'}^{q/N}\big),
\eeq
which are valid also for $q_{\rm rel.}^2\ne 0$, if $\Tr\beta=0$, and valid only for $q_{\rm rel.}^2= 0$, if $\Tr\beta\ne 0$. Note that above, we 
use the short-hand notation $ F_i^{q/N}=\text{diag}\,\big(F_i^{u/N}, F_i^{d/N}, F_i^{s/N}\big)$, $i=A,P,P'$.

Choosing $\beta=\text{diag}\,(1,-1,0)$ and $\beta=\text{diag}\,(1,1,-2)$ gives the PCAC relations, which are valid for any $q_{\rm rel.}^2$,
\begin{align}
\label{eq:PCAC:u-d}
F_P^{u/N}-F_P^{d/N}&=m_N\big(F_A^{u/N}-F_A^{d/N}\big)+\frac{q_{\rm rel.}^2}{4m_N}\big(F_{P'}^{u/N}-F_{P'}^{d/N}\big)
\\
\begin{split}
\label{eq:PCAC:u+d-2s}
F_P^{u/N}+F_P^{d/N}-2F_P^{s/N}&=m_N\big(F_A^{u/N}+F_A^{d/N}-2F_A^{s/N}\big)
\\
&\quad+\frac{q_{\rm rel.}^2}{4m_N}\big(F_{P'}^{u/N}+F_{P'}^{d/N}-2F_{P'}^{s/N}\big),
\end{split}
\end{align}
while for $\beta=\text{diag}\,(1,1,1)$ one obtains the relation, valid only for $q_{\rm rel.}^2=0$,
\begin{equation}
\label{eq:PCAC:u+d+s}
\sum_q F_P^{q/N}(0)= 3 F_{\tilde G}^{N}(0) + \sum_q m_N F_A^{q/N}(0)\,.
\end{equation}

Using the $q_{\rm rel.}^2$ expansion of form factors in \eqref{eq:Fi} and \eqref{eq:F_PP'} in the relations \eqref{eq:PCAC:u-d}, \eqref{eq:PCAC:u+d-2s}, matching the pole structures on LHS and RHS, and using the fact that $u-d$ ($u+d-2s$) current only has the $\pi$ ($\eta$) pole, 
gives the following relations for the residues,
\begin{align}
\label{eq:PCAC1}
a_{P,\pi}^{u/N}-a_{P,\pi}^{d/N}&=\frac{m_\pi^2}{4m_N} \big(a_{P',\pi}^{u/N}-a_{P',\pi}^{d/N}\big),
\\
\label{eq:PCAC2}
a_{P,\eta}^{u/N}+a_{P,\eta}^{d/N}-2 a_{P,\eta}^{s/N}&=\frac{m_\eta^2}{4m_N} \big(a_{P',\eta}^{u/N}+a_{P',\eta}^{d/N}-2a_{P',\eta}^{s/N}\big).
\end{align}
One can easily check that our expressions for $a_{P,M}^{q/N}$, $a_{P',M}^{q/N}$ satisfy the above relations. The $q_{\rm rel.}^2$ independent terms give
\begin{align}
\frac{1}{m_N}\big(b_P^{u/N}-b_P^{d/N}\big)&=F_{A}^{u/N}(0)-F_{A}^{d/N}(0)-\frac{1}{4} \big(a_{P',\pi}^{u/N}-a_{P',\pi}^{d/N}\big),
\\
\begin{split}
\frac{1}{m_N}\big(b_P^{u/N}+b_P^{d/N}-2b_P^{s/N}\big)&=F_{A}^{u/N}(0)+F_{A}^{d/N}(0)-2F_{A}^{s/N}(0)
\\
&\quad-\frac{1}{4} \big(a_{P',\eta}^{u/N}+a_{P',\eta}^{d/N}-2a_{P',\eta}^{s/N}\big).
\end{split}
\end{align}
Using the above two relations, together with \eqref{eq:PCAC:u+d+s} (setting $N=p$), as well as  $F_A^{q/p}(0)=\Delta q$ and the expression for $F_{\tilde G}^p$ in  \eqref{FGtilde:LO}, and for $F_P^{q/N}$ in \eqref{eq:F_PP'} with \eqref{eq:aPpi}, \eqref{eq:aPeta}, one can then solve for $b_P^{q/N}$, with the results given in Eqs. \eqref{eq:bPu} and \eqref{eq:bPs}.

\section{WET translation}\label{app:wet-translation}
In Tables \ref{tab:HHMRZ_JMS_basis} and \ref{tab:HHMRZ_JMS_translation} we list and outline explicitly the translation between a variant of the dimension-6 WET--3 flavor basis of Jenkins, Manohar, and Stoffer~\cite{Jenkins:2017jig} to our basis in Eqs.~(\ref{eq:dim6:Q1Q2:light})--(\ref{eq:dim6:Q9Q10:light}). These naming convention and translations are used within \texttt{MuonConverter} to interface with the existing SMEFT/WET software.
\begin{table*}[p!h!] 
\caption{The \texttt{WCxf}-compatible operators and Wilson coefficient naming schemes for \texttt{MuonConverter} (left) and the \texttt{WET-3 JMS} basis \cite{Aebischer:2017ugx} (right). The translation between the two bases can be found in Table \ref{tab:HHMRZ_JMS_translation}.}
\label{tab:HHMRZ_JMS_basis}
\centering
\def\arraystretch{1.0}
\begin{tabular}{|r|c|}
\multicolumn{2}{c}{\texttt{MuonConverter}} \\
\hline
WC name & Operator \\
\hline 
\texttt{Tegamma\_12} -- $\hat{\mathcal{C}}_1^{(5)}$ & ($\bar e \sigma^{\alpha\beta} \mu) F_{\alpha\beta}$ \\
\texttt{ATegamma\_12} -- $\hat{\mathcal{C}}_2^{(5)}$ & ($\bar e \sigma^{\alpha\beta} i\gamma_5 \mu) F_{\alpha\beta}$ \\
\texttt{VVeu\_1211} -- $\hat{\mathcal{C}}_{1,u}^{(6)}$ & $(\bar e \gamma_\alpha \mu) (\bar u \gamma^\alpha u)$\\
\texttt{VVed\_1211} -- $\hat{\mathcal{C}}_{1,d}^{(6)}$ & $(\bar e \gamma_\alpha \mu) (\bar d \gamma^\alpha d)$\\
\texttt{VVed\_1222} -- $\hat{\mathcal{C}}_{1,s}^{(6)}$ & $(\bar e \gamma_\alpha \mu) (\bar s \gamma^\alpha s)$\\
\texttt{AVVeu\_1211} -- $\hat{\mathcal{C}}_{2,u}^{(6)}$ & $(\bar e\gamma_\alpha\gamma_5 \mu)(\bar u \gamma^\alpha u)$\\
\texttt{AVVed\_1211} -- $\hat{\mathcal{C}}_{2,d}^{(6)}$ & $(\bar e\gamma_\alpha\gamma_5 \mu)(\bar d \gamma^\alpha d)$\\
\texttt{AVVed\_1222} -- $\hat{\mathcal{C}}_{2,s}^{(6)}$ & $(\bar e\gamma_\alpha\gamma_5 \mu)(\bar s \gamma^\alpha s)$\\
\texttt{VAVeu\_1211} -- $\hat{\mathcal{C}}_{3,u}^{(6)}$& $(\bar e \gamma_\alpha \mu)(\bar u \gamma^\alpha \gamma_5 u)$\\
\texttt{VAVed\_1211} -- $\hat{\mathcal{C}}_{3,d}^{(6)}$& $(\bar e \gamma_\alpha \mu)(\bar d \gamma^\alpha \gamma_5 d)$\\
\texttt{VAVed\_1222} -- $\hat{\mathcal{C}}_{3,s}^{(6)}$& $(\bar e \gamma_\alpha \mu)(\bar s \gamma^\alpha \gamma_5 s)$\\
\texttt{AVAVeu\_1211} -- $\hat{\mathcal{C}}_{4,u}^{(6)}$& $(\bar e \gamma_\alpha\gamma_5 \mu)(\bar u \gamma^\alpha \gamma_5 u)$\\
\texttt{AVAVed\_1211} -- $\hat{\mathcal{C}}_{4,d}^{(6)}$& $(\bar e \gamma_\alpha\gamma_5 \mu)(\bar d \gamma^\alpha \gamma_5 d)$\\
\texttt{AVAVed\_1222} -- $\hat{\mathcal{C}}_{4,s}^{(6)}$& $(\bar e \gamma_\alpha\gamma_5 \mu)(\bar s \gamma^\alpha \gamma_5 s)$\\
\texttt{SSeu\_1211} -- $\hat{\mathcal{C}}_{5,u}^{(6)}$& $(\bar e \mu)( \bar u u)$\\
\texttt{SSed\_1211} -- $\hat{\mathcal{C}}_{5,d}^{(6)}$& $(\bar e \mu)( \bar d d)$\\
\texttt{SSed\_1222} -- $\hat{\mathcal{C}}_{5,s}^{(6)}$& $(\bar e \mu)( \bar s s)$\\
\texttt{ASeu\_1211} -- $\hat{\mathcal{C}}_{6,u}^{(6)}$& $(\bar e i \gamma_5 \mu)( \bar u u)$\\
\texttt{ASed\_1211} -- $\hat{\mathcal{C}}_{6,d}^{(6)}$& $(\bar e i \gamma_5 \mu)( \bar d d)$\\
\texttt{ASed\_1222} -- $\hat{\mathcal{C}}_{6,s}^{(6)}$& $(\bar e i \gamma_5 \mu)( \bar s s)$\\
\texttt{SAeu\_1211} -- $\hat{\mathcal{C}}_{7,u}^{(6)}$& $(\bar e \mu) (\bar u i \gamma_5 u)$\\
\texttt{SAed\_1211} -- $\hat{\mathcal{C}}_{7,d}^{(6)}$& $(\bar e \mu) (\bar d i \gamma_5 d)$\\
\texttt{SAed\_1222} -- $\hat{\mathcal{C}}_{7,s}^{(6)}$& $(\bar e \mu) (\bar s i \gamma_5 s)$\\
\texttt{AAeu\_1211} -- $\hat{\mathcal{C}}_{8,u}^{(6)}$& $(\bar e i \gamma_5 \mu)(\bar u i \gamma_5 u)$\\
\texttt{AAed\_1211} -- $\hat{\mathcal{C}}_{8,d}^{(6)}$& $(\bar e i \gamma_5 \mu)(\bar d i \gamma_5 d)$\\
\texttt{AAed\_1222} -- $\hat{\mathcal{C}}_{8,s}^{(6)}$& $(\bar e i \gamma_5 \mu)(\bar s i \gamma_5 s)$\\
\texttt{TTeu\_1211} -- $\hat{\mathcal{C}}_{9,u}^{(6)}$& $(\bar e \sigma^{\alpha\beta} \mu) (\bar u \sigma_{\alpha\beta} u)$\\
\texttt{TTed\_1211} -- $\hat{\mathcal{C}}_{9,d}^{(6)}$ & $(\bar e \sigma^{\alpha\beta} \mu) (\bar d \sigma_{\alpha\beta} d)$\\
\texttt{TTed\_1222} -- $\hat{\mathcal{C}}_{9,s}^{(6)}$& $(\bar e \sigma^{\alpha\beta} \mu) (\bar s \sigma_{\alpha\beta} s)$\\
\texttt{ATTeu\_1211} -- $\hat{\mathcal{C}}_{10,u}^{(6)}$& $(\bar e  i \sigma^{\alpha\beta} \gamma_5 \mu)(\bar u \sigma_{\alpha\beta} u)$\\
\texttt{ATTed\_1211} -- $\hat{\mathcal{C}}_{10,d}^{(6)}$& $(\bar e  i \sigma^{\alpha\beta} \gamma_5 \mu)(\bar d \sigma_{\alpha\beta} d)$\\
\texttt{ATTed\_1222} -- $\hat{\mathcal{C}}_{10,s}^{(6)}$& $(\bar e  i \sigma^{\alpha\beta} \gamma_5 \mu)(\bar s \sigma_{\alpha\beta} s)$\\
\hline
\end{tabular}
\quad
\begin{tabular}{|c|c|}
\multicolumn{2}{c}{\texttt{WET-3 JMS} \cite{Jenkins:2017jig, Aebischer:2017ugx}} \\
\hline
WC name & Operator \\
\hline 
\texttt{egamma\_12} & $\bar e_L \sigma^{\alpha\beta} \mu_R F_{\alpha\beta}$\\
\texttt{VeuLL\_1211} & $(\bar e_L \gamma_\alpha \mu_L) (\bar u_L \gamma^\alpha u_L)$\\
\texttt{VedLL\_1211} & $(\bar e_L \gamma_\alpha \mu_L) (\bar d_L \gamma^\alpha d_L)$\\
\texttt{VedLL\_1222} & $(\bar e_L \gamma_\alpha \mu_L) (\bar s_L \gamma^\alpha s_L)$\\
\texttt{VeuRR\_1211} & $(\bar e_R \gamma_\alpha \mu_R) (\bar u_R \gamma^\alpha u_R)$\\
\texttt{VedRR\_1211} & $(\bar e_R \gamma_\alpha \mu_R) (\bar d_R \gamma^\alpha d_R)$\\
\texttt{VedRR\_1222} & $(\bar e_R \gamma_\alpha \mu_R) (\bar s_R \gamma^\alpha s_R)$\\
\texttt{VeuLR\_1211} & $(\bar e_L \gamma_\alpha \mu_L) (\bar u_R \gamma^\alpha u_R)$\\
\texttt{VedLR\_1211} & $(\bar e_L \gamma_\alpha \mu_L) (\bar d_R \gamma^\alpha d_R)$\\
\texttt{VedLR\_1222} & $(\bar e_L \gamma_\alpha \mu_L) (\bar s_R \gamma^\alpha s_R)$\\
\texttt{VueLR\_1112} & $(\bar e_R \gamma_\alpha \mu_R) (\bar u_L \gamma^\alpha u_L)$\\
\texttt{VdeLR\_1112} & $(\bar e_R \gamma_\alpha \mu_R) (\bar d_L \gamma^\alpha d_L)$\\
\texttt{VdeLR\_2212} & $(\bar e_R \gamma_\alpha \mu_R) (\bar s_L \gamma^\alpha s_L)$\\
\texttt{SeuRL\_1211} & $(\bar e_L \mu_R)( \bar u_R u_L)$\\
\texttt{SedRL\_1211} & $(\bar e_L \mu_R)( \bar d_R d_L)$\\
\texttt{SedRL\_1222} & $(\bar e_L \mu_R)( \bar s_R s_L)$\\
\texttt{SeuRR\_1211} & $(\bar e_L \mu_R)( \bar u_L u_R)$\\
\texttt{SedRR\_1211} & $(\bar e_L \mu_R)( \bar d_L d_R)$\\
\texttt{SedRR\_1222} & $(\bar e_L \mu_R)( \bar s_L s_R)$\\
\texttt{TeuRR\_1211} & $(\bar e_L \sigma^{\alpha\beta} \mu_R) (\bar u_L \sigma_{\alpha\beta} u_R)$\\
\texttt{TedRR\_1211} & $(\bar e_L \sigma^{\alpha\beta} \mu_R) (\bar d_L \sigma_{\alpha\beta} d_R)$\\
\texttt{TedRR\_1222} & $(\bar e_L \sigma^{\alpha\beta} \mu_R) (\bar s_L \sigma_{\alpha\beta} s_R)$\\
\hline
\end{tabular}
\end{table*}

\newpage
\begin{table*}[p!h!]
\caption{The translation between the \texttt{MuonConverter} basis and the \texttt{WET-3 JMS} basis.\\}
\label{tab:HHMRZ_JMS_translation}
\vspace{0.1in}
\centering
\def\arraystretch{1.12}
\begin{tabular}{|c|c|}
\hline
\texttt{MuonConverter} & \texttt{WET-3} translation \\
\hline 
\texttt{Tegamma\_12} &  $4 \pi^2$ \texttt{egamma\_12}\\
\texttt{ATegamma\_12} &  $-4 \pi^2 i$ \texttt{egamma\_12}\\
\texttt{VVeu\_1211} & (\texttt{VeuLL\_1211} + \texttt{VeuRR\_1211} + \texttt{VeuLR\_1211} + \texttt{VueLR\_1112})$ / 4$\\
\texttt{VVed\_1211} & (\texttt{VedLL\_1211} + \texttt{VedRR\_1211} + \texttt{VedLR\_1211} + \texttt{VdeLR\_1112})$ / 4$\\
\texttt{VVed\_1222} & (\texttt{VedLL\_1222} + \texttt{VedRR\_1222} + \texttt{VedLR\_1222} + \texttt{VdeLR\_2212})$ / 4$\\
\texttt{AVVeu\_1211} & ($-$\texttt{VeuLL\_1211} + \texttt{VeuRR\_1211} $-$ \texttt{VeuLR\_1211} $+$ \texttt{VueLR\_1112})$ / 4$\\
\texttt{AVVed\_1211} &($-$\texttt{VedLL\_1211} + \texttt{VedRR\_1211} $-$ \texttt{VedLR\_1211} $+$ \texttt{VdeLR\_1112})$ / 4$ \\
\texttt{AVVed\_1222} & ($-$\texttt{VedLL\_1222} + \texttt{VedRR\_1222} $-$ \texttt{VedLR\_1222} $+$ \texttt{VdeLR\_2212})$ / 4$ \\
\texttt{VAVeu\_1211} & ($-$\texttt{VeuLL\_1211} + \texttt{VeuRR\_1211} $+$ \texttt{VeuLR\_1211} $-$ \texttt{VueLR\_1112})$ / 4$\\
\texttt{VAVed\_1211} & ($-$\texttt{VedLL\_1211} + \texttt{VedRR\_1211} $+$ \texttt{VedLR\_1211} $-$ \texttt{VdeLR\_1112})$ / 4$\\
\texttt{VAVed\_1222} & ($-$\texttt{VedLL\_1222} + \texttt{VedRR\_1222} $+$ \texttt{VedLR\_1222} $-$ \texttt{VdeLR\_2212})$ / 4$\\
\texttt{AVAVeu\_1211} & (\texttt{VeuLL\_1211} + \texttt{VeuRR\_1211} $-$ \texttt{VeuLR\_1211} $-$ \texttt{VueLR\_1112})$ / 4$\\
\texttt{AVAVed\_1211} & (\texttt{VedLL\_1211} + \texttt{VedRR\_1211} $-$ \texttt{VedLR\_1211} $-$ \texttt{VdeLR\_1112})$ / 4$\\
\texttt{AVAVed\_1222} & (\texttt{VedLL\_1222} + \texttt{VedRR\_1222} $-$ \texttt{VedLR\_1222} $-$ \texttt{VdeLR\_2212})$ / 4$\\
\texttt{SSeu\_1211} & (\texttt{SeuRL\_1211} + \texttt{SeuRR\_1211})$ / 4$\\
\texttt{SSed\_1211} & (\texttt{SedRL\_1211} + \texttt{SedRR\_1211})$ / 4$\\
\texttt{SSed\_1222} & (\texttt{SedRL\_1222} + \texttt{SedRR\_1222})$/ 4$\\
\texttt{ASeu\_1211} & $-i$(\texttt{SeuRL\_1211} + \texttt{SeuRR\_1211})$/4$\\
\texttt{ASed\_1211} & $-i$(\texttt{SedRL\_1211} + \texttt{SedRR\_1211})$/4$\\
\texttt{ASed\_1222} & $-i$(\texttt{SedRL\_1222} + \texttt{SedRR\_1222})$/4$\\
\texttt{SAeu\_1211} & $i$(\texttt{SeuRL\_1211} $-$ \texttt{SeuRR\_1211})$/4$\\
\texttt{SAed\_1211} & $i$(\texttt{SedRL\_1211} $-$ \texttt{SedRR\_1211})$/4$\\
\texttt{SAed\_1222} & $i$(\texttt{SedRL\_1222} $-$ \texttt{SedRR\_1222})$/4$\\
\texttt{AAeu\_1211} & (\texttt{SeuRL\_1211} $-$ \texttt{SeuRR\_1211})$/4$\\
\texttt{AAed\_1211} & (\texttt{SedRL\_1211} $-$ \texttt{SedRR\_1211})$/4$\\
\texttt{AAed\_1222} & (\texttt{SedRL\_1222} $-$ \texttt{SedRR\_1222})$/4$\\
\texttt{TTeu\_1211} & \texttt{TeuRR\_1211}$/4$\\
\texttt{TTed\_1211} & \texttt{TedRR\_1211}$/4$\\
\texttt{TTed\_1222} & \texttt{TedRR\_1222}$/4$\\
\texttt{ATTeu\_1211} & $-i$\texttt{TeuRR\_1211}$/4$\\
\texttt{ATTed\_1211} & $-i$\texttt{TedRR\_1211}$/4$\\
\texttt{ATTed\_1222} & $-i$\texttt{TedRR\_1222}$/4$\\
\hline
\end{tabular}
\end{table*}

\clearpage

\bibliographystyle{JHEP}
\bibliography{mu2e_tower}
  
\end{document}